\documentclass[showkeys]{revtex4}
\usepackage{graphicx}
\usepackage{dcolumn}
\usepackage{amsmath}
\usepackage{color}
\usepackage{epstopdf}
\usepackage{float}
\usepackage{epsfig}
\usepackage{graphicx}
\usepackage{subfigure}
\usepackage{tikz}
\usepackage[utf8x]{inputenc}
\definecolor{coolblack}{rgb}{0.0, 0.18, 0.39}
\definecolor{darkred}{rgb}{0.5,0,0}
\definecolor{darkgreen}{rgb}{0,0.5,0}
\definecolor{darkblue}{rgb}{0,0,0.5}
\definecolor{lapislazuli}{rgb}{0.15, 0.38, 0.61}
\definecolor{venetianred}{rgb}{0.78, 0.03, 0.08}
\definecolor{bleudefrance}{rgb}{0.19, 0.55, 0.91}
\definecolor{dogwoodrose}{rgb}{0.84, 0.09, 0.41}
\usepackage[linktocpage,colorlinks]{hyperref}
\hypersetup{colorlinks=true, citecolor=darkgreen, linkcolor=blue,urlcolor = darkblue}
\makeatletter
\def\btt#1{\texttt{\@backslashchar#1}}
\DeclareRobustCommand\bblash{\btt{\@backslashchar}} \makeatother

\RequirePackage{fix-cm}

\begin{document}

\title{Shadow and weak gravitational lensing of a rotating regular black hole in a non-minimally coupled Einstein-Yang-Mills theory in the presence of plasma 
}
\author{Shubham Kala $^{a}$}\email{shubhamkala871@gmail.com}
\author{ Hemwati Nandan $^{b,c}$}\email{hnandan@associates.iucaa.in}
\author{Prateek Sharma $^{d}$}\email{prteeksh@gmail.com}

\affiliation{$^{a}$Department of Applied Sciences (Physics), Quantum University, Roorkee-247 167, Uttarakhand, India}
\affiliation{$^{b}$Department of Physics, Gurukula Kangri (Deemed to be University), Haridwar 249 404, Uttarakhand, India}
\affiliation{$^{c}$Center for Space Research, North-West University, Mahikeng 2745, South Africa}
\affiliation{$^{d}$Department of Physics, Chandigarh University, Mohali, Punjab 140413, India}

%
\begin{abstract}
The null geodesics of the regular and rotating magnetically charged black hole in a non-minimally coupled Einstein-Yang-Mills theory surrounded by a plasma medium is studied. The effect of magnetic charge and Yang-Mills parameter on the effective potential and radius of photon orbits has investigated. We then study the shadow of a regular and rotating magnetically charged black hole along with the observables in presence of the plasma medium. The presence of plasma medium affects the apparent size of the shadow of a regular rotating black hole in comparison to vacuum case. Variation of shadow radius and deformation parameter with Yang-Mills and plasma parameter has examined. Furthermore, the deflection angle of the massless test particles in weak field approximation around this black hole spacetime in presence of homogeneous plasma medium is also investigated. Finally, we have compared the obtained results with Kerr-Newman  and Schwarzschild black hole solutions in general relativity (GR).
%
\end{abstract}
\maketitle
\section{Introduction}
\label{sec:1}
Black holes (BHs) those make an appearance as the solution of Einstein's Field Equations (EFEs) are one of the most striking compact objects predicted by General Relativity (GR) and other alternative theories of gravity. \cite{Hartle:2003yu,joshi1993global,chandrasekhar1998mathematical}. The first direct image of a BH in the centre of galaxy M87 has been observed recently by Event Horizon Telescope (EHT) collaboration using a very long baseline interferometer (VLBI) \cite{Akiyama:2019cqa,Akiyama:2019brx,Akiyama:2019sww,Akiyama:2019bqs,Akiyama:2019fyp,Akiyama:2019eap}. The discussions on theoretical aspects of a BH shadow began with Synge who introduced the concept of escape cone \cite{synge1960relativity}. Synge has investigated the shape and size of a shadow for a non-rotating BH and also obtained the formula of angular radius assuming a static observer at infinity \cite{synge1966escape}. Unlike a non-rotating BH, the shadow of a rotating BH does not remain circular and it could be deformed in the presence of a rotation parameter. Bardeen was the first person who extended the concept of shadow to a rotating BH and obtained an accurate procedure to quantify the effect of rotation on a shadow considering Kerr BH. So far, the number of investigations have been performed to enhance our understanding of BH shadows in GR as well as other modified theories of gravity over the past four decades \cite{takahashi2005black,amir2016shapes,amir2018shadows,cunha2017fundamental,gralla2019black,stuchlik2018light,moffat2015modified,guo2020innermost,sharif2016shadow,zeng2020shadows,Papnoi:2014aaa,Sharma:2021von,Kala:2020prt,Peng:2020wun,Wang:2021ara,Junior:2021dyw,Hioki:2009na,Amarilla:2010zq,Kumar:2019pjp,khodadi2020black,contreras2019black,contreras2021geodesic,wang2020shadow,kumar2020black,jusufi2021black,atamurotov2016horizon,zhang2020optical,dastan2016shadow,jha2021bumblebee,atamurotov2021axion}.\\
The study of complex environments outside the BHs, such as plasma medium, the jets or the accretion disk has become the field of crucial interest for researchers in recent years. Plasma is a dispersive medium and the light rays are usually refracted by the dispersive medium before they reached our eyes. In this context, it is interesting to investigate the effect of plasma medium in the background of BHs as well as other compact objects. The study of the shadow of Schwarzschild and Kerr BH has been established a remarkable examination to probe their properties with a plasma medium \cite{perlick2015influence,atamurotov2015optical} by applying the Synge method \cite{synge1960relativity}. Further, the behavior of the plasma medium of the same was investigated using a specific approach \cite{perlick2017light}. In order to investigate whether the presence of plasma around any BH leads to any observational effects, several investigations have been carried out time and again \cite{yan2019influence,ahmedov2019optical,abdujabbarov2017shadow,babar2020optical,das2020shadow,zaman2020optical,huang2018revisiting,li2021gravitational,atamurotov2016observing}.\\
Gravitational lensing (GL) can provide useful information about the properties of a BH spacetime. Weak GL in which the geometry of spacetime is less favourable has been proven to be one of the powerful tools in astrophysics and cosmology, in testing theories of gravity and in exploring interaction beyond the Einstein-Maxwell theory \cite{liu2019probing,schneider1992gravitational,schneider2006gravitational,petters2012singularity,keeton2005formalism,collett2018precise,cao2018weak}. Synge was the first person to develop a self-consistent approach to the light propagation in the gravitational field, in presence of a plasma medium \cite{synge1965relativity}. The application of Synge’s general relativistic Hamiltonian theory for the geometrical optics have investigated by Bi$ \check{c}\grave{a} $k  and Hadrava \cite{bicak1975general}. Thereafter, Perlick implements Synge's theory and introduce a ray optics method in a plasma medium, and obtained the deflection angle of Schwarzschild BH and Kerr BH in presence of spherically distributed plasma medium \cite{perlick2015influence}. Bisnovatyi-Kogan and Tsupko further, investigated  non-linear effects connected with the combined action of gravity and plasma medium by using Synge's theory \cite{bisnovatyi2010gravitational,bisnovatyi2017gravitational}. Moreover, the effect of plasma medium around BHs on lensing effects has been studied in \cite{crisnejo2018weak,javed2020weak,atamurotov2021weak,matsuno2021light,babar2020optical,babar2021gravitational,babar2021particle,jha2021optical,tsupko2021deflection,ahmedov2019optical,yan2019influence,rogers2015frequency,jin2020strong,fathi2020gravitational,hensh2019gravitational,kimpson2019spatial} and references therein.
The study of modified theories of gravity (MGT) has been of great interest because it provides numerous static and spherically symmetric BH solutions \cite{rezzolla2014new,sullivan2020numerical,nashed2019charged,moffat2015modified,aghmohammadi2010spherical,carames2009spherically,moffat2015black}. However, due to the complexity of non-linear partial differential equations, an exact solution of a rotating BH by solving the coupled field equations in any MGT model is still not achieved.  In particular, one can obtain the metric of stationary and axis-symmetric BHs by applying the Newman-Janis algorithm (NJA) \cite{newman1965note} and its implementation by starting with any static and spherically symmetric spacetime \cite{azreg2014generating}. The non complexification procedure can be obtained by implementing these modifications to NJA. Subsequently, this method has been extensively used to obtain rotating BH solutions \cite{li2021gravitational, azreg2014regular,azreg2014static,toshmatov2017rotating,xu2017kerr,toshmatov2017comments,shaikh2019black,contreras2020black,kumar2020rotating,jusufi2020rotating,benavides2020rotating}.\\
In addition to Einstein-Maxwell theory of gravity where the $U(1)$ electromagnetic fields are used to describe the matter \cite{yang1954conservation}, a more generalized and geometrically rich picture is however emerged in Einstein-Yang-Mills (EYM) theory of gravity where the solutions of EYM equations with non-Abelian gauge fields (viz $SU(2)$, $SU(3)$) to describe matter are studied in diverse contexts \cite{bartnik1988particlelike,bjoraker2000monopoles,jaffe2006quantum,Altas:2021htf}. There also exists a number of BH solutions in EYM theory of gravity \cite{Kleihaus:1996vk,Kleihaus:1997rb,Kleihaus:2002ee,Nandan:2009kt,Bezares-Roder:2009lns} with an essentially non-Abelian gauge structure. Furthermore, the gauge group $SU(2)$, a shred of observational evidence about the existence of various BH solutions to the EYM theory for any event horizon is provided in \cite{smoller1993existence}. The numerically rotating BHs in the minimally coupled EYM theory along with non-static spherically symmetric EYM BHs and the slowly rotating non-abelian BHs have been studied previously in \cite{volkov1997slowly,kleihaus2002rotating,kleihaus2011rotating,ghosh2010radiating}. Introducing NGA formalism, a new regular rotating BH solution has been also derived recently with a YM electromagnetic theory \cite{jusufi2021quasinormal}. Our aim here is to study the shadow and WL to observe the effect of various BH parameters involved on the shadow and lensing in the background of a homogeneous plasma medium.\\
The organization of the present paper is as follows. In  Sect.~\ref{sec:3}, we have obtained the equations of motion by the Hamilton-Jacobi method around a regular and rotating magnetically charged BH in a non-minimally coupled EYM theory surrounded by a plasma medium. In Sect.~\ref{sec:4}, we have examined the effective potential and radius of the photon sphere with their pictorial representation. We have obtained the necessary analytic expression for the radius of shadow and obtained various shadow images from different inclination angles and also examined observables in Sect.~\ref{sec:5}. The analysis of the energy emission rate by taking into account a non-rotating charged BH in EYM theory is also incorporated in subsection of Sect.~\ref{sec:5}. Sect.~\ref{sec:6}, includes a detailed analysis of the deflection angle in a weak-field approximation. Finally, in Sect.~\ref{sec:7}, we have concluded our results obtained. 
\section{Properties of a rotating regular BH in non-minimally coupled EYM theory}\label{sec:2}
In Boyer-Lindquist coordinates, a regular and rotating magnetically charged BH with a YM electromagnetic source in the non-minimal EYM theory is characterized by the line element \cite{jusufi2021quasinormal},\\

	\begin{equation}
		ds^{2} = -c^2\left(1-\frac{2 P(r) r}{\Sigma}\right)dt^{2} -2ac sin^{2}{\theta}\frac{2 P(r) r}{\Sigma} dt d{\phi} + \frac{\Sigma}{\Delta} dr^{2} + \Sigma d{\theta}^{2} +  \frac{\left[(r^{2}+a^{2})^{2}-a^{2} \Delta sin^{2}{\theta} \right] sin^{2}{\theta}}{\Sigma} d{\phi}^{2}, \label{e1}
	\end{equation}

where,
\begin{equation}
	P(r) = \frac{r(1-g(r))}{2} ,               \label{e2}
\end{equation}

\begin{equation}
	\Sigma = r^{2} + a^{2} cos^{2} \theta   ,           \label{e3}
\end{equation}

\begin{equation}
	g(r) = 1 + \left(\frac{r^{4}}{r^{4} + 2\lambda}\right) \left(-\frac{2GM}{c^{2}r} + \frac{GQ^{2}}{4\pi{\epsilon}_{0}c^{4}r^{2}}\right)  ,             \label{e4}
\end{equation}
and
\begin{equation}
	\Delta(r) = r^{2} -\left(\frac{r^{4}}{r^{4} + 2\lambda}\right) \left(-\frac{2GM}{c^{2}r} +\frac{GQ^{2}}{4\pi{\epsilon}_{0}c^{4}r^{2}}\right) + a^{2}.            \label{5}
\end{equation}
Here $\lambda = \sigma Q^{2}$ is the YM parameter while Q is the magnetic charge parameter. The parameters $M$ and $a$ represents the total mass and spin of the BH, respectively. The BH metric in Eq. \ref{e1} represents an exact solution of the EFEs in Einstein YM theory \cite{jusufi2021quasinormal} and reduced to the Kerr-Newman BH with a magnetic charge if $\lambda = 0$ and it further reduces to Schwarzschild BH in GR if $\lambda = 0$ and $Q=0$. The horizons of this BH spacetime can be obtained by solving $\Delta(r) = 0$ and its ergo-surface can be calculated via $g_{tt} =0$. The behavior of event horizon and ergo-surface of this BH spacetime recently has been studied detail in \cite{jusufi2021quasinormal}. The YM parameter and charge parameter has an evident influence on horizon and ergoregion of the BH. However, it has observed that this solution represents a compact object without horizons and singularities at the centre since beyond the critical value of angular momentum parameter, the horizons of BH no longer  exists.\\

\begin{figure*}[h]
	\centering
		\subfigure[]{\includegraphics[width=7cm,height=6cm]{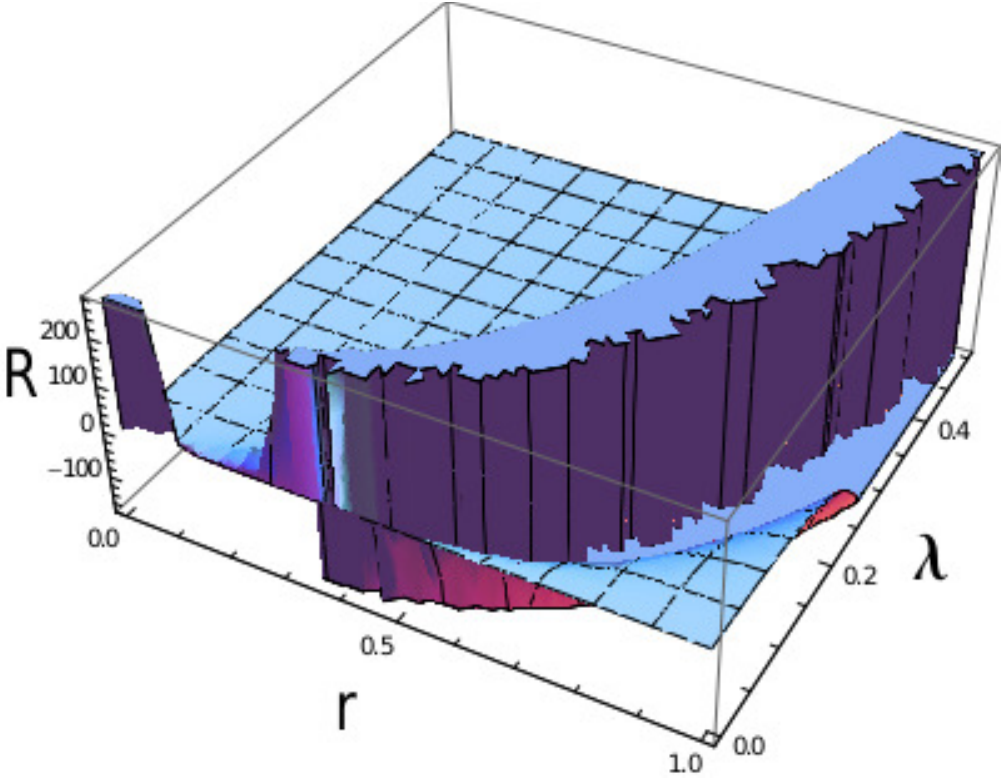}} 
		\subfigure[]{\includegraphics[width=7cm,height=6cm]{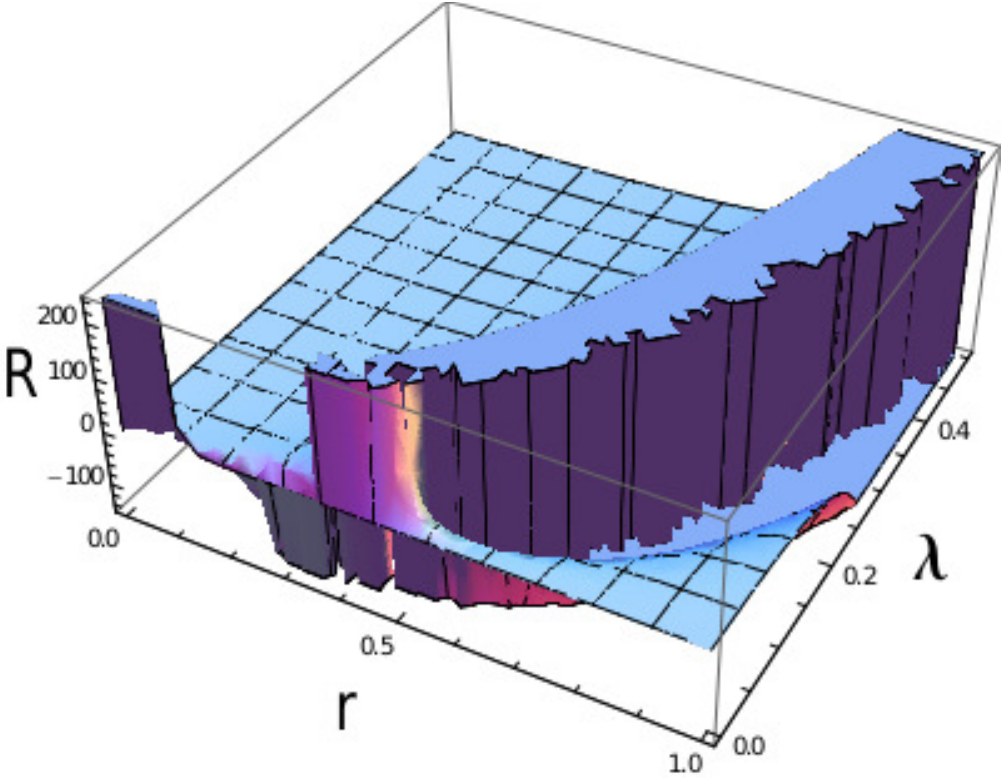}}
	\caption{The variation of Ricci Scalar with radial distance and YM parameter for (a) $Q=0.3$ and (b) $Q=0.6$.} \label{f1.01}
\end{figure*}

The curvature invariants play an important role to understand the geometrical properties of BH spacetime. In order to investigate the curvature singularity in this BH spacetime, we obtain the curvature invariants of a rotating regular BH in a non-minimally coupled EYM theory. The curvature invariant is thus given as
\begin{equation}
	R = \frac{8 \lambda r^2\left[Q^2\left(5 r^4-6\lambda\right)+ M \left(-6 r^5+20\left(5 r^4-6\lambda r\right)\lambda\right)\right]}{\left(r^{2} + a^{2} cos^{2} \theta \right)\left(r^4 + 2 \lambda\right)^3},
\end{equation}
and
\begin{equation}
	R^{\alpha\beta\mu\nu}R_{\alpha\beta\mu\nu} = \frac{P\left(r,cos^{2}, M, Q^2, a^2, \lambda \right)}{\left(r^{2} + a^{2} cos^{2} \theta \right)^6\left(r^4 + 2 \lambda\right)^6}.
\end{equation}
Here, $P$ represents a finite polynomial of its arguments. Further, the behavior of  Ricci scalar with radial distance and YM parameter is shown in \figurename{ \ref{f1.01}}, for two distinct values of charge parameter. From \figurename{ \ref{f1.01}}, it is clearly seen that the curvature scalar is regular and BH spacetime is singularity free for $\lambda > 0$.\\
In order to investigate the effect of various parameter on the event horizon, one can analyze the structure of horizon and static limit surface. The corresponding horizon radius of this BH spacetime can be obtained by calculating roots of the following equation
\begin{equation}
	\Delta(r) = r^{2} -\left(\frac{r^{4}}{r^{4} + 2\lambda}\right) \left(-\frac{2M}{r} +\frac{Q^{2}}{r^{2}}\right) + a^{2} =0, 
\end{equation}
and the corresponding static limit surface can also be obtained by $g_{tt}=0$,  i.e.,
\begin{equation}
	r^{2} g(r) + a^{2} cos^{2} \theta =0 .
\end{equation}

\begin{figure*}[h]
	\centering
		\subfigure[]{\includegraphics[width=7cm,height=6cm]{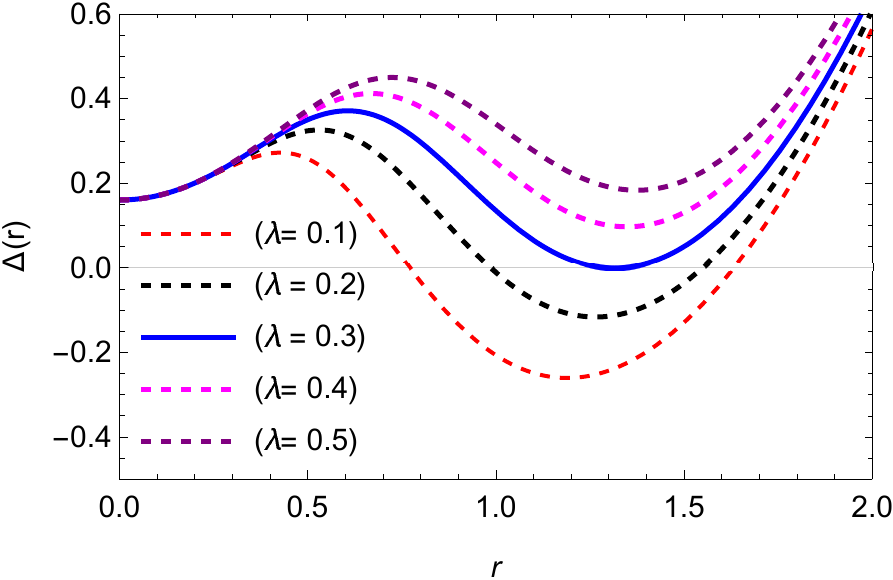}} 
		\subfigure[]{\includegraphics[width=7cm,height=6cm]{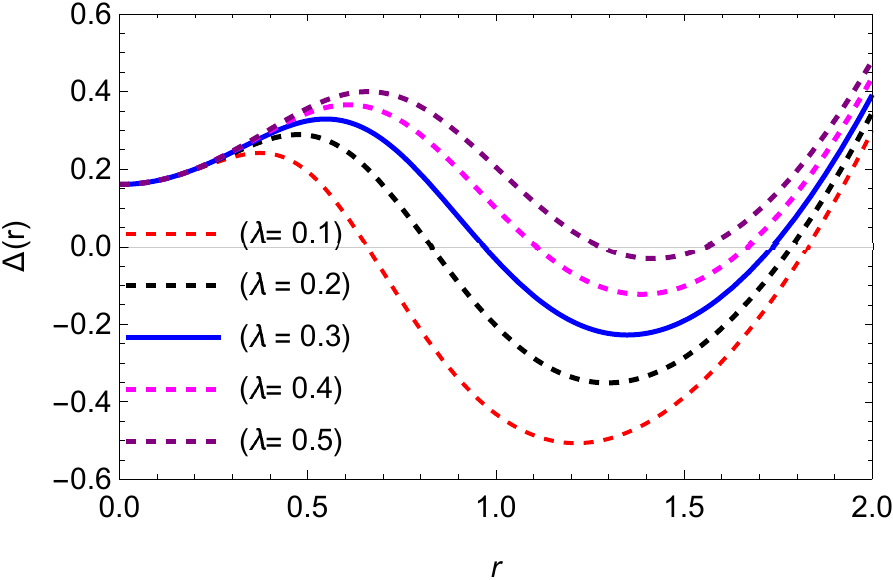}}\\
		\subfigure[]{\includegraphics[width=7cm,height=6cm]{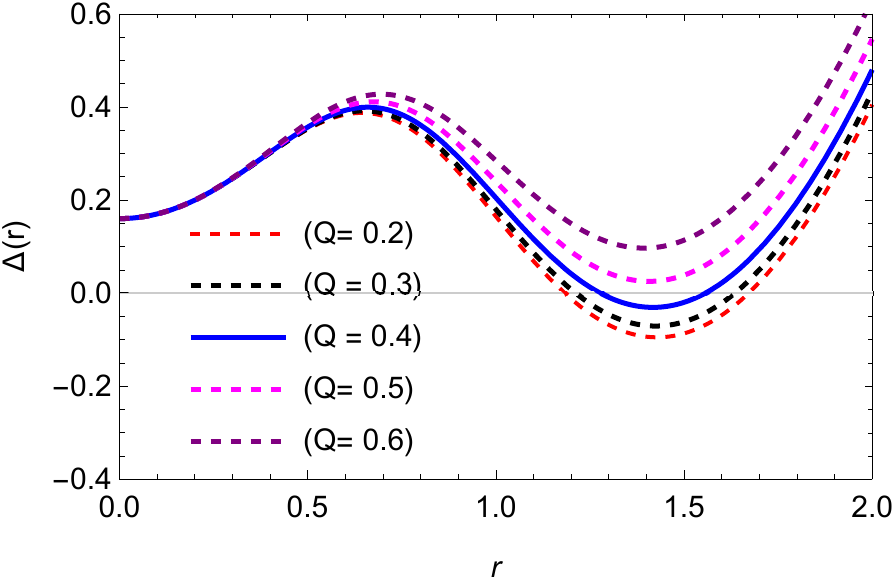}} 
		\subfigure[]{\includegraphics[width=7cm,height=6cm]{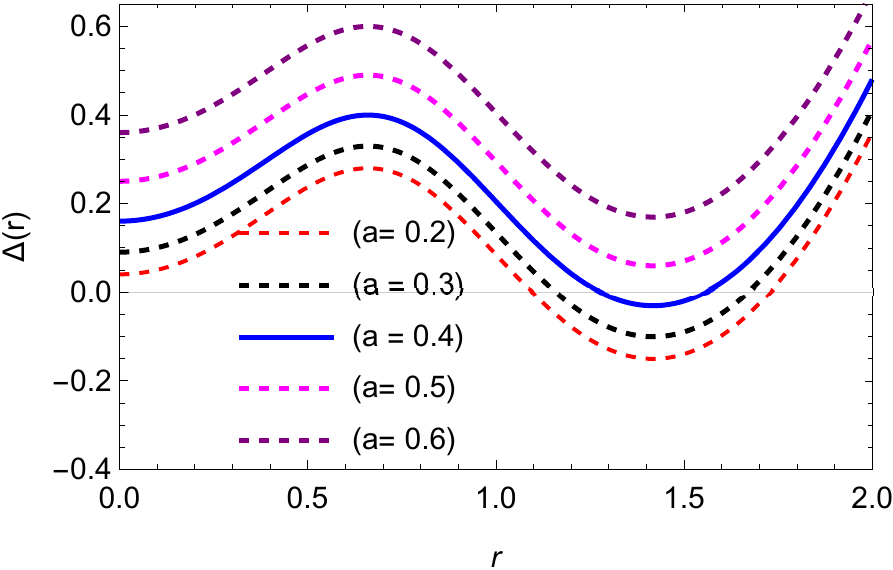}}
	\caption{Variation of $\Delta(r)$ with radial distance for different values of parameters involved in the BH spacetime.} \label{f1.1}
\end{figure*}
\begin{figure*}[h]
	\centering
		\subfigure[]{\includegraphics[width=7cm,height=6cm]{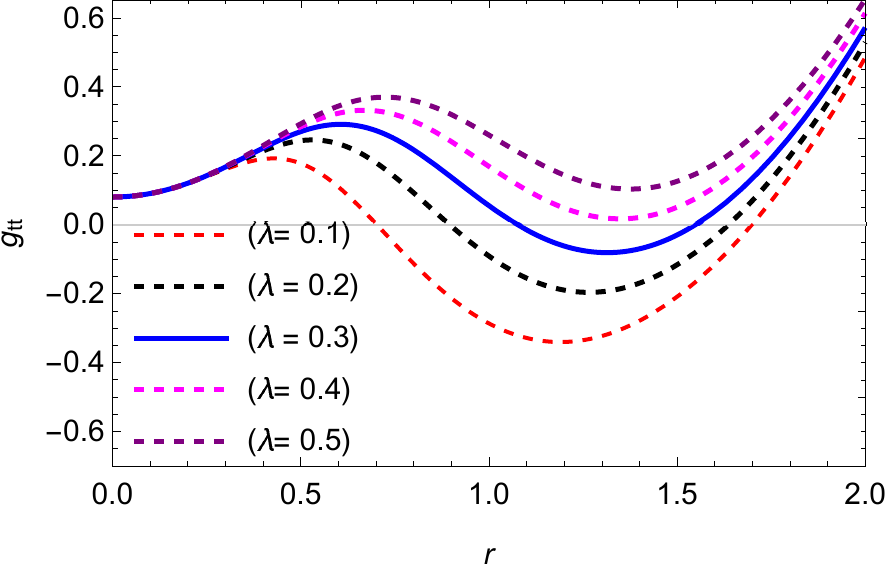}} 
		\subfigure[]{\includegraphics[width=7cm,height=6cm]{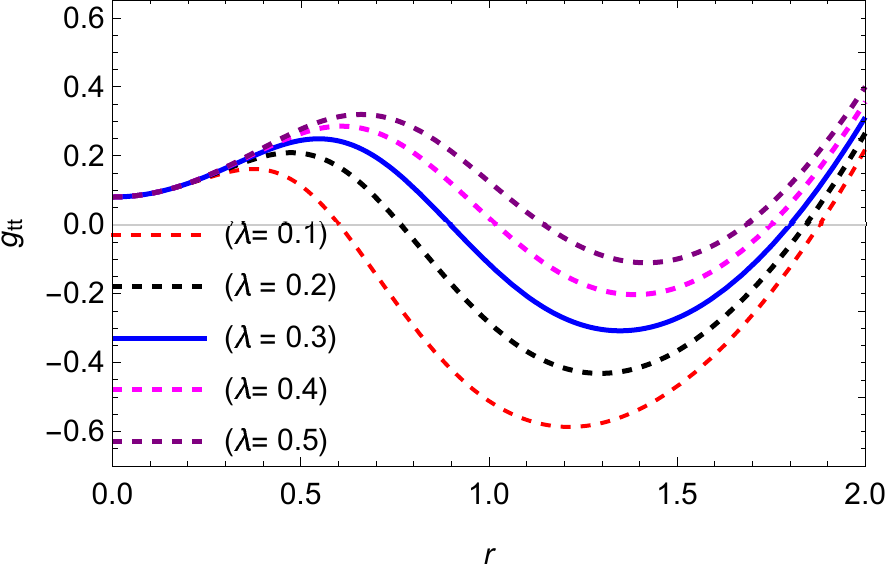}}\\
		\subfigure[]{\includegraphics[width=7cm,height=6cm]{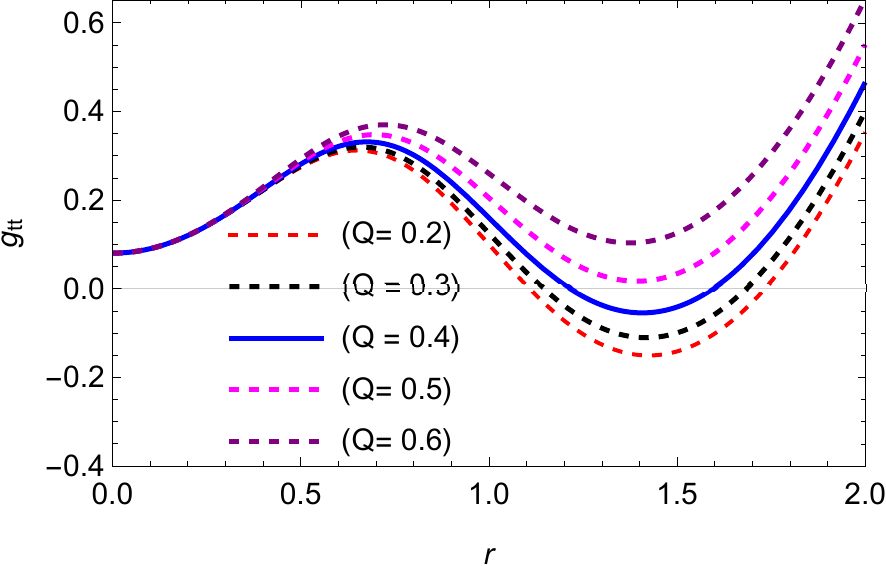}} 
		\subfigure[]{\includegraphics[width=7cm,height=6cm]{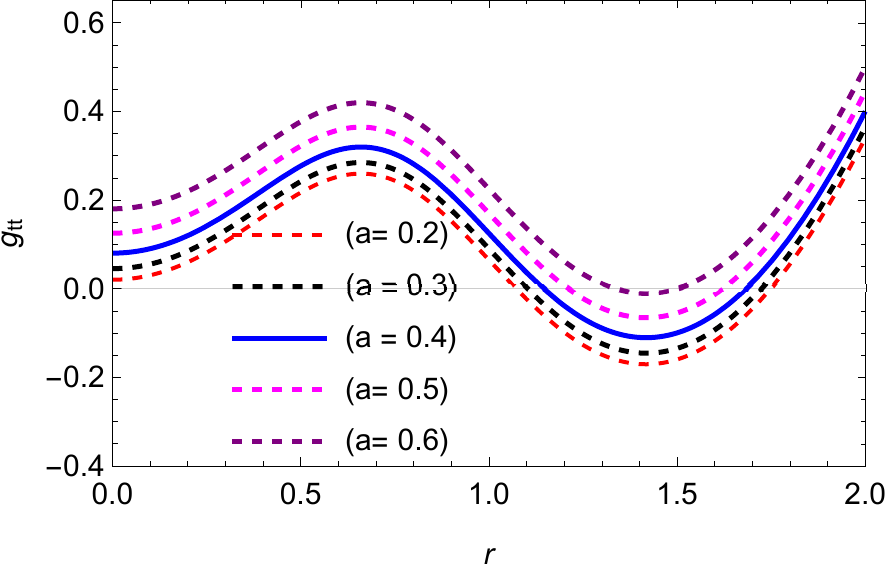}}
	\caption{The variation of static surface limit with radial distance for different values of parameters involved in the BH spacetime.} \label{f1.2}
\end{figure*}

In \figurename{ \ref{f1.1}}, the behavior of $\Delta(r)$ is depicted with radial distance for different values of YM, plasma, charge and spin parameter respectively. From \figurename{ \ref{f1.1}} (a), it is clearly observed that for a fix value of $Q$ and $a$ one can obtain two horizons if $\lambda < \lambda_{c}$ (critical value). However, one can have an extremal BH if the both horizons coincide to each other {\it i.e.}, $\lambda = \lambda_{c}$. One of the important aspects of BH geometry is the singularity and in case of going beyond to critical value, the horizon no longer exists and singularity is observed at the center. Similarly, the effect of other parameters and there corresponding critical value can also examined from the pictorial representation. It is interesting to see that apart from spin parameter beyond any fix value of radial distance, there is no variation observed for different values  of $\lambda$ and $Q$. The variation of static limit surface with radial distance for different values of various parameter is depicted in \figurename{ \ref{f1.2}}. It turns out that the behavior of static limit surface is almost similar to the horizon.  
\section{Photon motion around the rotating regular BH in non-minimally coupled EYM theory in presence of plasma}
\label{sec:3}
In this section, we have considered a static inhomogeneous plasma for our further calculations with a refractive index $n$ and the expression for this refractive index was formulated by Synge \cite{synge1966escape} as expressed below,
\begin{equation}
	n^{2} = 1+ \frac{p_{\mu}p^{\mu}}{(p_{\nu}u^{\nu})^{2}},   \label{e6}
\end{equation}
here, $p_{\mu}$ and $u^{\nu}$ correspond to the four momentum and four velocity of the massless test particle respectively. All these expressions reduce to vacuum case when we takes $n=1$.
In order to find out the null geodesic equations for this BH spacetime surrounded by plasma medium, we have used the Hamilton-Jacobi equation below,
\begin{equation}\label{e7}
		H(x^{\mu}p_{\mu}) = \frac{1}{2} \left[g^{\mu\nu}p_{\mu} p_{\nu}- (n^{2}-1)(p_{\nu}u^{\nu})^{2}\right].  
\end{equation}
Now the equation of motions of the photons for a given spacetime geometry can be defined by the Hamilton-Jacobi equation which is given as
\begin{equation}
	H(x^{\mu}p_{\mu}) = \frac{1}{2} \left[g^{\mu\nu}p_{\mu}p_{\nu}-(n^{2}-1)(p_{0}\sqrt{g^{00}})^{2}\right].  \label{e8}
\end{equation}
The equation $\dot{x}^{\mu}=\partial H/\partial p_{\mu}$ and $\dot{p}^{\mu}=\partial H/\partial x_{\mu}$ define the trajectories of massless particles in the plasma medium as below, 
\begin{equation}
	\Sigma \frac{dt}{d\sigma} = \frac{r^{2}+a^{2}}{\Delta} \left[n^{2}E(r^{2}+a^{2})-aL\right]-a(an^{2}E sin^{2}\theta-L),  \label{e9}
\end{equation}
\begin{equation}
	\Sigma \frac{d\phi}{d\sigma} = \frac{a}{\Delta} \left[E(r^{2}+a^{2})-aL\right] - \left(aE-\frac{L}{sin^{2}\theta}\right),   \label{e10}
\end{equation}
\begin{equation}
	\Sigma \frac{dr}{d\sigma} = \pm \sqrt{\mathcal{R}(r)},  \label{e11}
\end{equation}
and
\begin{equation}
	\Sigma \frac{d\theta}{d\sigma} = \pm \sqrt{\Theta(\theta)},   \label{e12}
\end{equation}
where,
\begin{equation}
	\mathcal{R}(r) = \left[X(r)E-aL\right]^{2} - \Delta(r)\left[\mathcal{K}+(L-aE)^{2}\right] \\+ X^{2}(r)(n^{2}-1)E^{2},  \label{e13}
\end{equation}
and
\begin{equation}
	\Theta(\theta) = \mathcal{K} + (n^{2}-1)a^{2}E^{2} - L^{2} cot^{2} \theta,  \label{e14}
\end{equation}
with $X(r) = (r^{2}+a^{2})$. Here, $\mathcal{K}$ is the separation constant popularly known as Carter constant while the function $\Delta(r)$ is defined by Eq. \ref{5}. The conserved quantities $E$ and $L$, along the axix of symmetry represent the energy and angular momentum of massless particle at infinity respectively.

\section{Effective Potential and Photon Sphere}
\label{sec:4}
Here, first we introduce the refractive index parameter in terms of plasma frequency since the frequency of plasma medium affects the geodesics of a photon passing by a compact object. The refractive index $n$ is related to the plasma frequency $\omega_{p}$ in a specific form which reads as,  
\begin{equation}
	n^{2} = 1- \left(\frac{\omega_{p}}{\omega}\right)^{2},   \label{e15}
\end{equation}
where $\omega_{p}$ has the form
\begin{equation}
	\omega_{p} = \frac{4 \pi e^{2} N(r)}{m_{e}}.   \label{e16}
\end{equation}
In this equation, $e$, $N(r)$ and $m_{e}$ represent the charge, number density and mass of electron in plasma medium respectively. One can obtain the physically relevant form of $N(r)$ by the implication of radial law density which is given as,
\begin{equation}
	\left(\frac{\omega_{p}}{\omega}\right)^{2} = \frac{k}{r^{h}}, \hspace{1cm} k \geq 0.    \label{e17}
\end{equation} 
The refractive index may therefore read as
\begin{equation}
	n = \sqrt{1-\frac{k}{r^{h}}}.   \label{e18}
\end{equation}
It is well known that the power of radial distance characterizes the different properties of the plasma medium but for the simplicity, here we choose to work with $h=1$. So that the expression of refractive index takes the simple form
\begin{equation}
	n = \sqrt{1-\frac{k}{r}}.     \label{e19}
\end{equation}
The effective radial potential can be directly evaluated by radial photon motion and it is written as,
\begin{equation}
	V_{eff} = \frac{1}{r^{4}} \big[\Delta(r)\left[\mathcal{K}+(L-aE)^{2}\right]\\ - X^{2}(r)(n^{2}-1)E^{2} - \left[X(r)E-aL\right]^{2}\big].   \label{e20}
\end{equation}
The condition for the unstable circular orbits are given by
\begin{equation}
	V_{eff}(r) \vert_{r=r_{p}} =0, \hspace{1cm} V^{''}_{eff}(r) \vert_{r=r_{p}} =0.   \label{e21}
\end{equation}
The condition for maximizing $V_{eff}(r)$ can however be interpreted as,
\begin{equation}
	V^{''}_{eff}(r) \vert_{r=r_{p}} < 0.        \label{e22}
\end{equation} 
The first condition in Eq.21 leads to,
\begin{equation}
	\big[\Delta(r)\left[\mathcal{K}+(L-aE)^{2}\right] \\- X^{2}(r)(n^{2}-1)E^{2} - \left[X(r)E-aL\right]^{2}\big] \vert_{r=r_{p}} = 0,   \label{e23}
\end{equation}
while the second condition leads to

	\begin{multline}
		\frac{8\left[\Delta(r)\left[\mathcal{K}+(L-aE)^{2}\right] - 4X^{2}(r)(n^{2}-1)E^{2} - 4\left[X(r)E-aL\right]^{2}\right]}{r^{5}}  \\  - \frac{\left[X(r)E-aL\right] X'(r) + \Delta'(r)\left[\mathcal{K}+(L-aE)^{2}\right]}{r^{4}}   -\frac{2X(r)X'(r)(n^{2}-1)E^{2}-2nn'X^{2}(r)E^{2}}{r^{4}} \arrowvert_{r=r_{p}}  = 0.    \label{e24}
	\end{multline}

The solution of above equation provides the radii of photon orbits. However, the analytical solution of this equation might be a bit  cumbersome, so we have solved the above mentioned equation by numerical method to investigate the effect of plasma on the photon sphere.\\

\begin{figure*}[h]
	\centering
		\subfigure[]{\includegraphics[width=7cm,height=6cm]{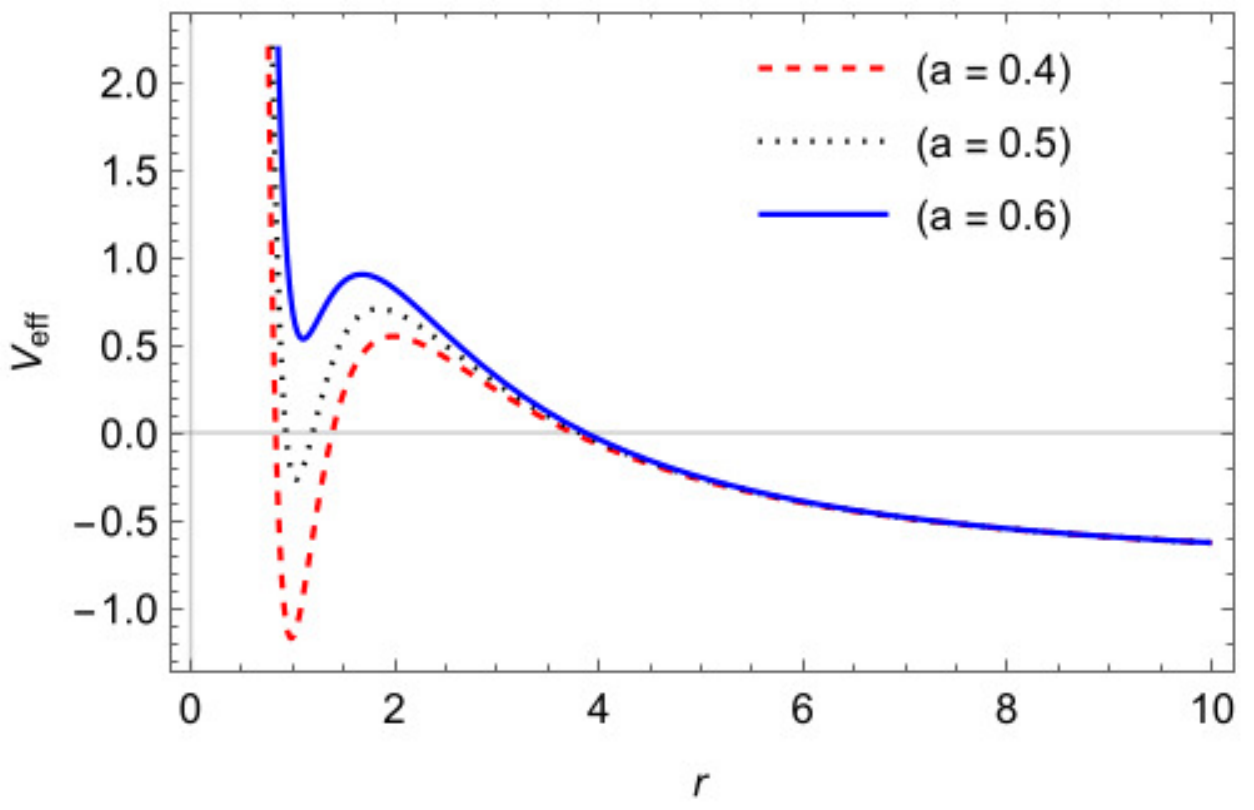}} 
		\subfigure[]{\includegraphics[width=7cm,height=6cm]{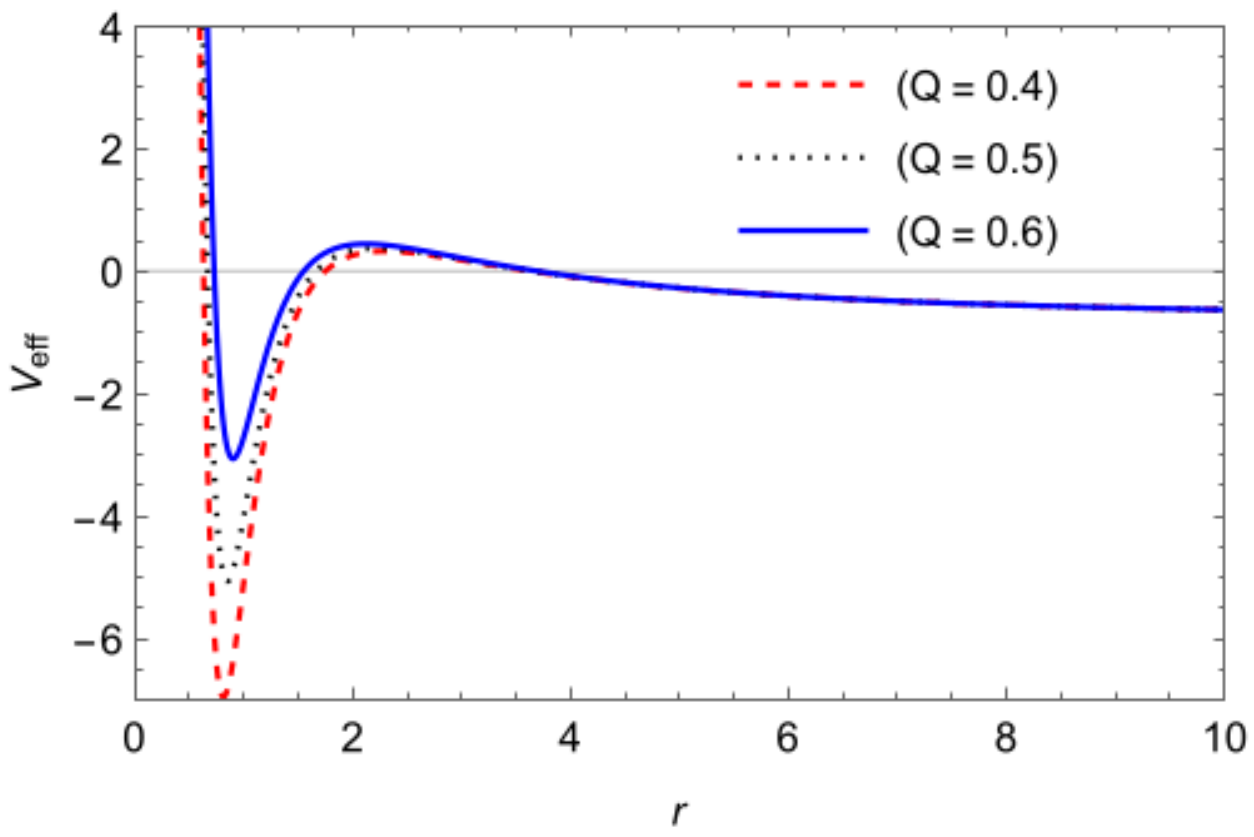}}\\
		\subfigure[]{\includegraphics[width=7cm,height=6cm]{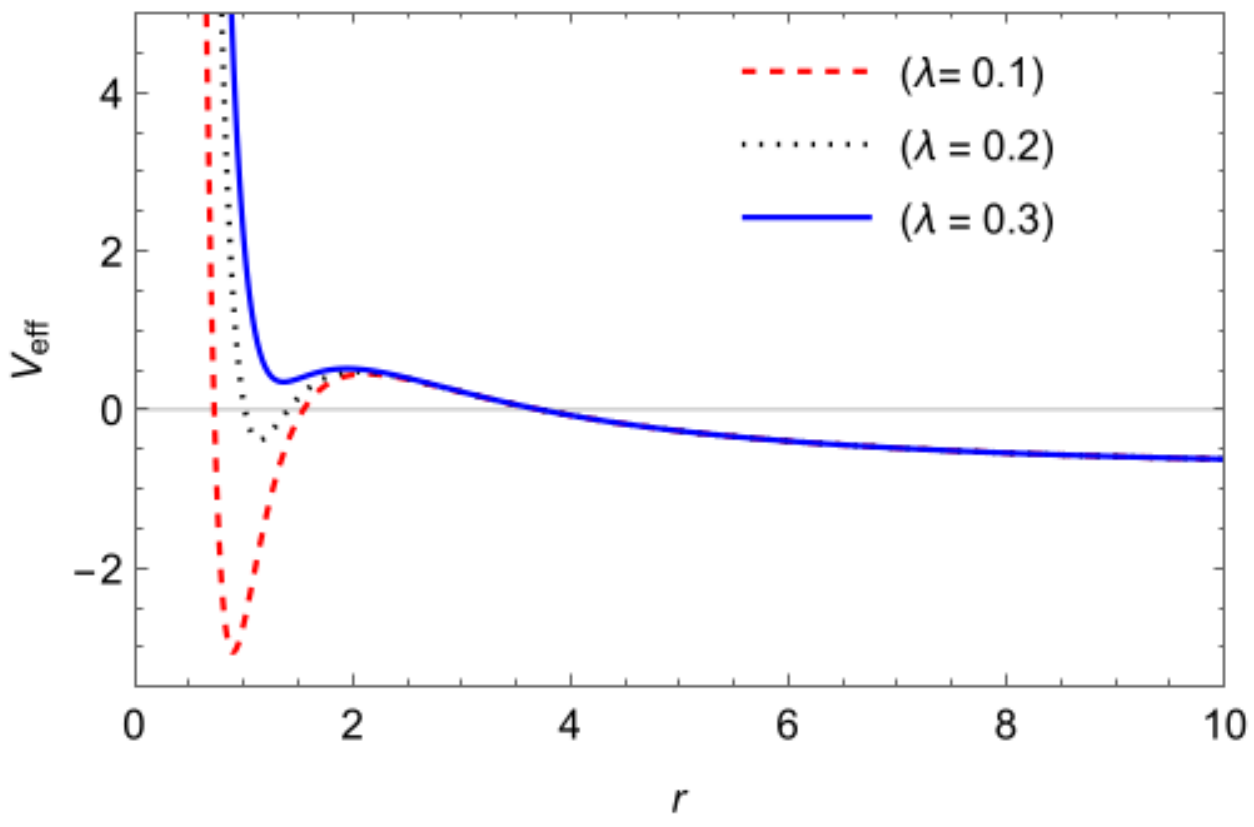}} 
		\subfigure[]{\includegraphics[width=7cm,height=6cm]{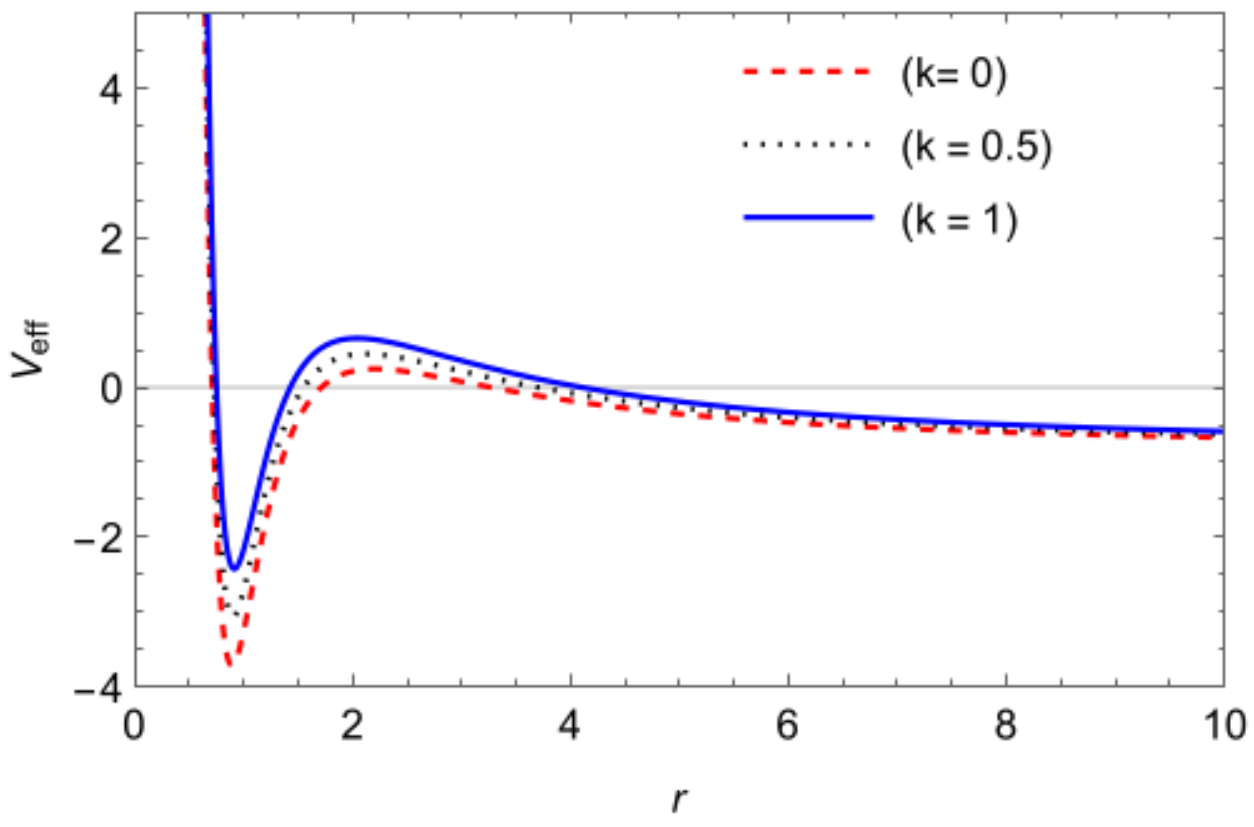}} 
	\caption{Variation of effective potential with radial distance for different values of various parameters for $L=4$ and $E=0.9$.} \label{f1}
\end{figure*}

\begin{figure*}[h]
	\centering
		\subfigure[]{\includegraphics[width=7cm,height=6cm]{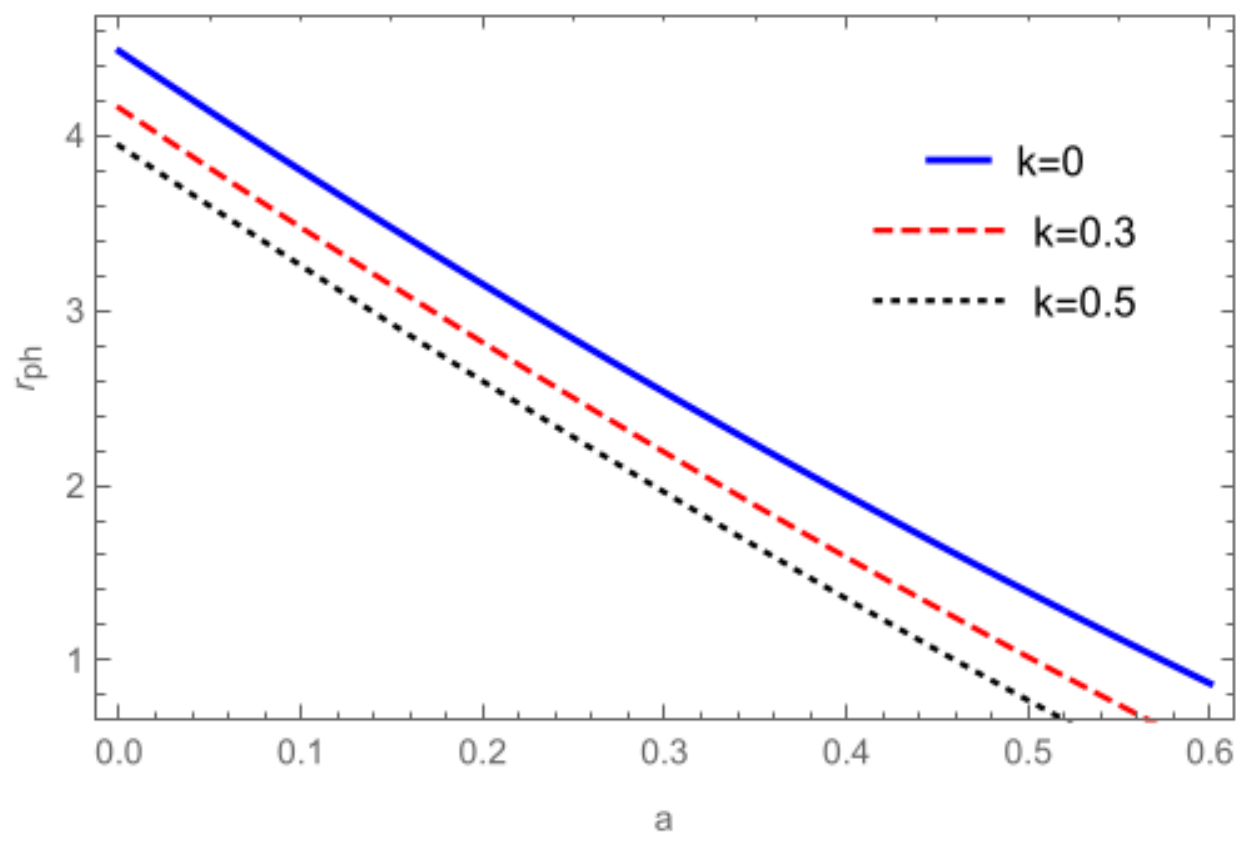}} 
		\subfigure[]{\includegraphics[width=7cm,height=6cm]{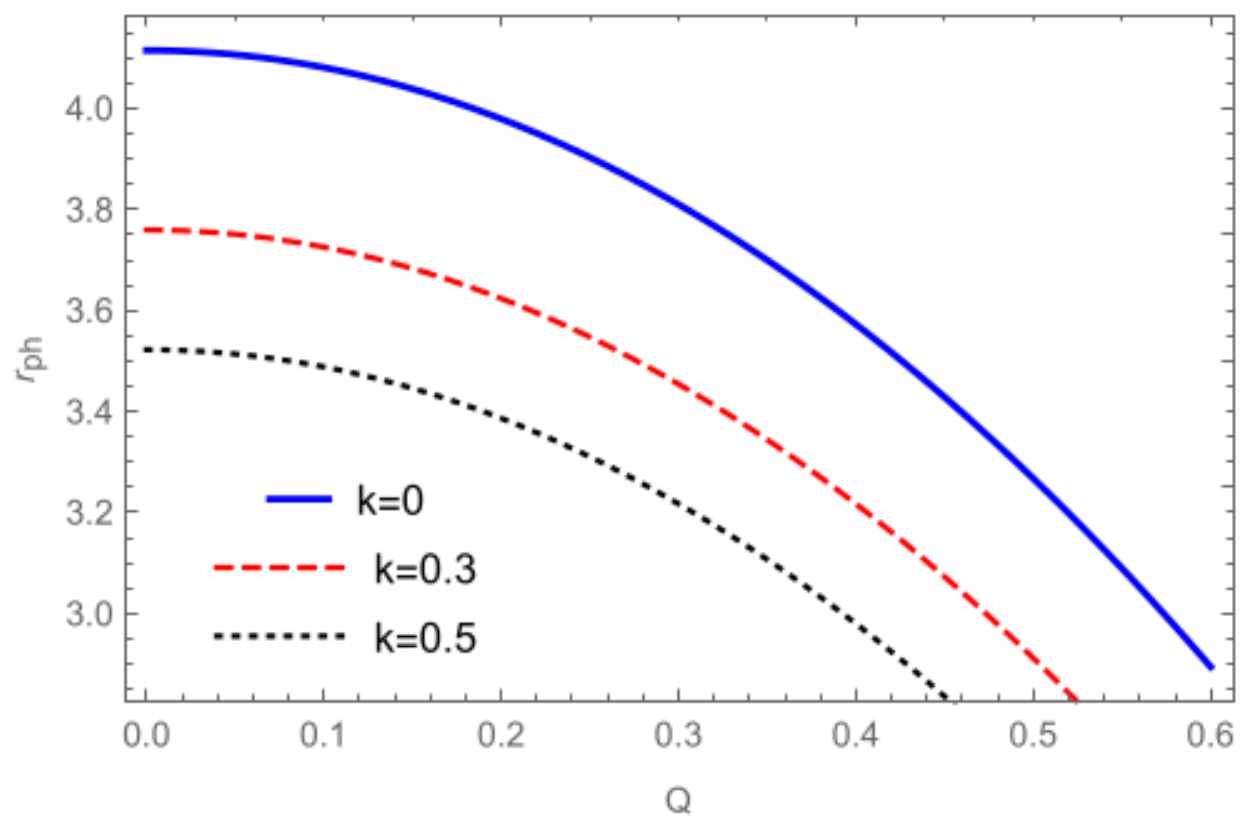}}
	\caption{Variation of photon orbit with spin parameter (left panel) and charge parameter (right panel) for different values of plasma parameter}. \label{f2}
\end{figure*}

The pictorial representation of the variation of effective potential with radial distance for different values of various BH parameters has depicted in \figurename{ \ref{f1}}. It is observed that  the effective potential attains its maximum value with increasing values of the spin and charge parameters. YM parameter also show the similar behavior however the effective potential decreases with decrease in the value of $\lambda$. One of the interesting results is that the effective potential attains its maximum value such that the turning point does not depend on the different values of spin, charge as well as $\lambda$. However, in the last case, it can be observed that after the potential  attains maxima, as the value of plasma parameter decreases, also the turning point shifts towards the left. \figurename{ \ref{f2}}, represents the behaviour of radius of photon orbits with spin and charge parameter for different values of plasma parameter. We have examined that the radii of photon sphere decreases with an increase in spin parameter and same effect has also observed in case of charge parameter. However, in case of charge parameter, the variation of photon orbit is observed maximum as compared to spin parameter.

\section{Shadow in Presence of Plasma}
\label{sec:5}
\begin{figure*}[h]
	\centering
		\subfigure[]{\includegraphics[width=5cm,height=5cm]{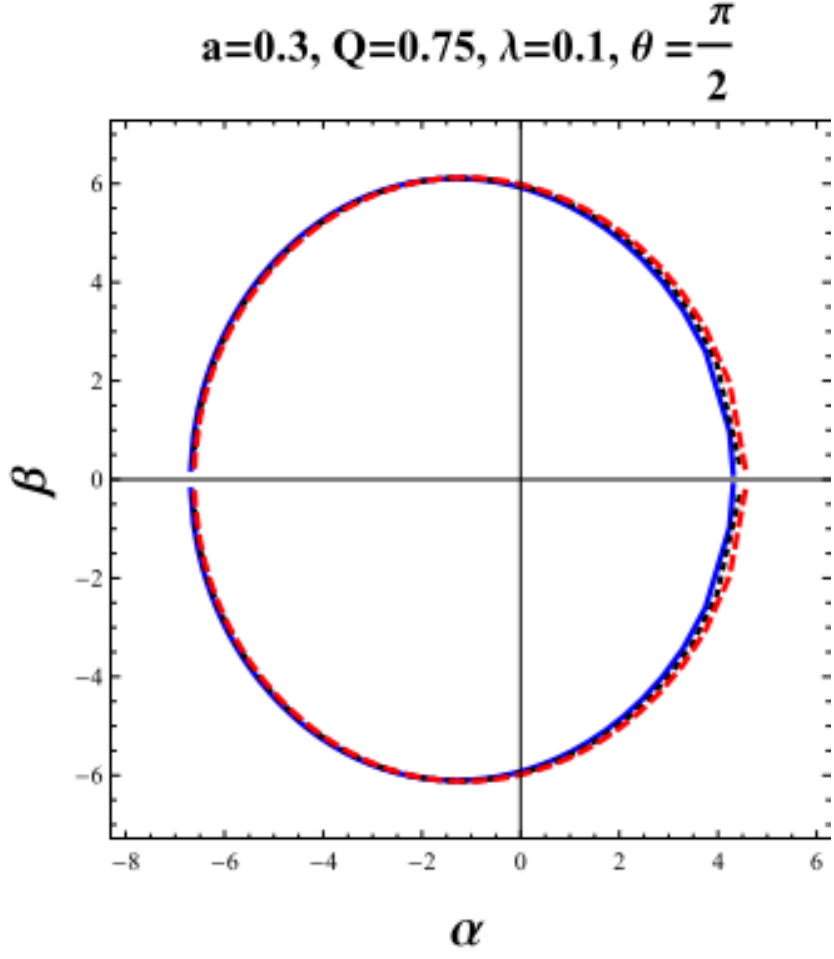}} 
		\subfigure[]{\includegraphics[width=5cm,height=5cm]{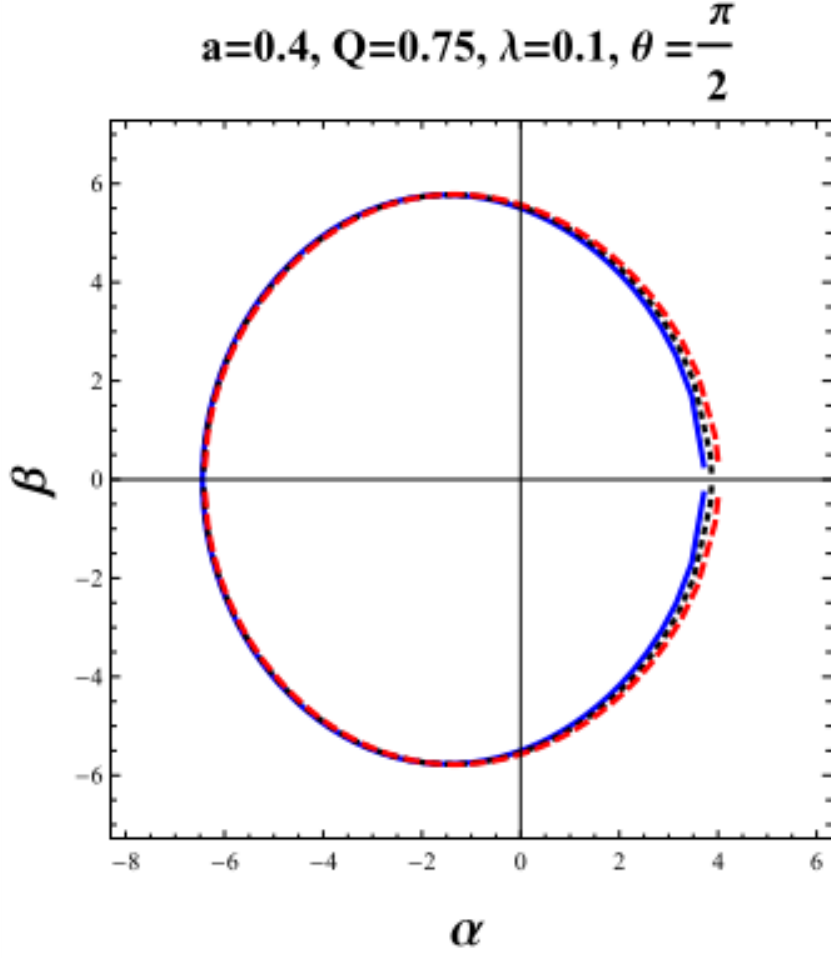}} 
		\subfigure[]{\includegraphics[width=5cm,height=5cm]{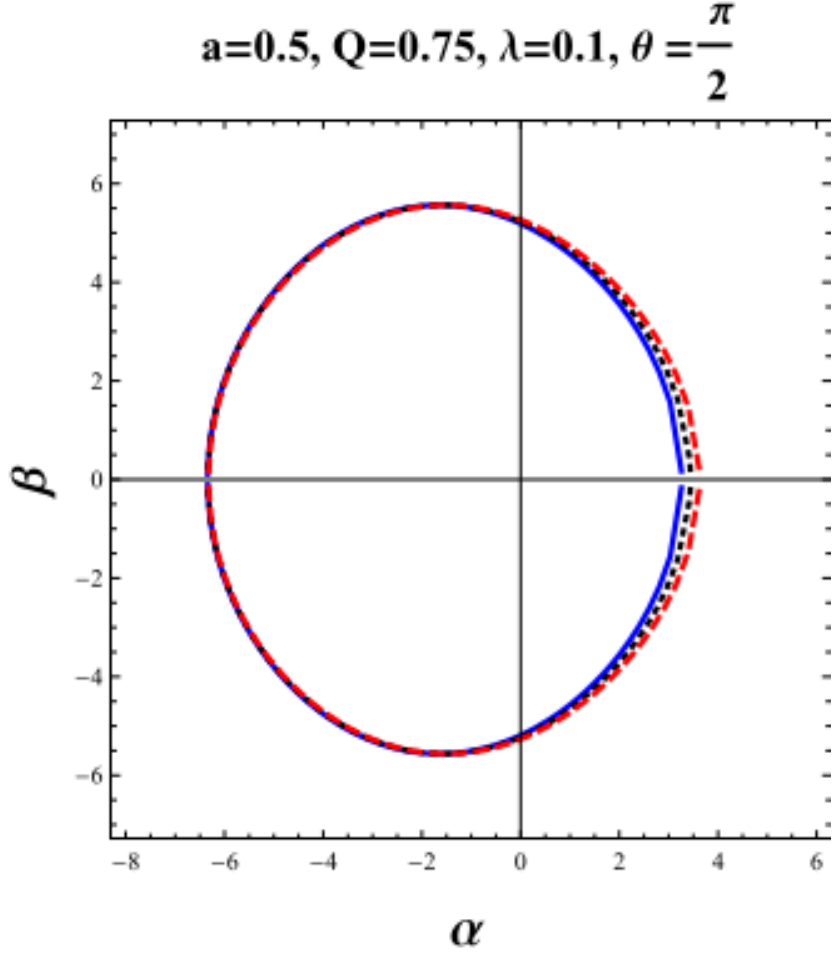}} \\
		\subfigure[]{\includegraphics[width=5cm,height=5cm]{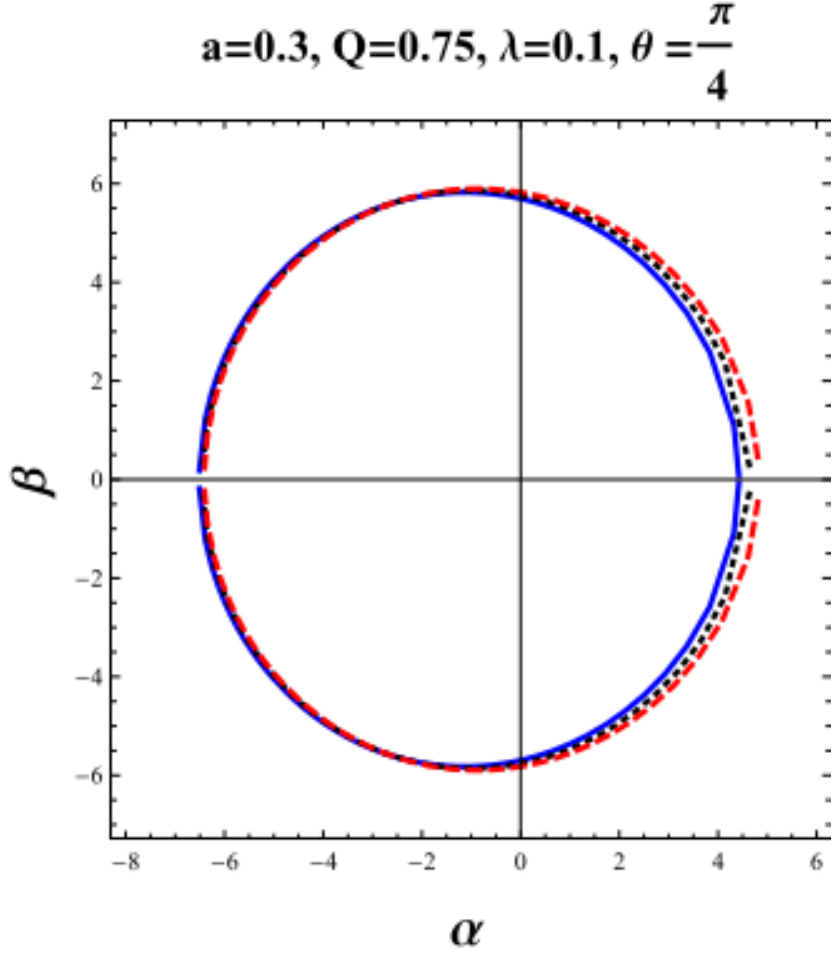}} 
		\subfigure[]{\includegraphics[width=5cm,height=5cm]{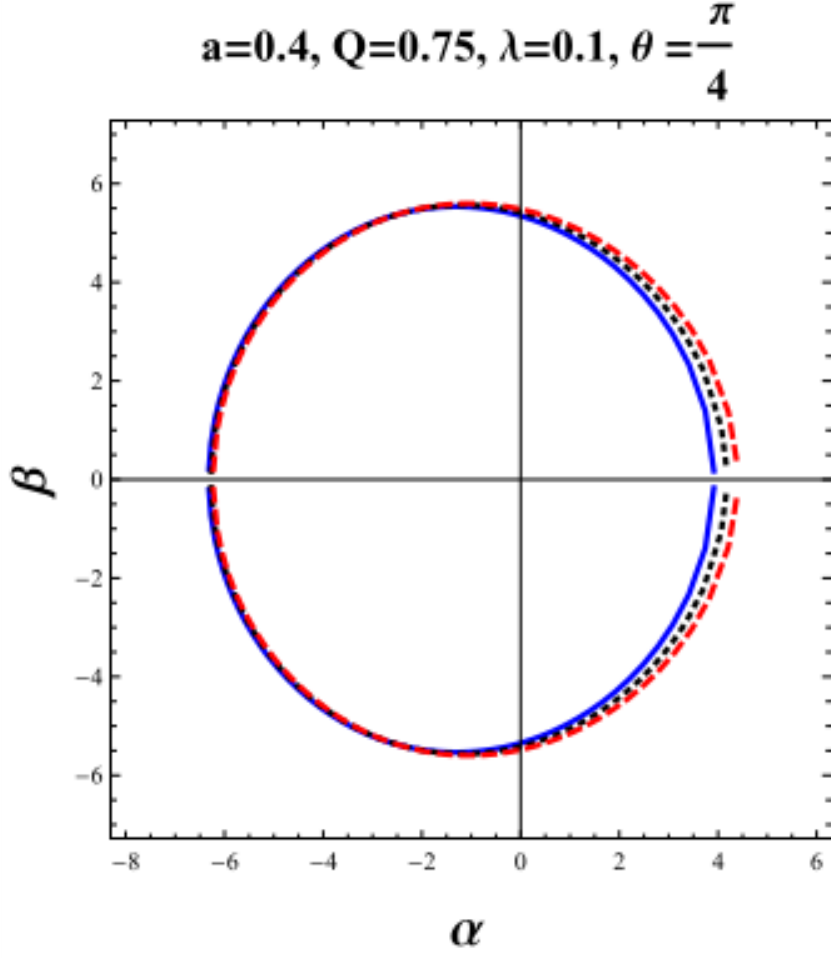}} 
		\subfigure[]{\includegraphics[width=5cm,height=5cm]{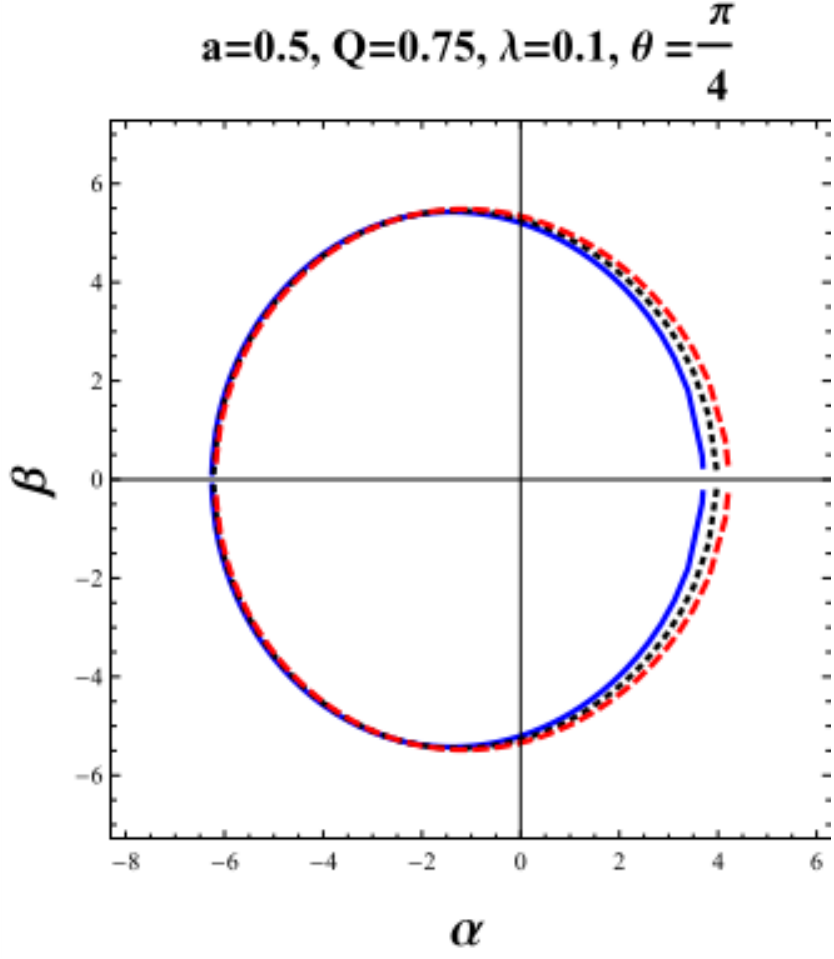}} \\
		\subfigure[]{\includegraphics[width=5cm,height=5cm]{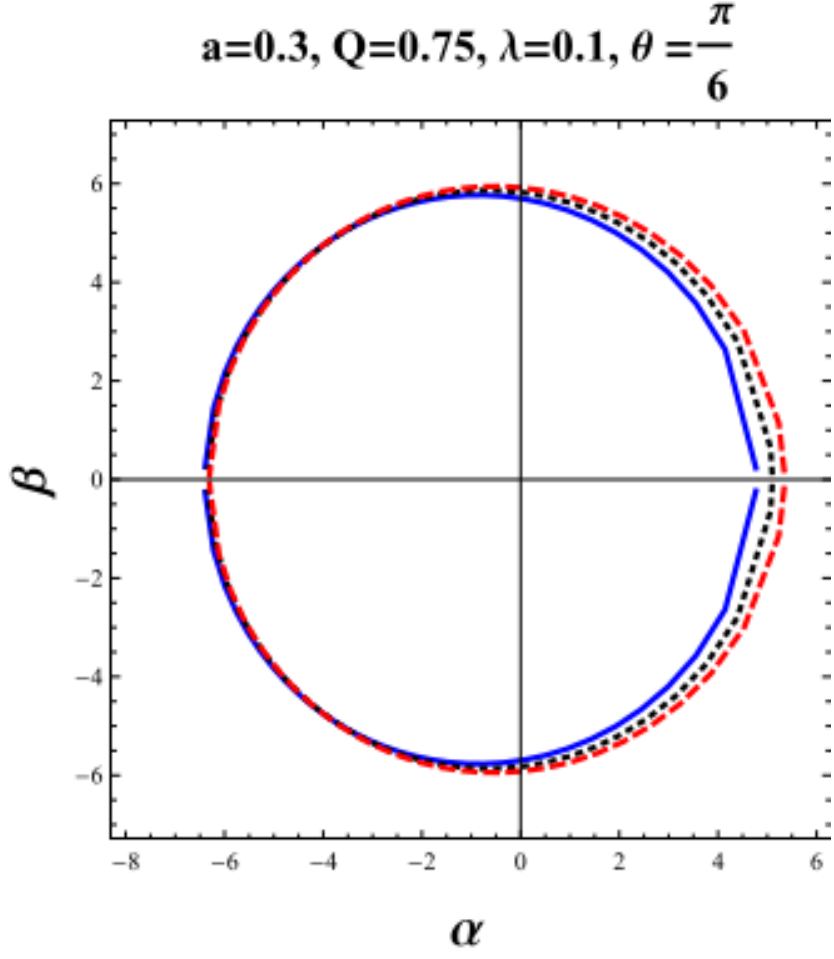}} 
		\subfigure[]{\includegraphics[width=5cm,height=5cm]{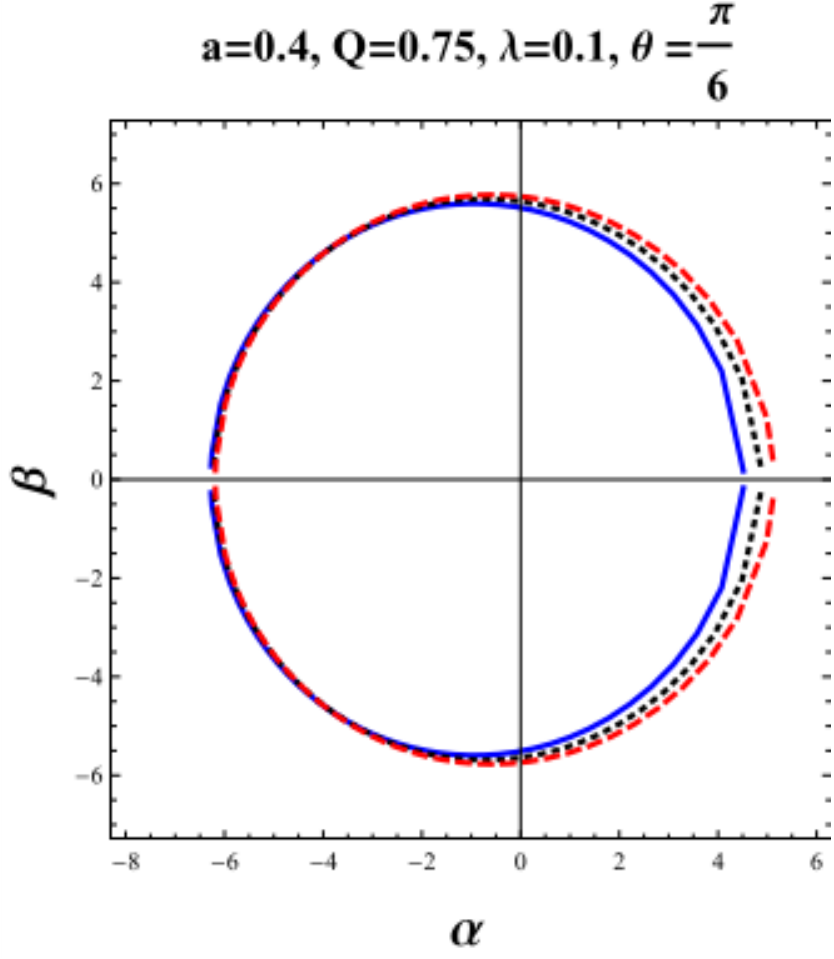}} 
		\subfigure[]{\includegraphics[width=5cm,height=5cm]{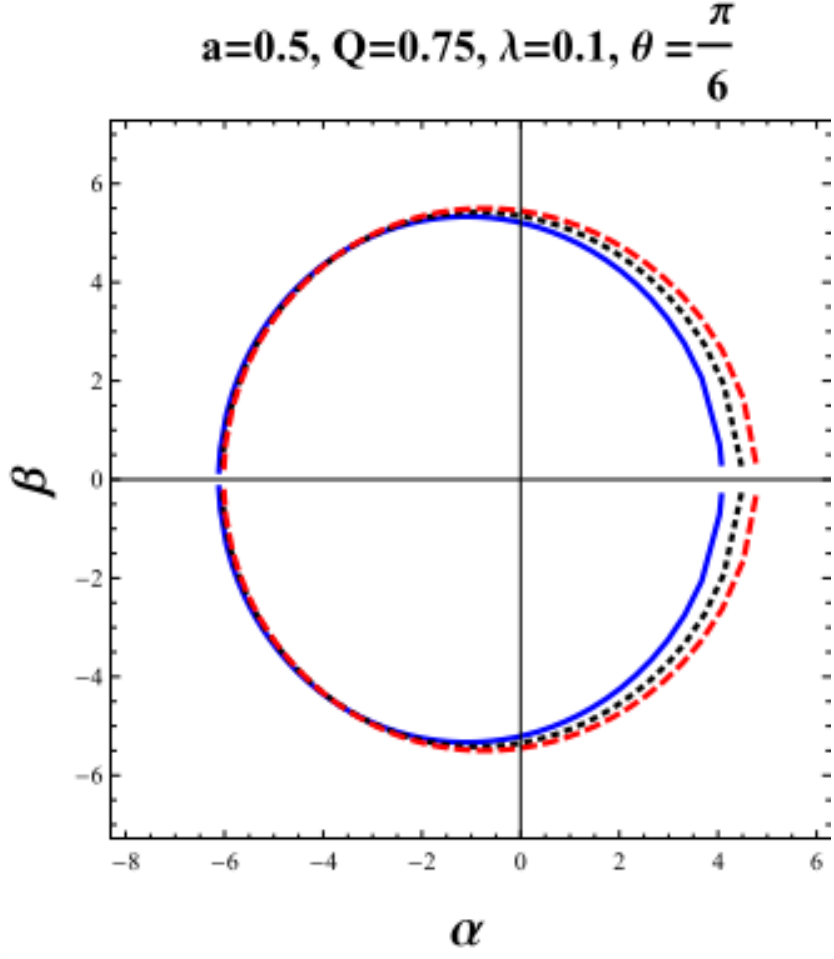}}\\
	\caption{Shape of shadows casted by a rotating regular BH in a non-minimally coupled EYM theory surrounded by plasma medium for the different values of rotation parameter and the refractive index of homogeneous plasma. The solid (blue) lines represent the vacuum case, while the dotted (black) and dashed (red) lines correspond to plasma parameter $k=0.5$ and $k=1.0$ respectively.  } \label{f3}
\end{figure*}

\begin{figure*}[h]
	\centering
		\subfigure[]{\includegraphics[width=5cm,height=5cm]{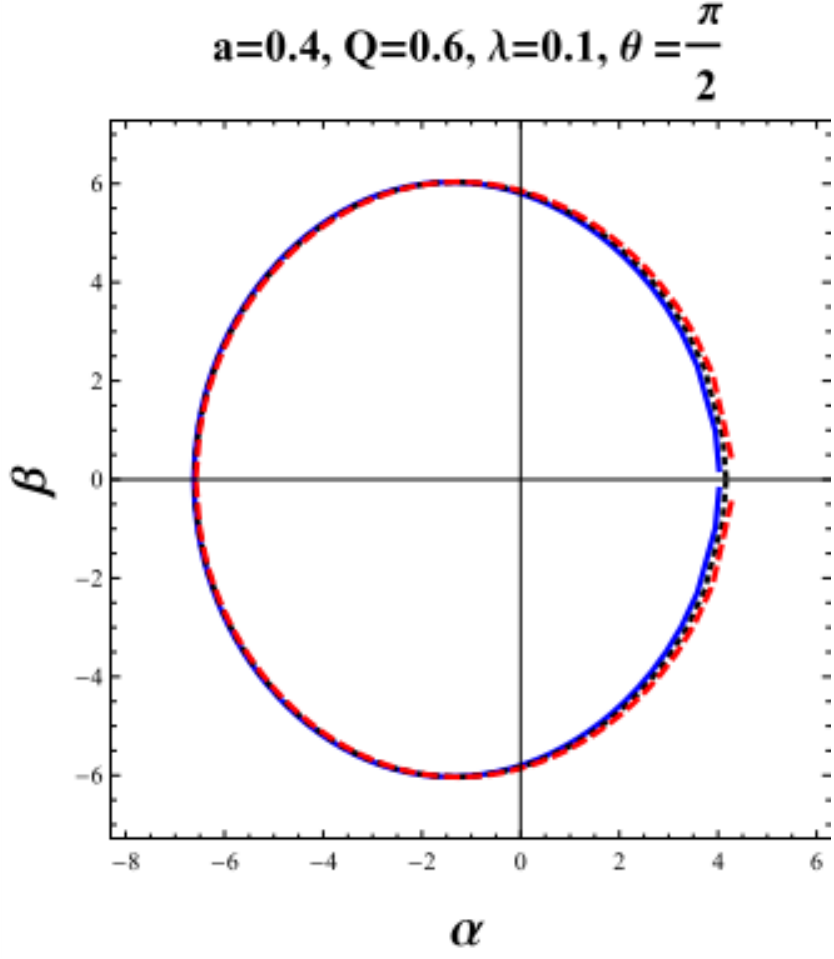}}  
		\subfigure[]{\includegraphics[width=5cm,height=5cm]{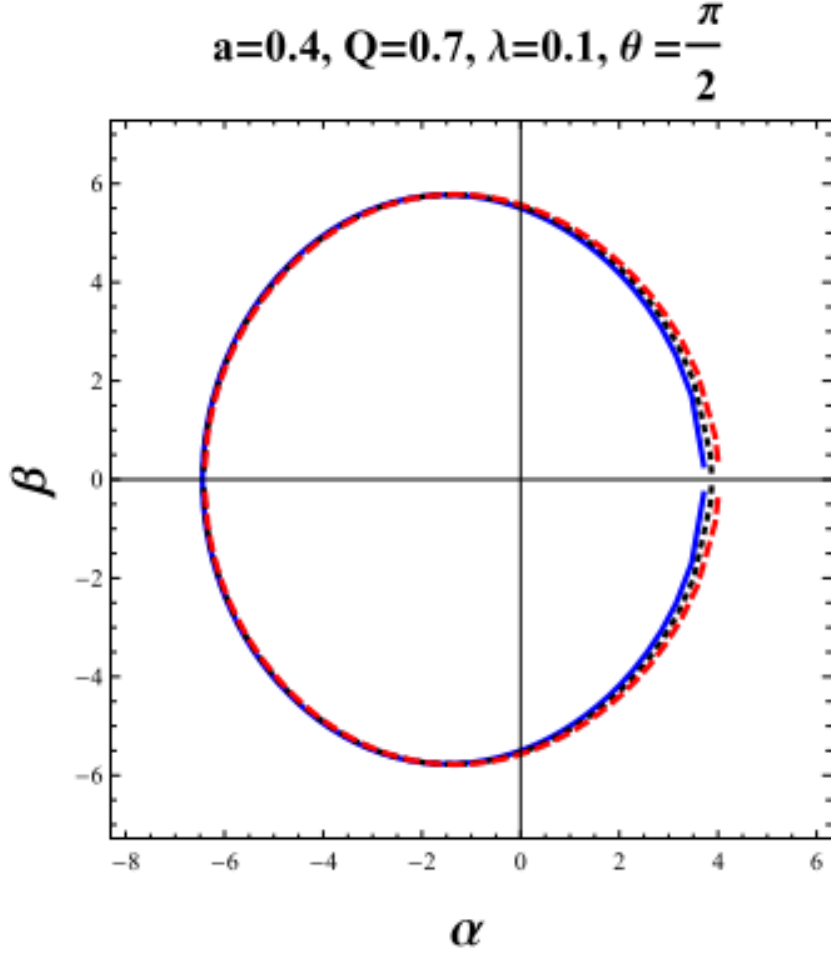}} 
		\subfigure[]{\includegraphics[width=5cm,height=5cm]{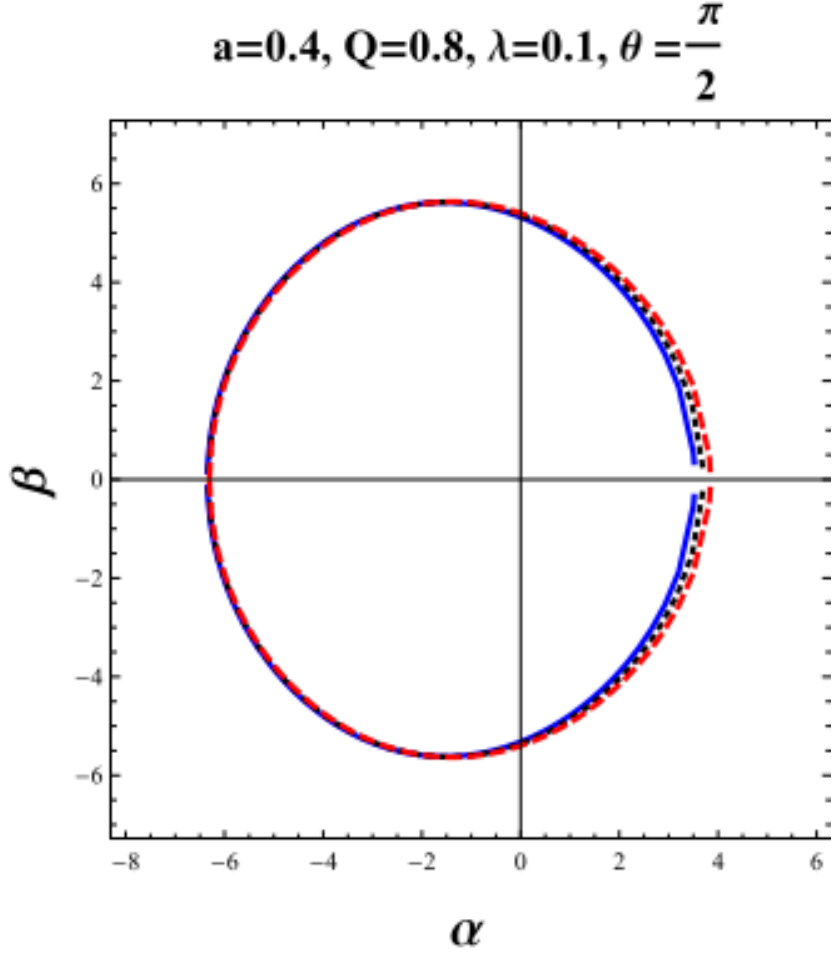}} \\
		\subfigure[]{\includegraphics[width=5cm,height=5cm]{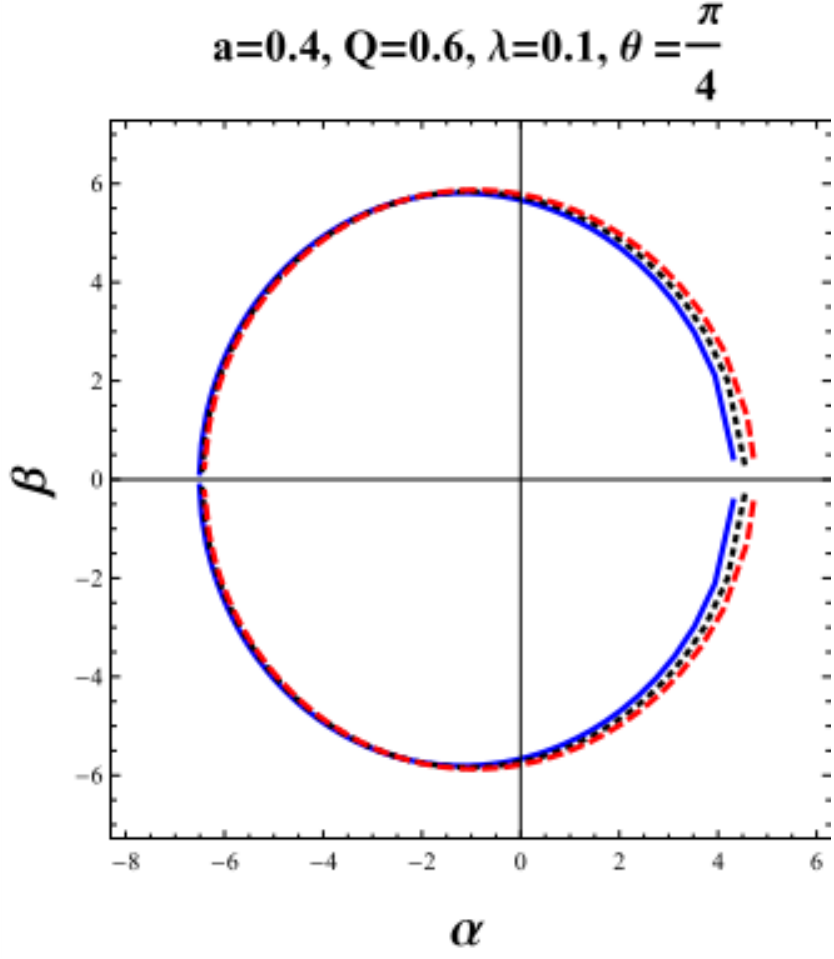}} 
		\subfigure[]{\includegraphics[width=5cm,height=5cm]{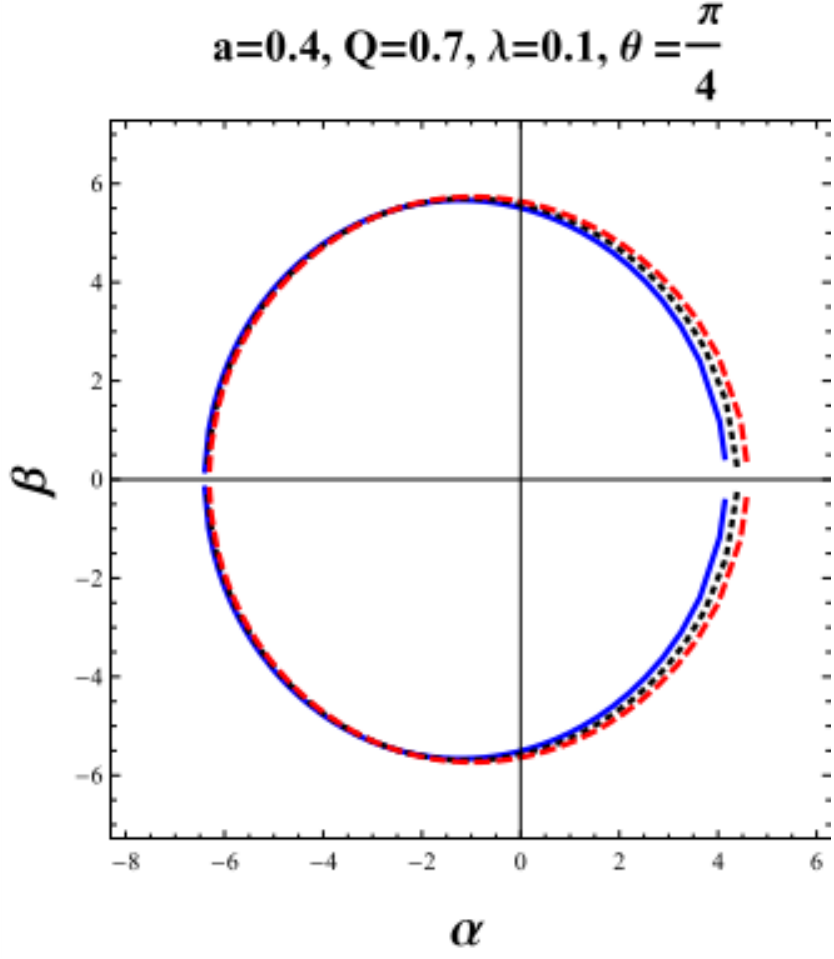}}
		\subfigure[]{\includegraphics[width=5cm,height=5cm]{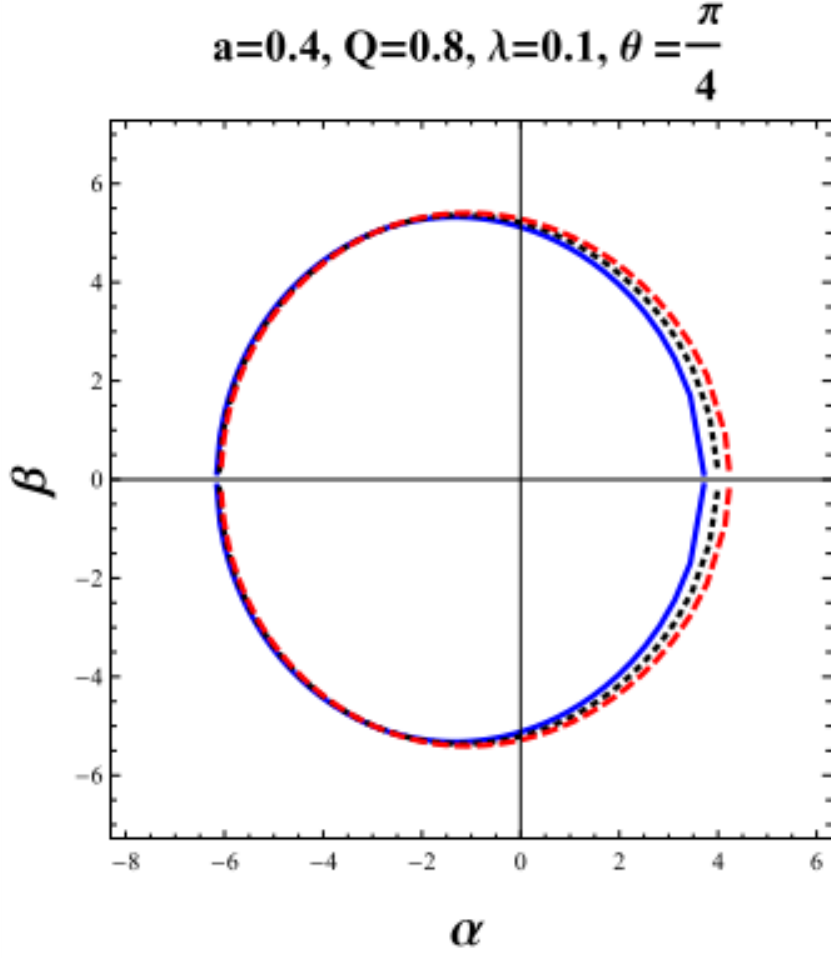}} \\
		\subfigure[]{\includegraphics[width=5cm,height=5cm]{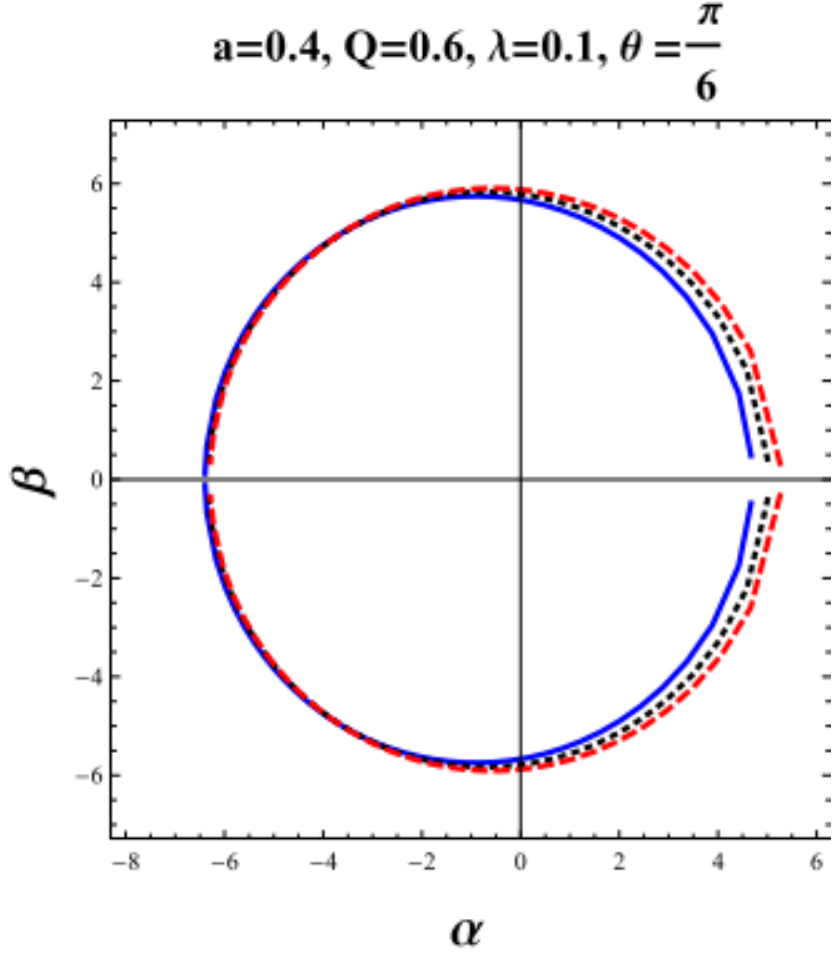}} 
		\subfigure[]{\includegraphics[width=5cm,height=5cm]{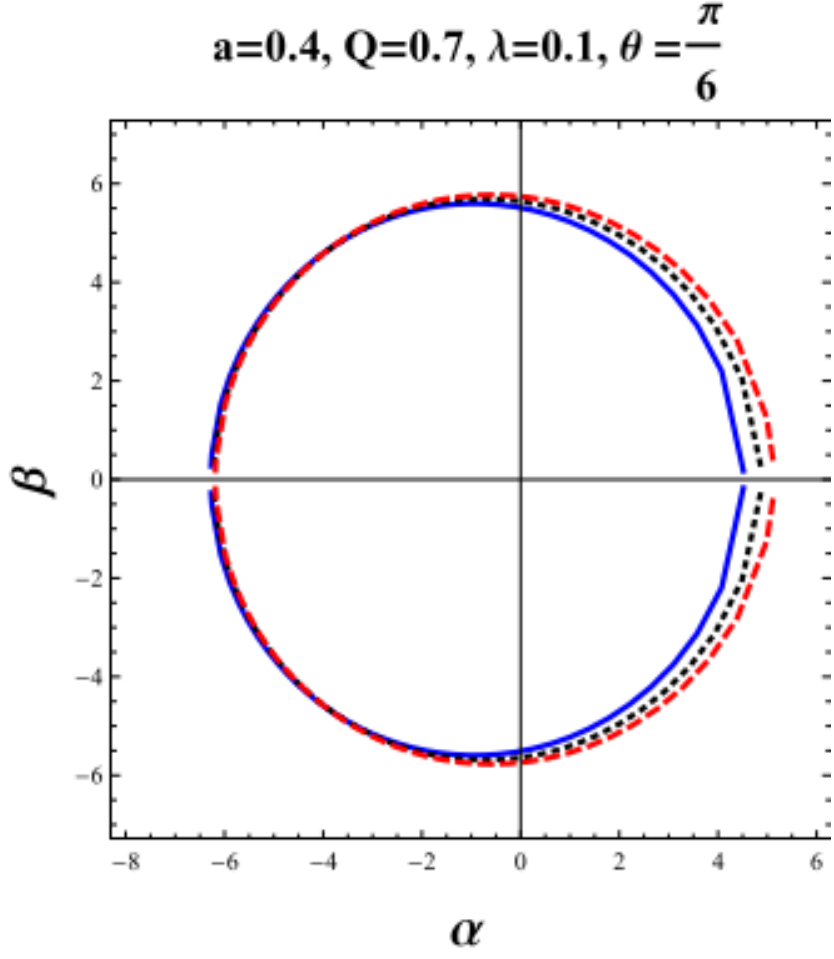}} 
		\subfigure[]{\includegraphics[width=5cm,height=5cm]{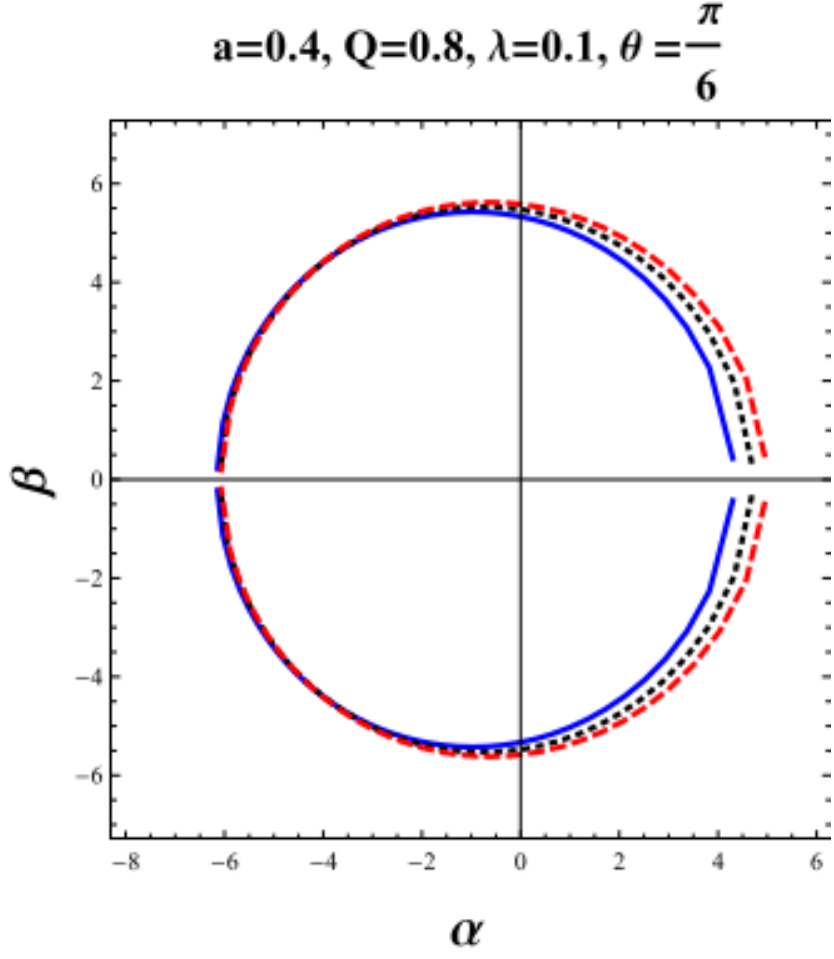}}
		\caption{Shape of shadows casted by a rotating regular BH in a non-minimally coupled EYM theory surrounded by plasma medium for the different values of charge parameter and the refractive index of homogeneous plasma. The solid (blue) lines represent the vacuum case, while the dotted (black) and dashed (red) lines correspond to plasma parameter $k=0.5$ and $k=1.0$ respectively.  } \label{f4}
\end{figure*}

\begin{figure*}[h]
	\centering
		\subfigure[]{\includegraphics[width=5cm,height=5cm]{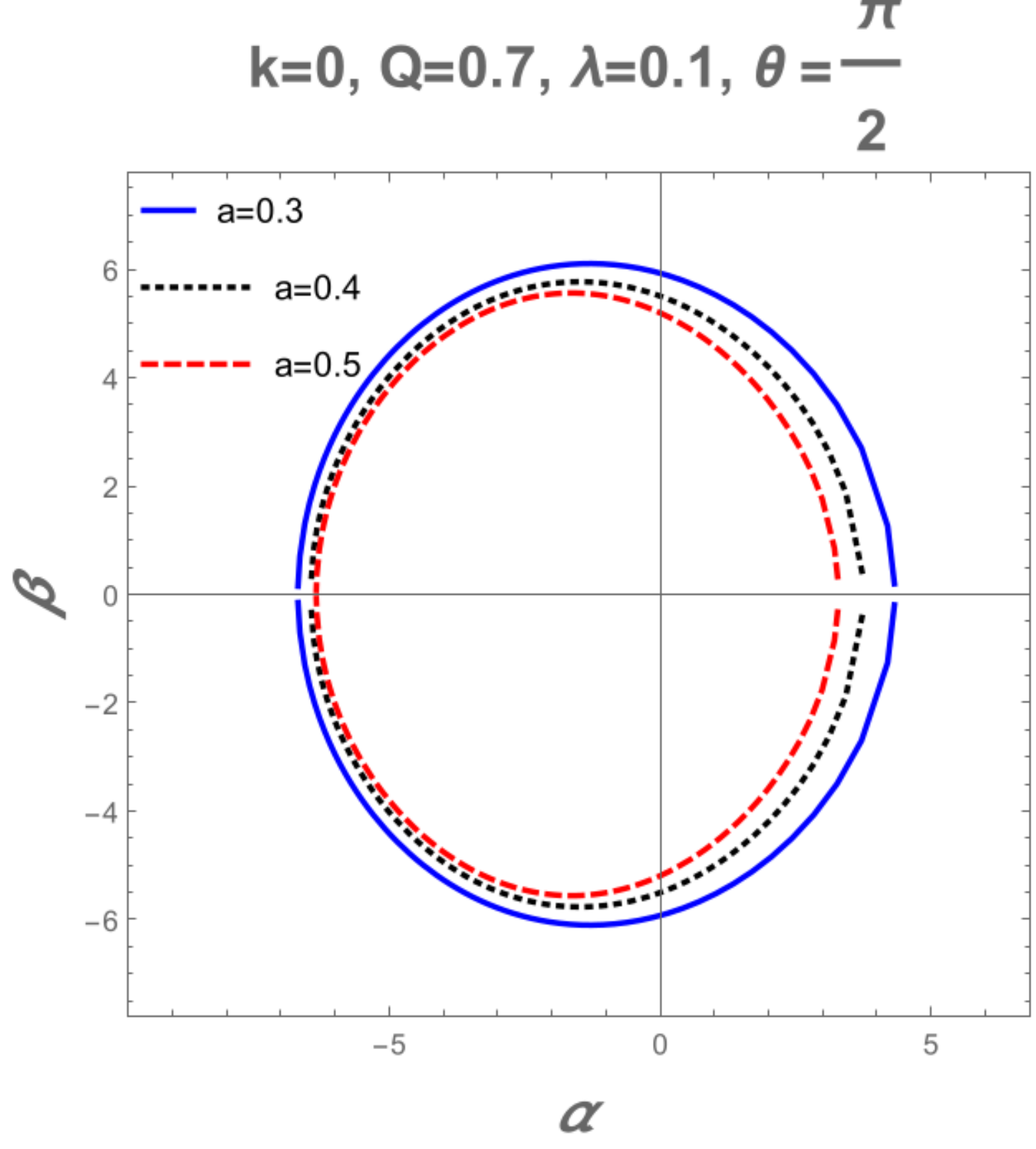}} 
		\subfigure[]{\includegraphics[width=5cm,height=5cm]{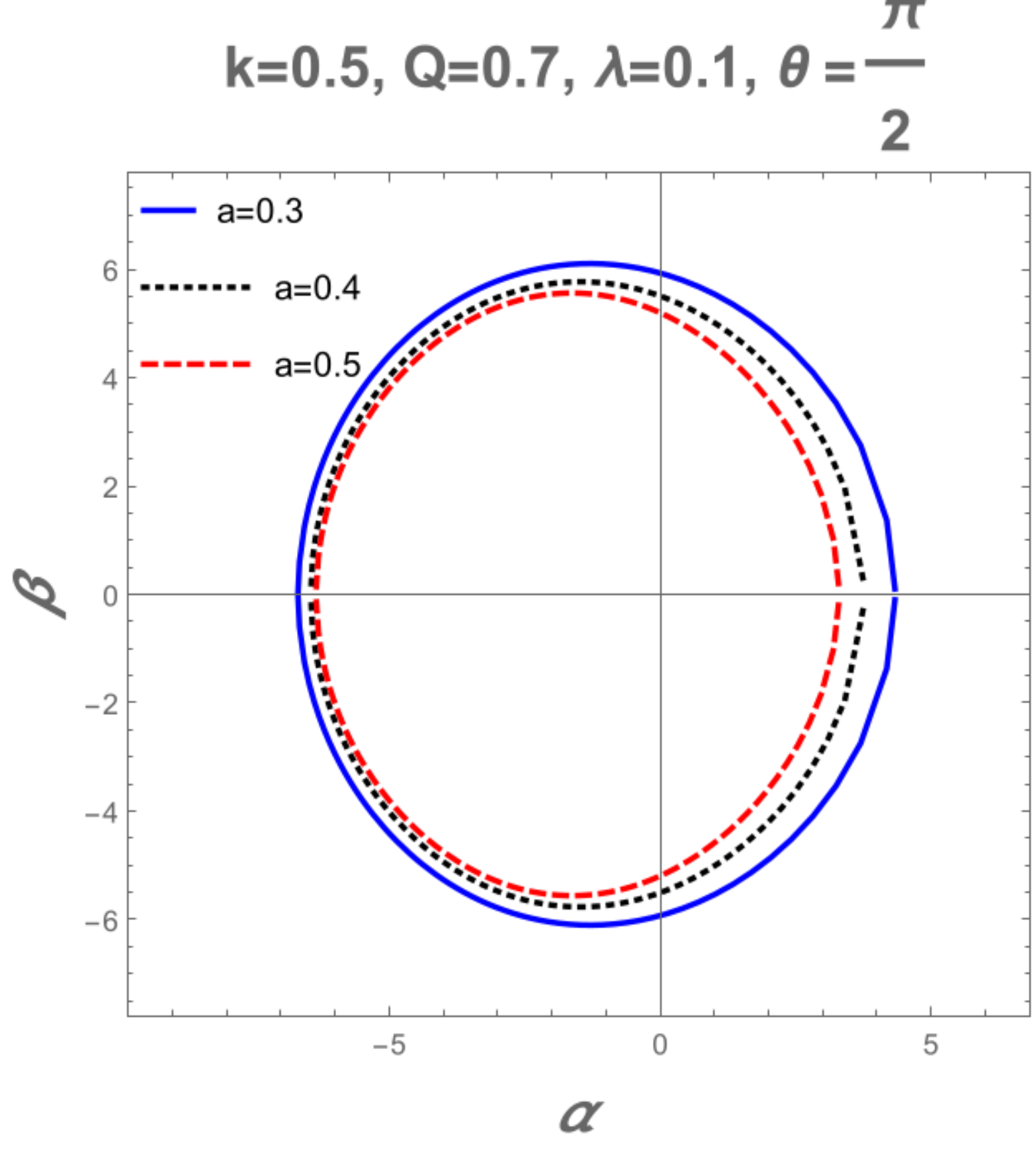}} 
		\subfigure[]{\includegraphics[width=5cm,height=5cm]{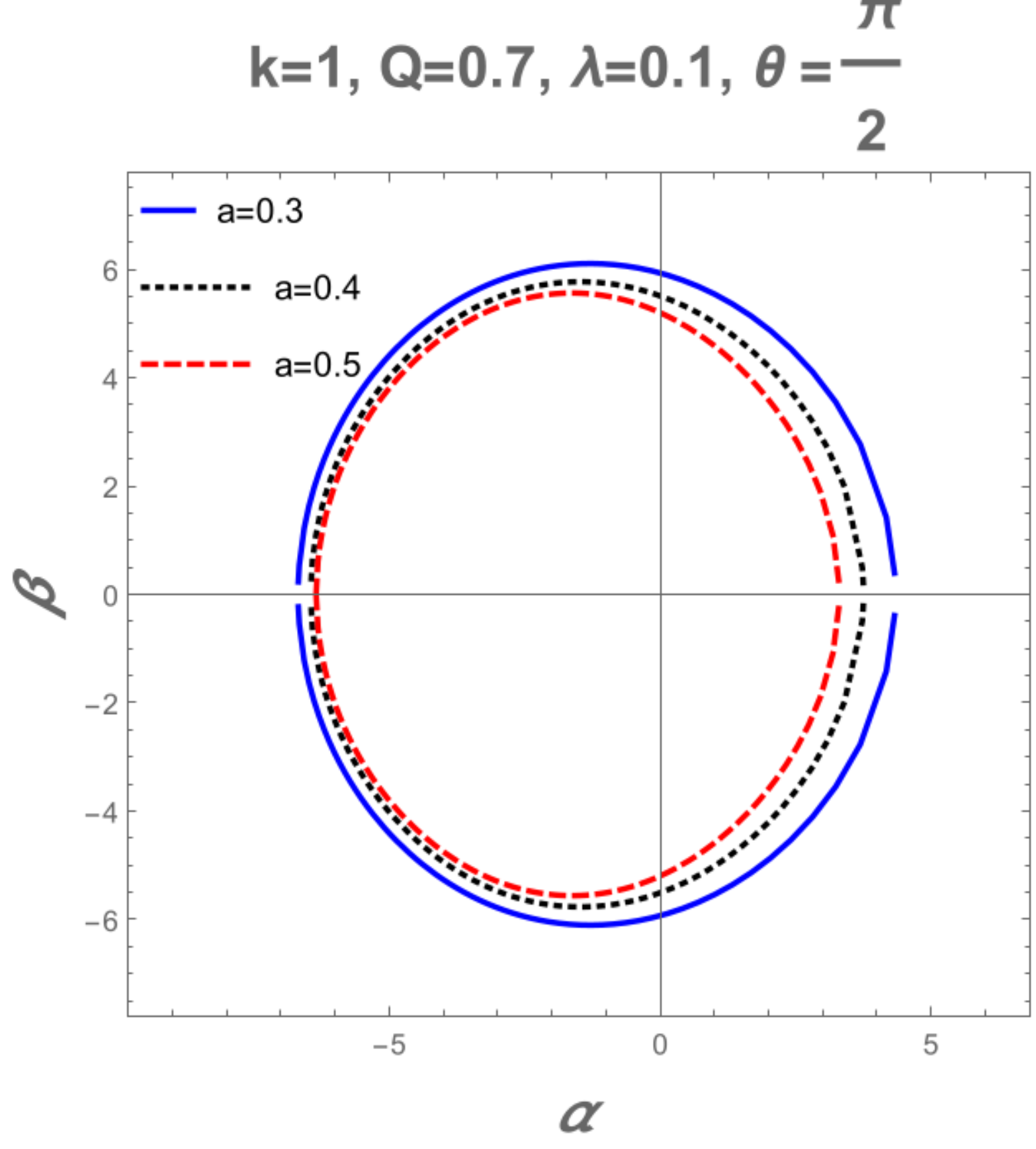}}\\
		\subfigure[]{\includegraphics[width=5cm,height=5cm]{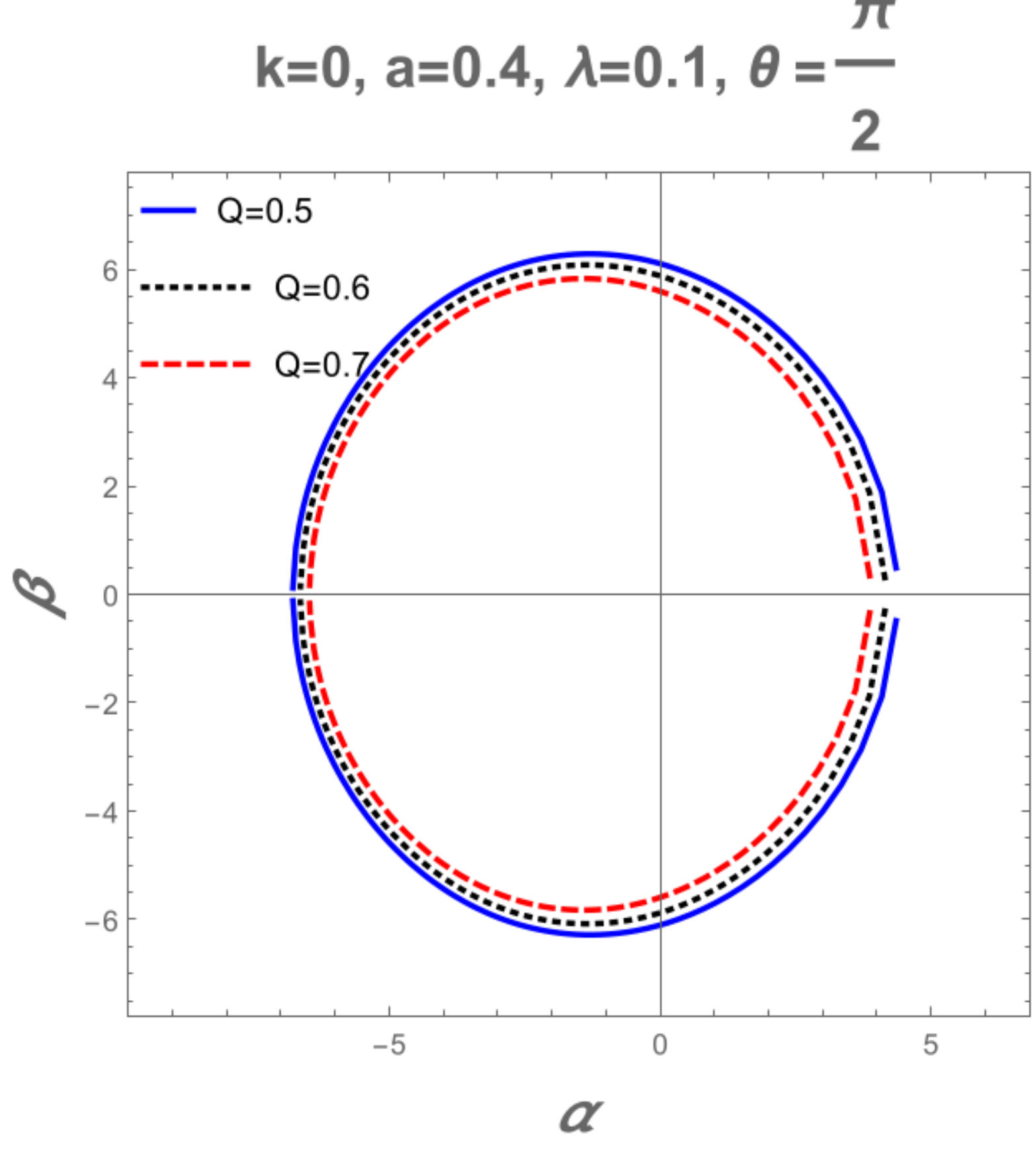}} 
		\subfigure[]{\includegraphics[width=5cm,height=5cm]{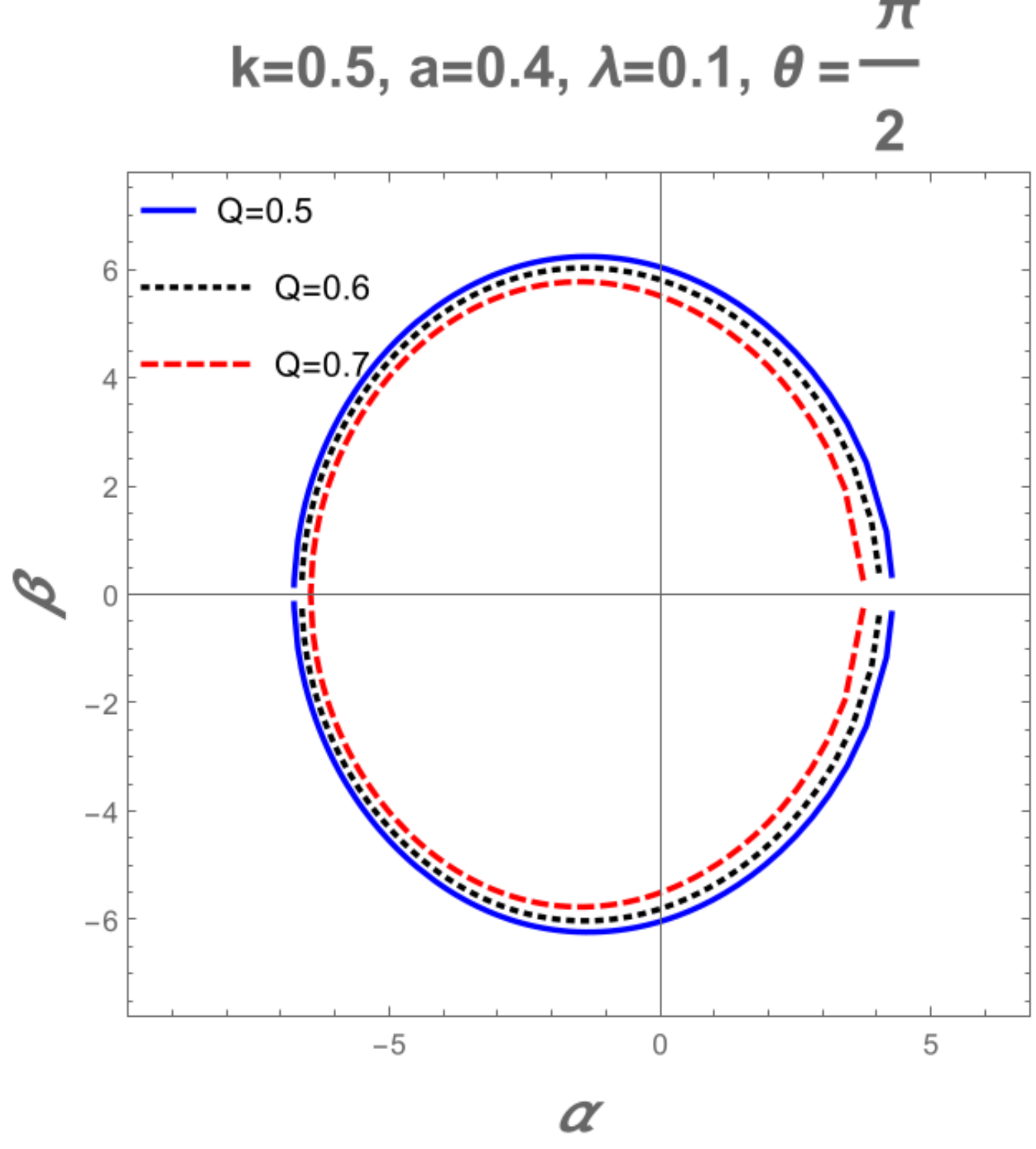}} 
		\subfigure[]{\includegraphics[width=5cm,height=5cm]{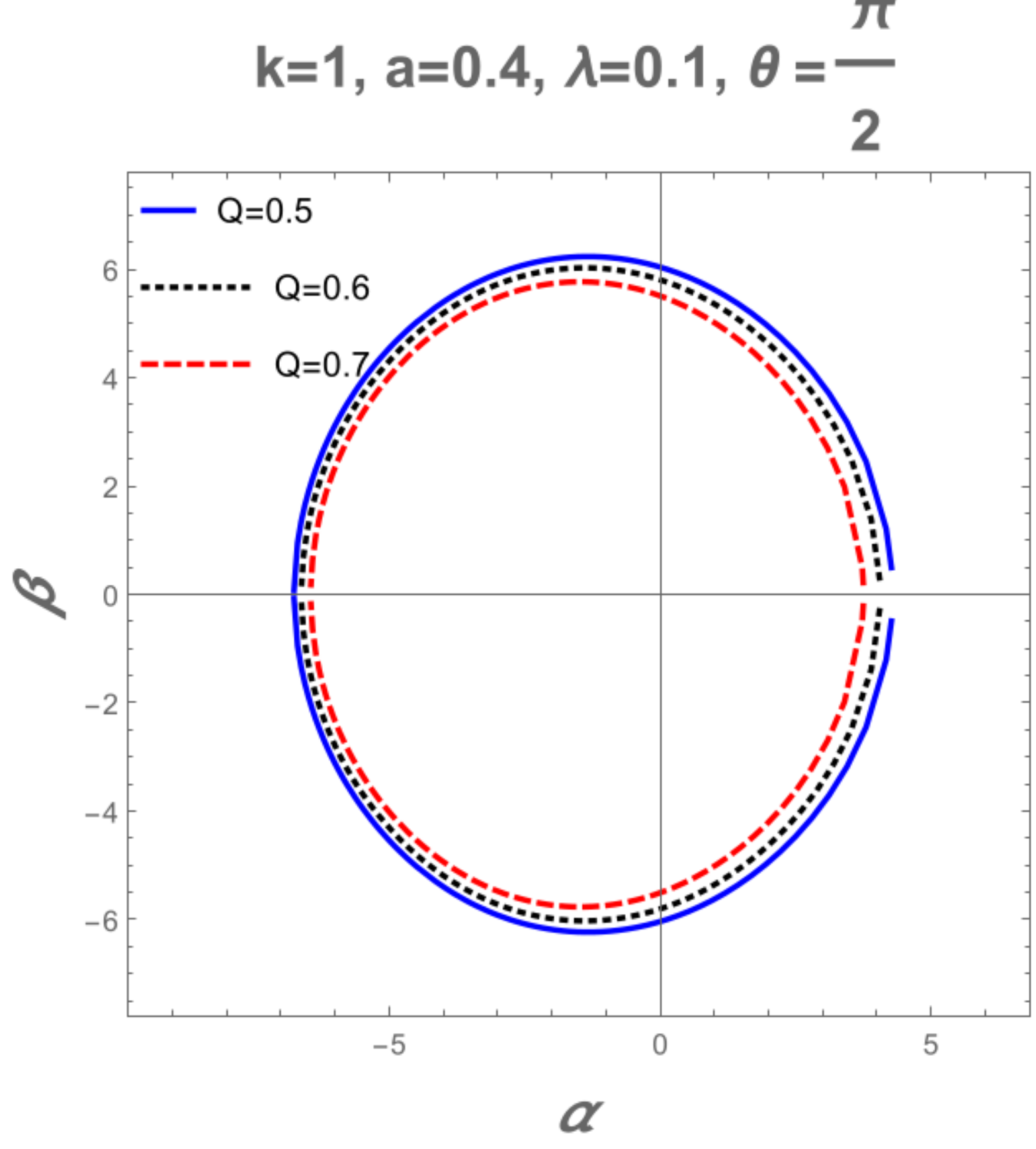}}\\
		\subfigure[]{\includegraphics[width=5cm,height=5cm]{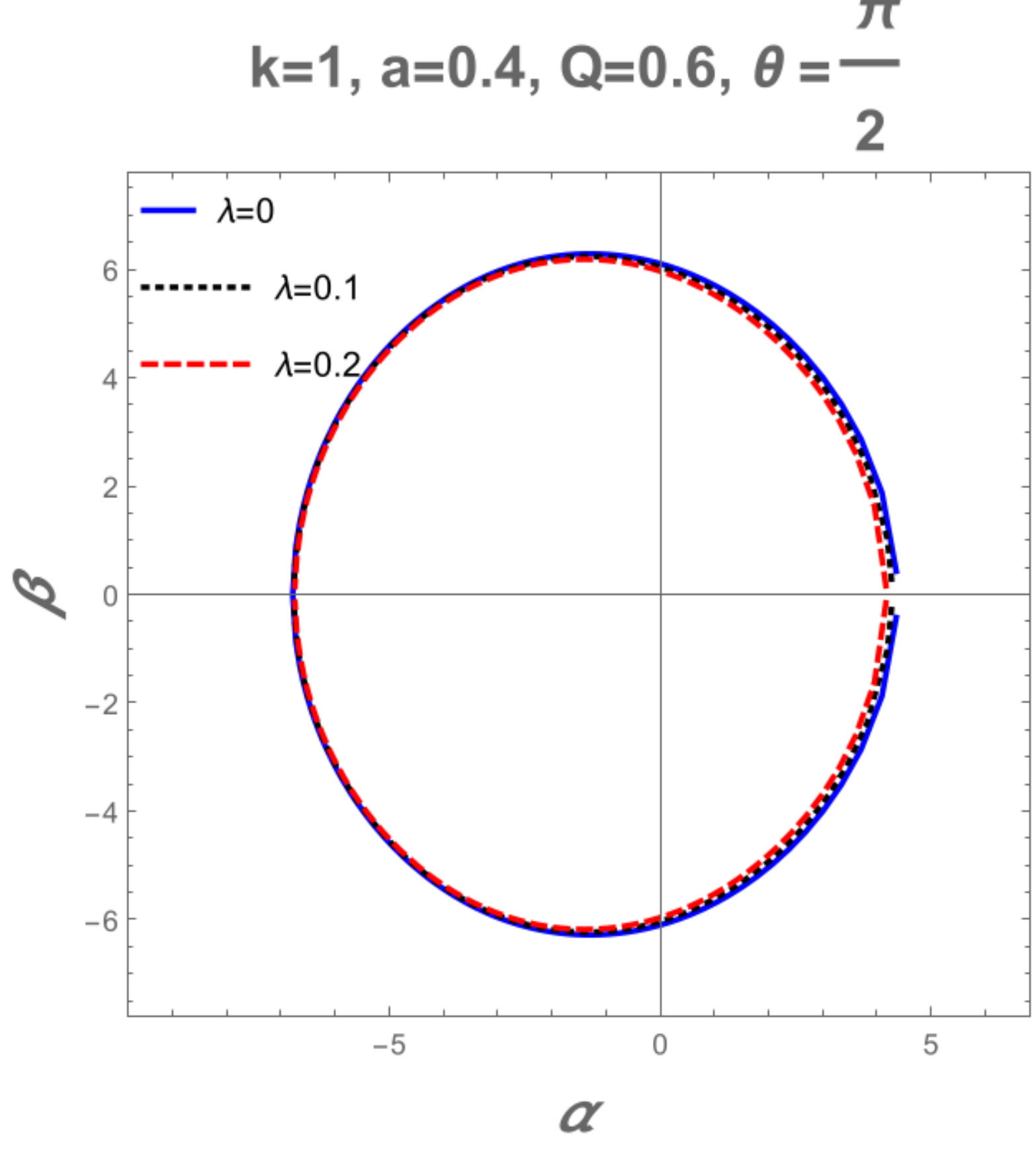}}  
		\subfigure[]{\includegraphics[width=5cm,height=5cm]{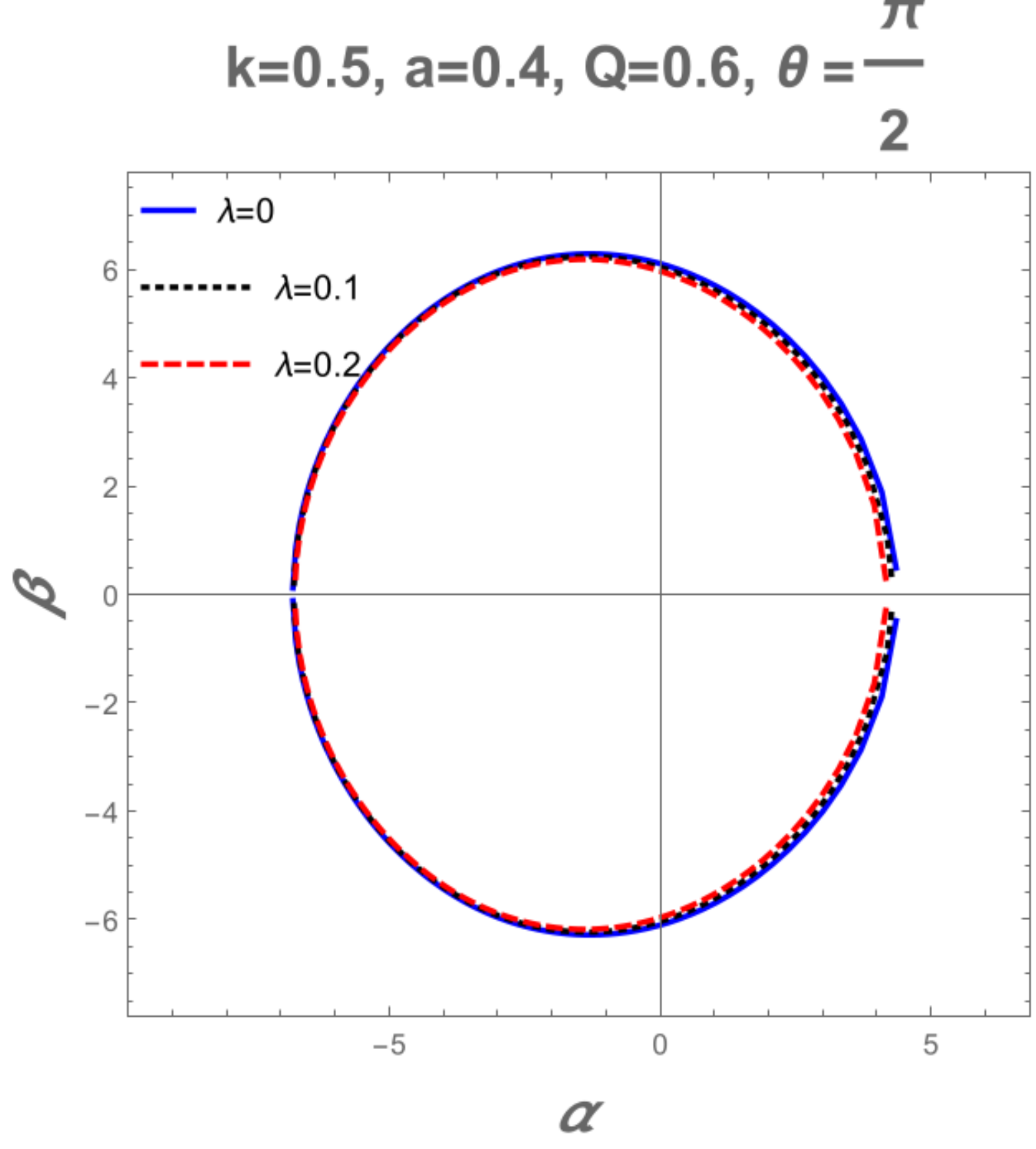}} 
		\subfigure[]{\includegraphics[width=5cm,height=5cm]{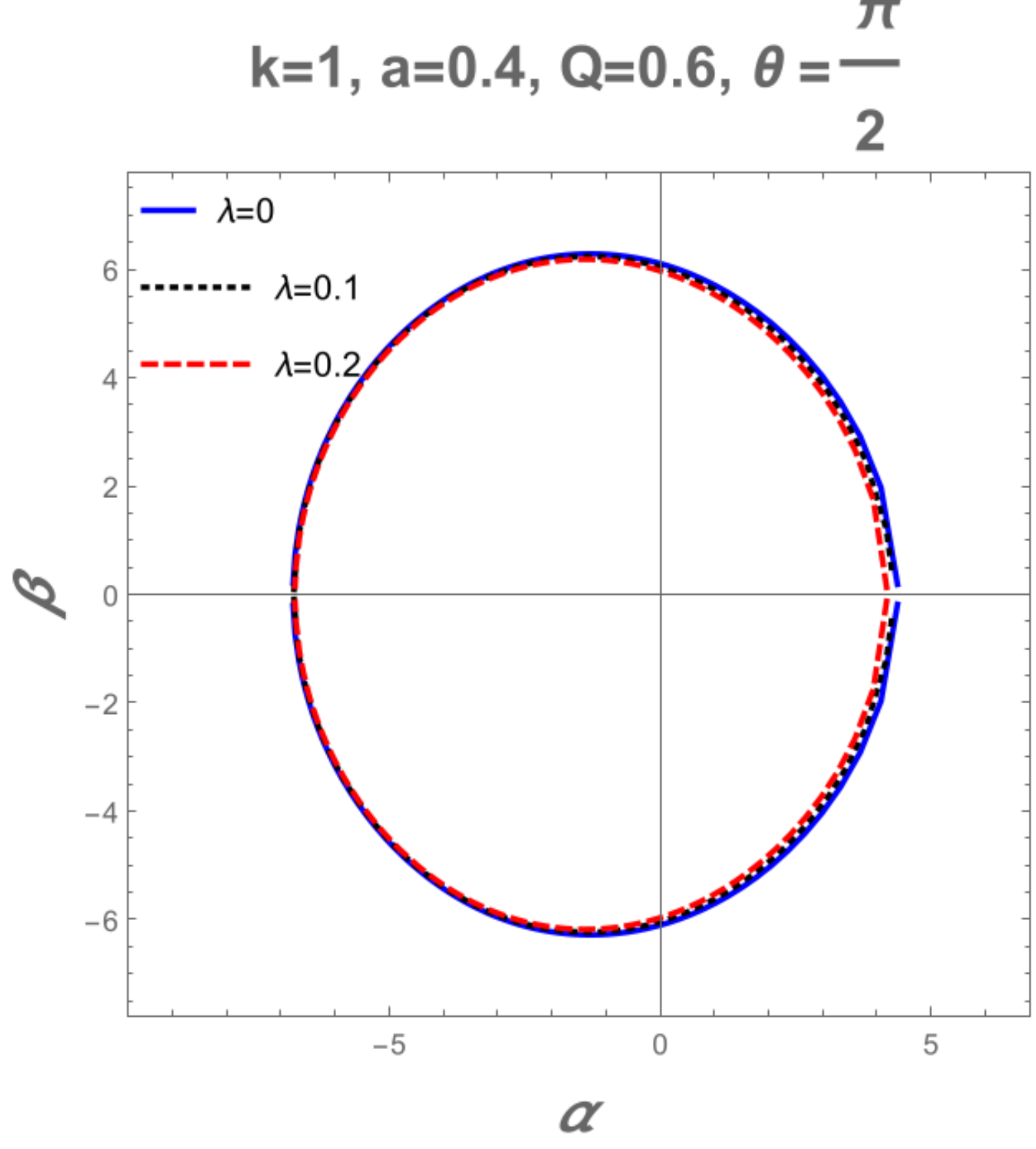}}
	\caption{Variation of radius of a BH shadow for different values of various parameters. The left panel represents the vacuum case (k=0), middle panel represents plasma medium with $k=0.5$ and right panel represents plasma medium with $k=1$.  } \label{f5}
\end{figure*}

The position of light source and observer play a crucial role to understand the shape of the shadow cast by any compact object. The BH shadow is defined as the reason of observer's sky which is left completely dark if there are light sources continuously distributed everywhere. This property of BH shadow remains unchanged when we examine the same effect from behind the observer \cite{perlick2021calculating}. One can consider the light rays coming from a source and an observer is situated at the coordinate $(r_{0}, \theta_{0})$, where $\theta_{0}$ is the inclination angle between the rotation axis of the BH and the line of sight of the observer with limit $r_{0} \rightarrow \infty$. If we define the conserved quantities {\it i.e.} angular momentum and energy parameter along with the separation constant $\mathcal{K}$, one can introduce the parameters $\xi = L/E$ and $\eta = \mathcal{K}/E^{2}$. By using the condition for unstable circular orbits, {\it i.e.} $	\mathcal{R}(r) =0$ and 	$d\mathcal{R}(r)/dr =0$, we have the following expressions,
\begin{equation}
	\left[X(r)-a\xi\right]^{2} - \Delta(r)\left[\eta+(\xi-a)^{2}\right]  \\ + X^{2}(r)(n^{2}-1) =0,   \label{e25}
\end{equation}
\begin{equation}
	2\left[X(r)-a\xi\right]X'(r) - \Delta'(r)\left[\eta+(\xi-a)^{2}\right] \\+ 2X(r)X'(r)(n^{2}-1)+2nn'X^{2}(r) =0.     \label{e26}
\end{equation}
By solving these two equations simultaneously, one can easily obtain the parameters $\xi$ and $\eta$ as,
\begin{equation}
	\xi = \frac{\mathcal{F}-\mathcal{G}}{\mathcal{H}} + \sqrt{\frac{\mathcal{G}^2}{\mathcal{H}^2} + \frac{\mathcal{I}}{\mathcal{H}}},  \label{e27}
\end{equation}
\begin{equation}
	\eta = \frac{\left[X(r)-a\xi\right]^{2}  + X^{2}(r)(n^{2}-1) - \Delta(r)(\xi-a)^{2}}{\Delta(r)},   \label{28}
\end{equation}
where the notations are defined as follows,
\begin{equation}
	\mathcal{H} = \frac{a^{2}}{\Delta(r)},    \label{e29}
\end{equation}
\begin{equation}
	\mathcal{F} = \frac{a^{2}}{\Delta(r)}X(r),  \label{e30}
\end{equation}
\begin{equation}
	\mathcal{G} = \frac{a^{2}}{\Delta(r)} \frac{\Delta(r)X'(r)}{\Delta'(r)},  \label{e31}
\end{equation}
and
\begin{equation}
	\mathcal{I} = \frac{n^{2}X(r)}{\Delta(r)} + \frac{2n^{2}X(r)-nn'X^{2}(r)}{\Delta'(r)}.  \label{e32}
\end{equation}
Here, prime denotes the differentiation with respect to radial coordinate $r$. one can obtain the apparent shape of the BH shadow by using the celestial coordinates as defined below,
\begin{equation}
	\alpha = \lim_{r_{0} \rightarrow \infty} \left(-r_{0}^{2} sin \theta_{0}\frac{d\phi}{dr}\arrowvert_{(r_{0}, \theta_{0})}\right),  \label{e33}
\end{equation}
\begin{equation}
	\beta = \lim_{r_{0} \rightarrow \infty} \left(r_{0}^{2} \frac{d\theta}{dr}\arrowvert_{(r_{0}, \theta_{0})}\right),  \label{e34}
\end{equation}
where, $(r_{0}, \theta_{0})$ are the position coordinates of the observer's plane. After taking on account the limit, we obtain,
\begin{equation}
	\alpha = -\frac{\xi}{n sin \theta_{0}},   \label{e35}
\end{equation}
\begin{equation}
	\beta = \pm \frac{\sqrt{\eta + a^{2}- n^{2} a^{2} sin^{2} \theta_{0} - \xi^{2}cot^{2} \theta_{0}}}{n}.   \label{e36}
\end{equation}
These equations represent a direct relationship between celestial coordinates $(\alpha, \beta)$ and the parameters $\xi$, $\eta$. We have studied the variation of BH shadow for the particular choices of parameters involved in this BH spacetime which can be visualized in \figurename{ \ref{f3}} to \figurename{ \ref{f5}} respectively. The behaviour of BH shadow from different inclination angle and different values of spin and charge parameter in presence of plasma medium  represented in \figurename{ \ref{f3}} and \figurename{ \ref{f4}} respectively. From these figures, it is observed that the presence of plasma medium affects the shape and size of BH shadow to be increased gradually. The radius of shadow also depends on the inclination angle and we have examined that the shadow radius increases with the decrease in inclination angle. In \figurename{ \ref{f4}}, we have simultaneously shown the effect of spin, magnetic charge and YM parameter by varying these parameters for vacuum case as well as for different values of plasma parameter. It is also investigated that the shadow radius decreases with the increase in spin and magnetic charge parameter while with the increase of YM parameter, the shadow radius decreases monotonically.

\begin{figure*}[h]
	\centering
		\includegraphics[width=9cm,height=7cm]{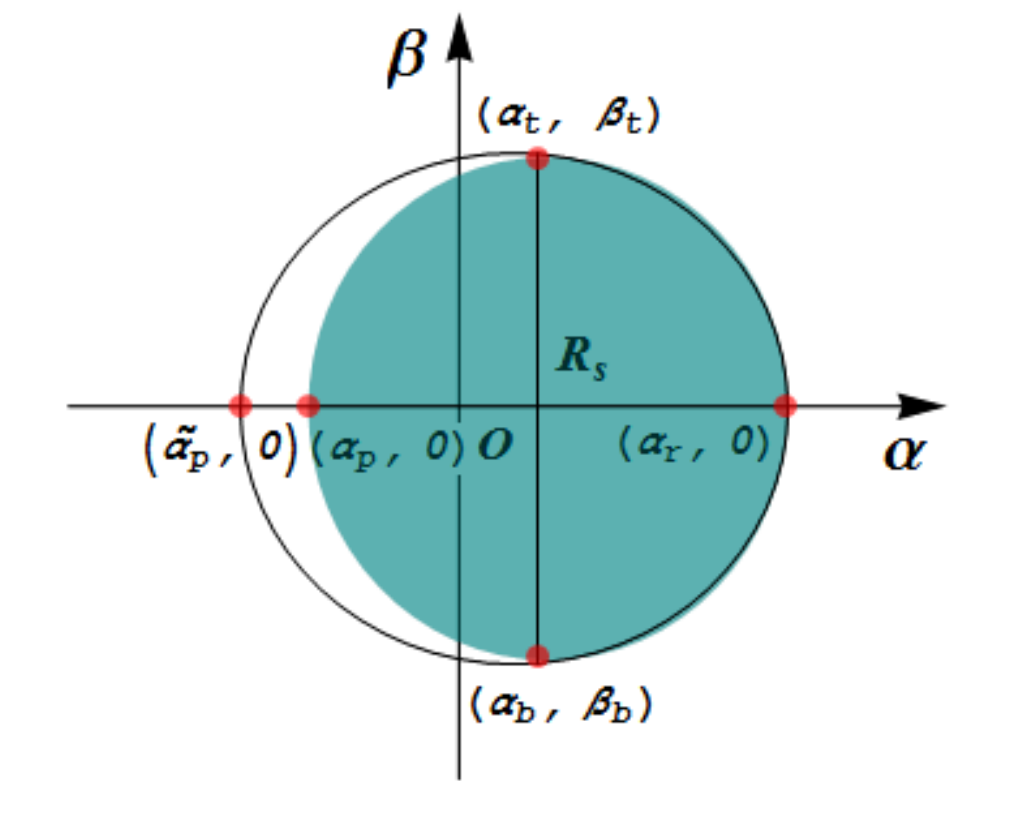}
	\caption{Illustration of the observables of the BH Shadow \cite{abdujabbarov2017shadow}. } \label{f6}
\end{figure*}

\begin{figure*}[h]
	\centering
		\subfigure[]{\includegraphics[width=7cm,height=4cm]{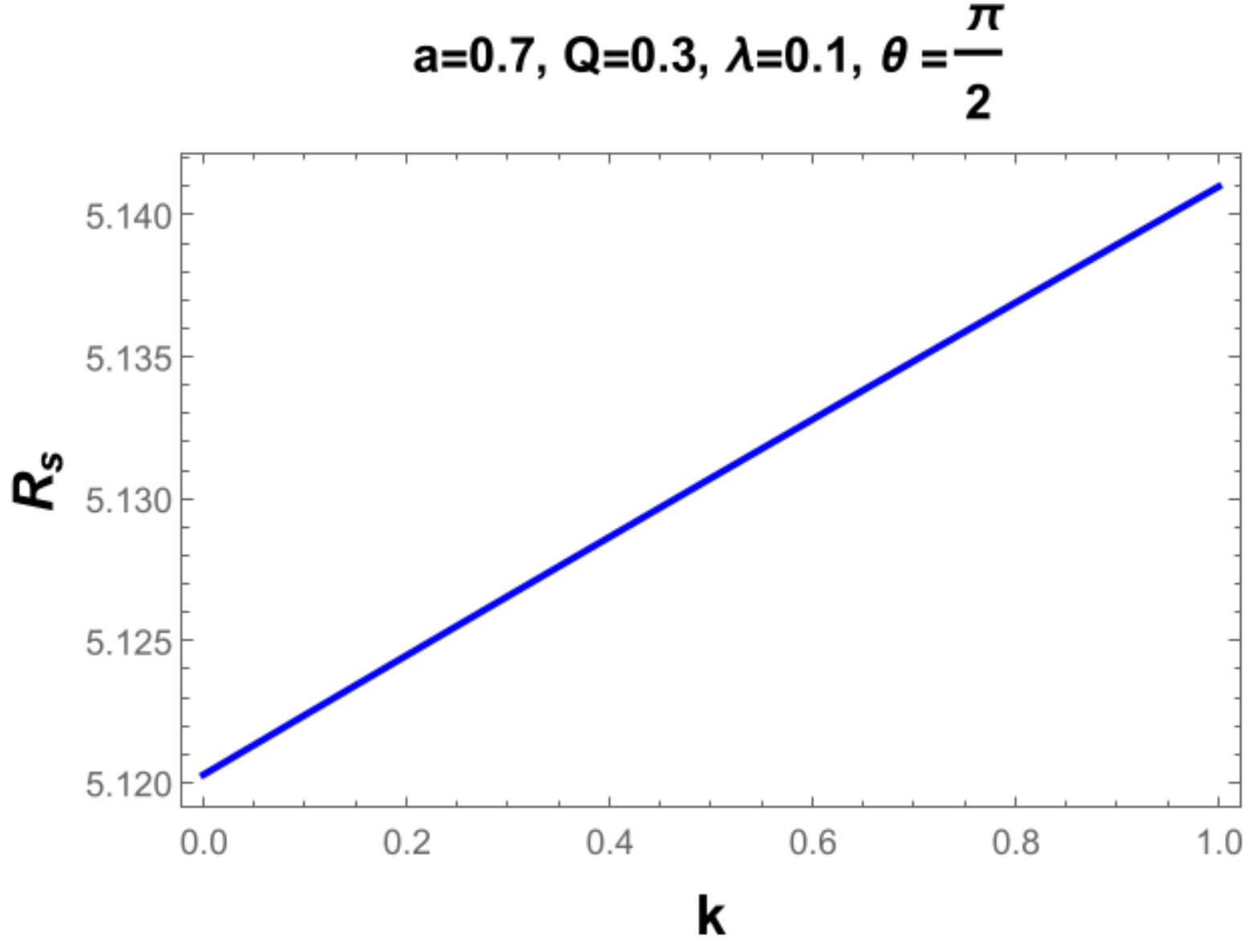}} 
		\subfigure[]{\includegraphics[width=7cm,height=4cm]{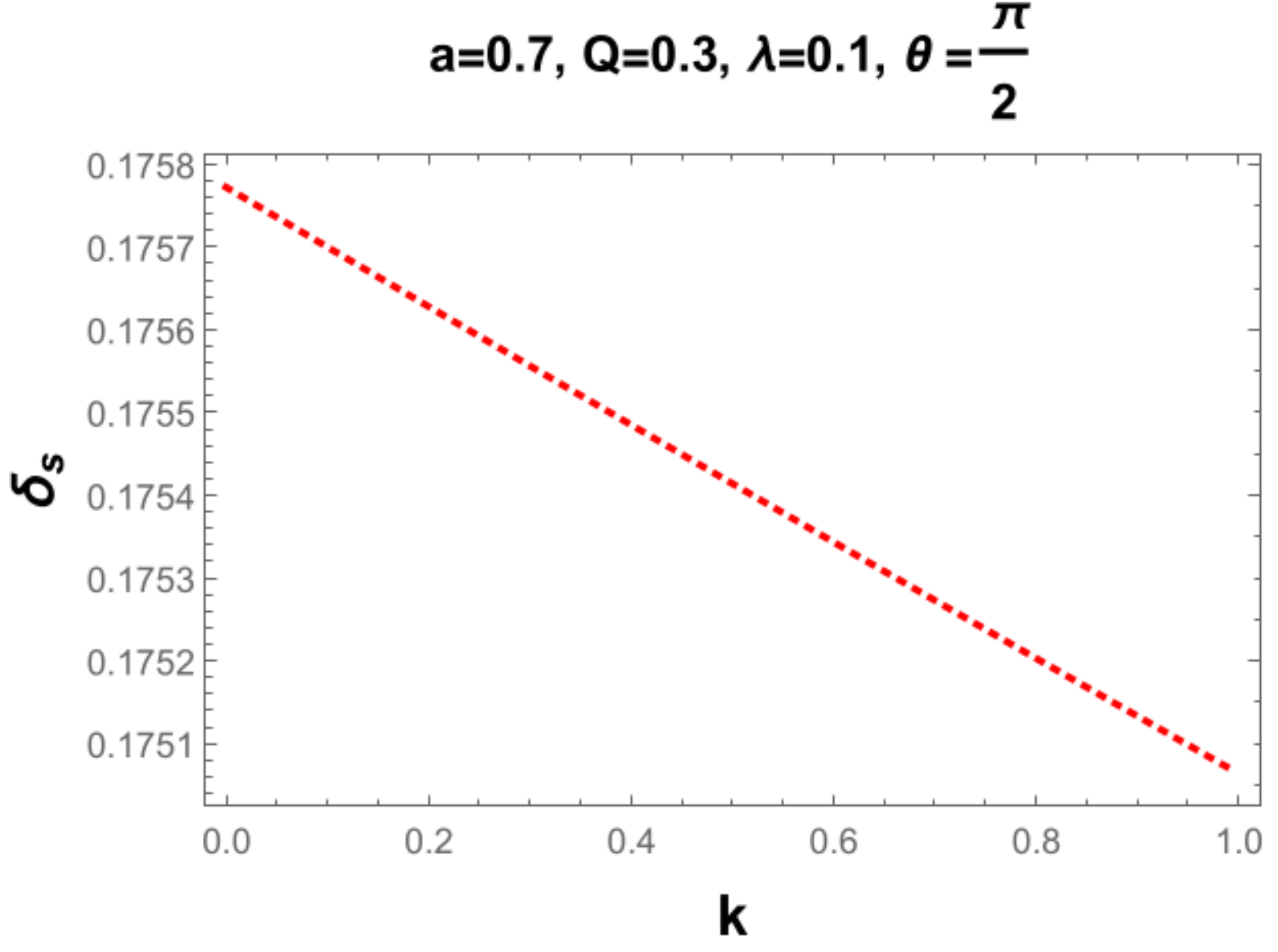}}\\
		\subfigure[]{\includegraphics[width=7cm,height=4cm]{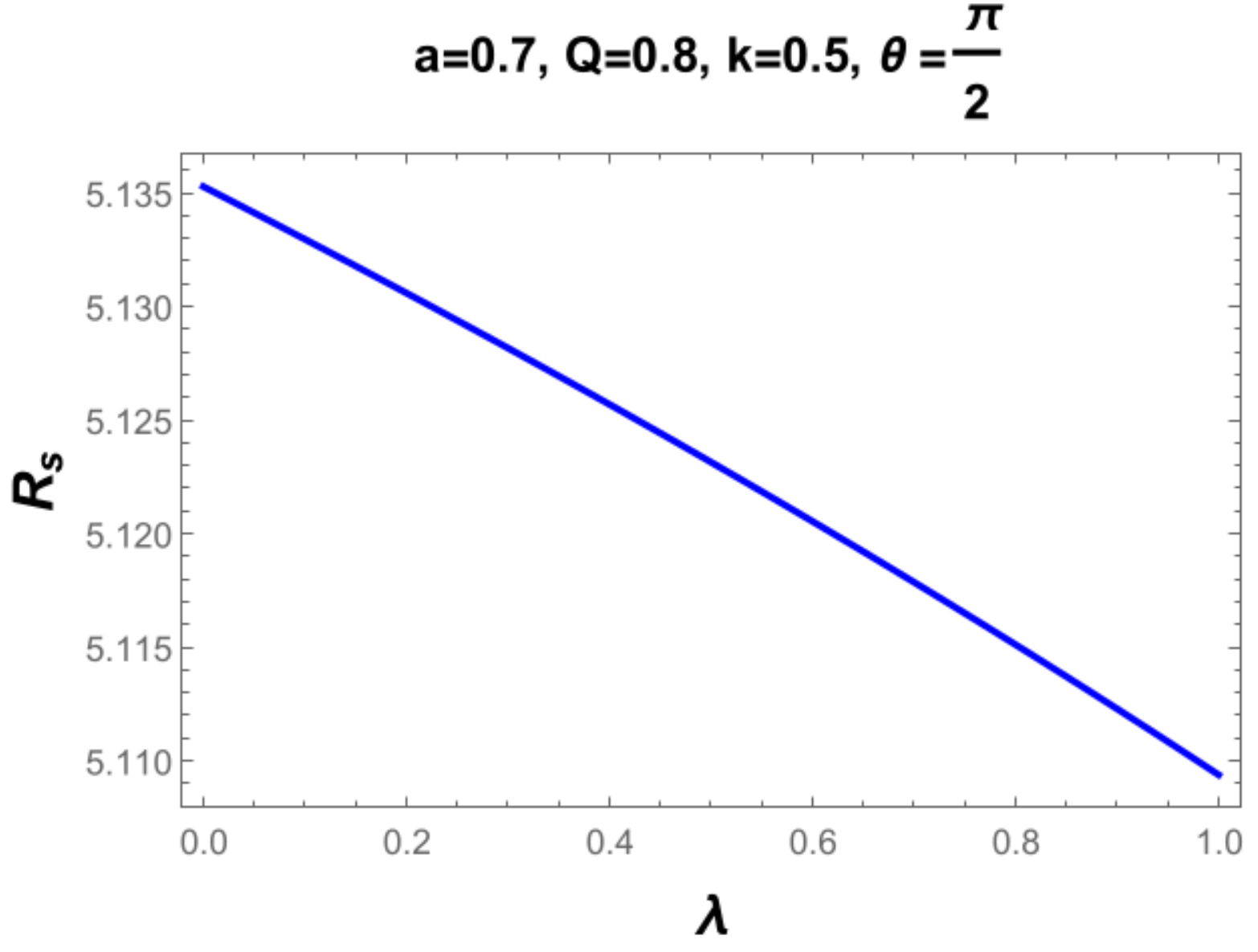}} 
		\subfigure[]{\includegraphics[width=7cm,height=4cm]{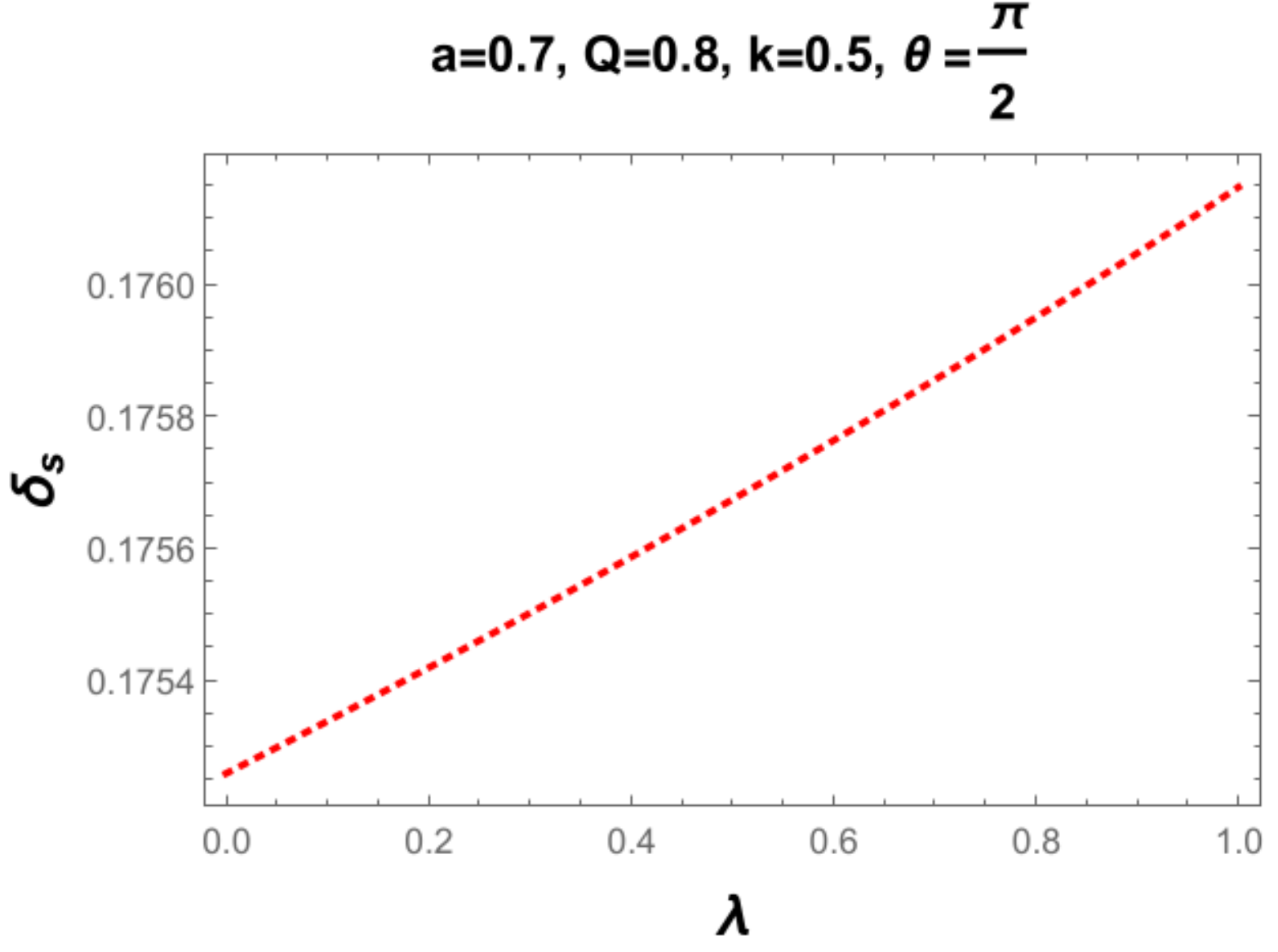}}
	\caption{Variation of radius of a BH shadow (left panel) and deformation parameter (right panel) with plasma and YM parameter respectively.} \label{f7}
\end{figure*}

\noindent The shadow radius and distortion parameter are two astronomical observables those are useful to analyse the shape of BH shadow in detail. The approximate size of the shadow of BH can be described with the help of shadow radius while distortion parameter measures the distortion appearing in shadow. In order to examine the shadow radius, we assume a circle passing through the different points,  $(\alpha_{t},\beta_{t})$,  $(\alpha_{b},\beta_{b})$ and  $(\alpha_{r},0)$ respectively. The representation of all these points is depicted in \figurename{ \ref{f6}}. The radius of BH shadow can be calculated through
\begin{equation}
	R_{s} = \frac{(\alpha_{t}-\alpha_{r})^2 + \beta_{t}^2}{2 \lvert \alpha_{t}-\alpha_{r} \rvert },   \label{e37}
\end{equation}
while the distortion parameter as given as
\begin{equation}
	\delta_{s} = \frac{(\tilde{\alpha_{p}}-\alpha_{p})}{R_{s}}.   \label{e38}
\end{equation}
In \figurename{ \ref{f6}}, the points $(\tilde{\alpha_{p}},0)$ and $(\alpha_{p},0)$ intersect the coordinate plane at the opposite side of $(\alpha_{r},0)$. However, the distance between reference circle and the left point of the shadow has represented by the point $(\tilde{\alpha_{p}}-\alpha_{p})$. The variation of the observables $R_{s}$ and $\delta_{s}$ with plasma parameter and YM parameter can be seen in \figurename{ \ref{f7}}. It is observed that the radius of shadow increases with an increase of plasma parameter and the distortion is reduced with the increase of plasma parameter. On other side, in case of YM parameter, the shadow radius decreases with increasing distortion parameter. Furthermore, in \figurename{ \ref{f5}}, if $\lambda=0$ then the results automatically reduce to the Kerr-Newman BH case with magnetic charge instead of electric charge and the significant change in shadow radius is clearly visible.

\subsection{Energy Emission Rate of a non-rotating BH in EYM theory}
\label{subsec:5.1}
In this section, we study the energy emission rate of a regular and rotating magnetically charged BH with a YM electromagnetic source in the non-minimal coupled EYM theory. The expression of energy emission rate reads as, 
\begin{equation}
	\frac{d^{2} Z(\omega)}{d\omega dt} = \frac{2 \pi^{2} \sigma_{lim}}{\exp(\frac{\omega}{T_{H}})-1} \omega^{3},
\end{equation}
where the parameters $Z(\omega)$, $\omega$ and $T_{H}$ represent the energy, frequency and Hawking temperature respectively corresponding to the BH. The expression of limiting constant value $\sigma_{lim}$ for rotating charged accelerating BH can be expressed as
\begin{equation}
	\sigma_{lim} = \pi 	R_{s}^{2},
\end{equation}
here, $R_{s}^{2}$ is the shadow radius of the BH. Therefore, the expression of energy emission rate for this BH spacetime becomes
\begin{equation}
	\frac{d^{2} Z(\omega)}{d\omega dt} = \frac{2 \pi^{3} 	R_{s}^{2}} {\exp(\frac{\omega}{T_{H}})-1} \omega^{3}.
\end{equation} 

\begin{figure*}[h]
	\centering
		\subfigure[]{\includegraphics[width=7cm,height=4cm]{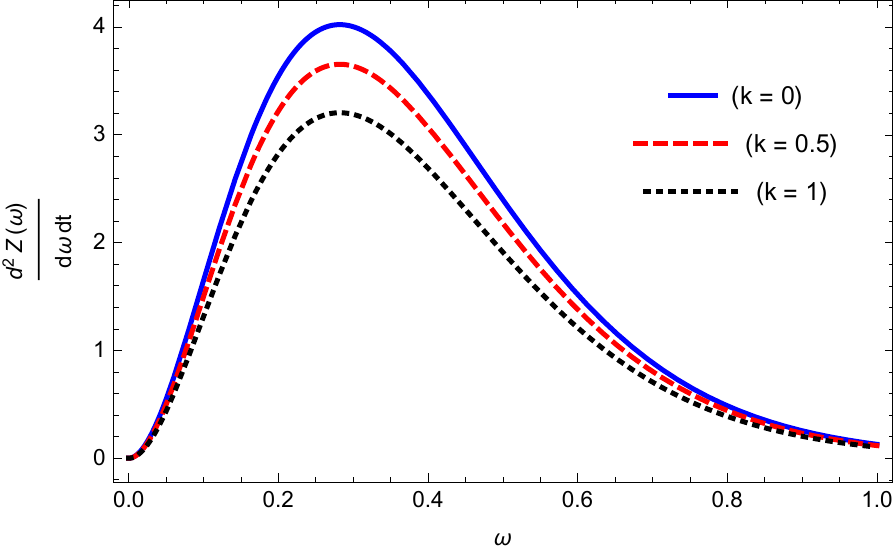}}
		\subfigure[]{\includegraphics[width=7cm,height=4cm]{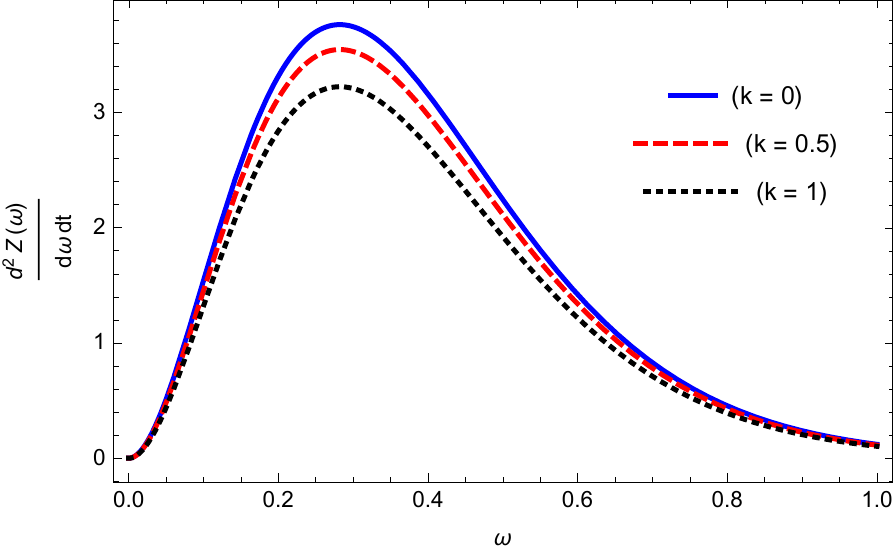}}\\
		\subfigure[]{\includegraphics[width=7cm,height=4cm]{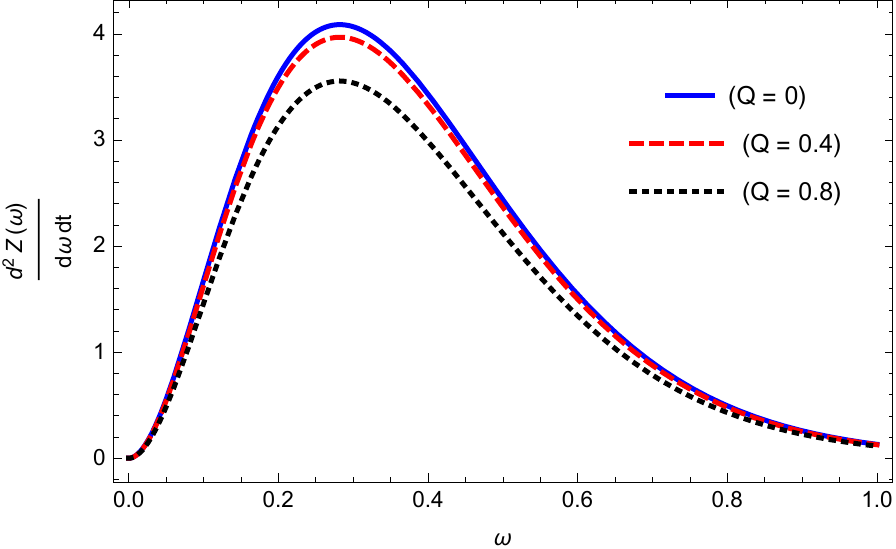}} 
		\subfigure[]{\includegraphics[width=7cm,height=4cm]{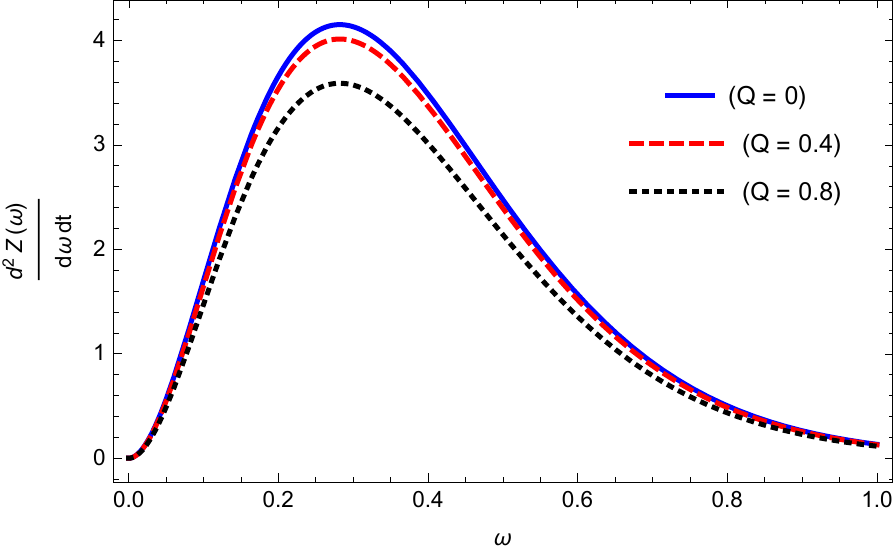}}
	\caption{Variation of energy emission rate with frequency for different values of plasma and charge parameter. Here, $\lambda=0$ for left panel and $\lambda=0.5$ for right panel.} \label{ee}
\end{figure*}
\figurename{ \ref{ee}}, shows the variation of energy emission rate with frequency for different values of charge and plasma parameter. It has been observed that as plasma parameter increases, the rate of energy emission decreases which marks the opposite behavior of plasma when we compare it to Kerr-Newman BH spacetime case in GR. This opposite behavior of rate of energy emission is due to the presence of magnetic charge unlike the usual Kerr-Newman BH which contains electric charge. Further, we have also observed that in the absence of charge parameter the energy emission rate of BH is maximum and when we consider the charge parameter, it decreases with an increase in the charge parameter. Apart from these, one of the noticeable point observed here is that the energy emission rate of BH is unaffected due the presence of YM parameter.  \\ 

\section{Weak field lensing in presence of plasma}
\label{sec:6}
In this section, the study of lensing in weak field limit is performed by considering the non-rotating analogue of this BH spacetime in the weak field approximation which is expressed by the following relation,
\begin{equation*}
	g_{\mu\nu} = \eta_{\mu\nu} + h_{\mu\nu}, \hspace{1cm} \eta_{\mu\nu} = (-1,1,1,1), \hspace{1cm}	h_{\mu\nu} << 1,
\end{equation*}
\begin{equation}
	h_{\mu\nu} \rightarrow 0 \hspace{1cm} \text{under} \hspace{1cm}  x^{i} \rightarrow \infty.
\end{equation}
Here, $g_{\mu\nu}$ is the metric tensor defined as, $ds^{2}=g_{\mu\nu}dx^{\mu}dx^{\nu}$ and $h_{\mu\nu}$ is the flat spacetime metric $(-1,1,1,1)$ with a small perturbation. The contravariant tensors of the metric are given as 
\begin{equation}
	g^{\mu\nu} = \eta^{\mu\nu} - h^{\mu\nu}, \hspace{1cm} h^{\mu\nu} = h_{\mu\nu}.
\end{equation}
In the weak plasma strength, taking into account the weak field approximation for propagation of photons along $z$ direction, one can introduce the deflection angle in the following form
\begin{equation}
	\hat{\delta_{p}} = -\frac{1}{2} \int_{-\infty}^{\infty} \left(h_{33}+\frac{h_{00}\omega^{2}-\omega_{e}^{2}}{\omega^{2}-\omega_{e}^{2}}\right)_{,k} dz.
\end{equation}
In the limit of large radial distance, the metric for BH spacetime can be written as, \\
\begin{equation}
	ds^{2} = ds_{0}^{2} +\left(\frac{2 M}{r}-\frac{Q^2}{r^2} +\frac{2(2 M r-Q^2)\lambda}{r^6}\right) dt^{2} \\+ \left(\frac{2 M}{r} -\frac{Q^2}{r^2}+\frac{2(2 M r-Q^2)\lambda}{r^6}\right) dr^{2},
\end{equation} 
where the flat part of above metric is $ ds_{0}^{2} = -dt^{2} + dr^{2} + r^{2} (d\theta^{2} + \sin^{2}\theta d\phi^{2})$. The components $h_{\mu\nu}$ in terms of cartesian coordinates are written as,
\begin{eqnarray}
	\nonumber	h_{00} &=& \left(\frac{R_{g}}{r}-\frac{Q^2}{r^2}+\frac{2(R_{g} r-Q^2)\lambda}{r^6}\right),\\
	\nonumber	h_{ik} &=& \left(\frac{R_{g}}{r}-\frac{Q^2}{r^2}+\frac{2(R_{g} r-Q^2)\lambda}{r^6}\right)n_{i} n_{k},\\
	h_{33} &=& \left(\frac{R_{g}}{r}-\frac{Q^2}{r^2}+\frac{2(R_{g} r-Q^2)\lambda}{r^6}\right) \cos^{2} \chi.	
\end{eqnarray} 
Here, $R_{g} = 2 M$, $n_{i}$ is the unit vector of three-radius vector $r_{i}=x_{1},x_{2},x_{3}$, the angle $\chi$ represents the polar angle between three-vector $r^{i}=r_{i}$ at $z$-axis, and $n_{3} = \cos \chi = z/r = z/\sqrt{b^2+z^2}$. The impact parameter of BH is represented by $b$ and in this geometry, it is defined as $b^{2}=x_{1}^2+x_{2}^2$. Now, in the weak filed limit, the deflection angle of BH surrounded by a homogeneous plasma medium can be obtained in the following form,
\begin{equation}
	\hat{\delta_{p}} = -\frac{1}{2} \int_{-\infty}^{\infty} \frac{\partial}{\partial b} \left[I_{1}(b) + I_{2}(b) \right] dz,
\end{equation}
where,
\begin{equation}
	I_{1}(b) = \left[\frac{R_{g}}{\sqrt{b^2+z^2}} -\frac{Q^2}{b^2+z^2}+\frac{2(R_{g} r-Q^2)\lambda}{(b^2+z^2)^3}\right]\frac{z^2}{b^2+z^2},
\end{equation}
and
\begin{equation}
	I_{2}(b) = \frac{1}{1-\omega_{e}^2/\omega^2} \left[\frac{R_{g}}{\sqrt{b^2+z^2}} -\frac{Q^2}{b^2+z^2}+\frac{2(R_{g} r-Q^2)\lambda}{(b^2+z^2)^3}\right].
\end{equation}
The photon frequency at large radial distance is given as, 
\begin{equation}
	\omega^{2} = \frac{\omega_{\infty}^{2}}{\left[1-\frac{R_{g}}{r}-\frac{Q^2}{r^2}+\frac{2(R_{g} r-Q^2)\lambda}{r^6}\right]}.
\end{equation}
Here, $\omega_{\infty}$ represents the photon frequency in terms of asymptotic value. One can expand the expression of refractive index in series on the powers of $1/r$ within the approximation of the YM, charge and large distance, we have,
\begin{equation}
	n^{2} = \left(1-\frac{\omega_{e}^{2}}{\omega^{2}}\right)^{-1} \simeq 1 + \frac{4 \pi e^2 N_{0} r_{0}}{m \omega_{\infty}^2 r} - \frac{4 \pi e^2 N_{0} r_{0} R_{g}}{m \omega_{\infty}^2 r^2}.
\end{equation}
\\ \vspace{0.1cm}
\noindent Using the above approximation, one can calculated the deflection angle  of light for this BH spacetime in presence of plasma medium as, \\

	\begin{eqnarray}
		\hat{\delta}_{EYM(P)} &=& \frac{2 R_{g}}{b} \left[1+ \frac{ \pi^2 e^2 N_{0} r_{0}}{m \omega_{\infty}^2 b} - \frac{4 \pi e^2 N_{0} r_{0}R_{g}}{m \omega_{\infty}^2 b^2}\right] -\frac{Q^2}{4 b^2} \left[3\pi + \frac{4 \pi e^2 N_{0} r_{0}}{m \omega_{\infty}^2 b} \left(8-\frac{3 \pi R_{g}}{b}\right)\right]
		\nonumber \\&& \hspace{2.5cm} + \frac{\lambda (Q^2-R_{g})}{8 b^3} \left[5 \pi + \frac{4 \pi e^2 N_{0} r_{0}}{m \omega_{\infty}^2 b^2} \left(15-\frac{5 \pi R_{g}}{b}\right) \right]. \label{dangle}
	\end{eqnarray} 

     	\begin{table}
     		\centering
     	\begin{tabular}{|c|c|c|c|c|c|c|c|}
     		\hline 
     		\multicolumn{2}{|c|}{$\lambda$}   & \multicolumn{1}{|c|}{0} & \multicolumn{1}{|c|}{0.1} & \multicolumn{1}{|c|}{0.2} &  \multicolumn{1}{|c|}{0.3} &  \multicolumn{1}{|c|}{0.4} & \multicolumn{1}{|c|}{0.5} \\
     		\hline
     		$ \delta $ & Q=0 & 1.14286 & 1.07874 & 1.01463 & 0.950515 & 0.886401& 0.822286  \\
     		\hline
     		$ \delta $ & Q=0.2 & 1.13516 & 1.07233 & 1.0095 & 0.946668 & 0.883836 & 0.821004   \\
     		\hline
     		$ \delta $ & Q=0.4 & 1.11208 & 1.0531 & 0.994112 & 0.935127 & 0.876142 & 0.817157  \\
     		\hline
     		$ \delta $ & Q=0.6 &1.07361& 1.02104& 0.968467& 0.915893& 0.86332& 0.810746   \\
     		\hline
     		$ \delta $ & Q=0.8 &1.01976 & 0.97616 & 0.932563 & 0.888965 & 0.845368 & 0.80177  \\
     		\hline
     	\end{tabular}
     	\caption{Variation of deflection angle for M = 1, b =3.5 and k = 0.}\label{table1}
     	\end{table}

	\begin{table}
		\centering
		\begin{tabular}{|c|c|c|c|c|c|c|c|}
			\hline 
			\multicolumn{2}{|c|}{$\lambda$}   & \multicolumn{1}{|c|}{0} & \multicolumn{1}{|c|}{0.1} & \multicolumn{1}{|c|}{0.2} &  \multicolumn{1}{|c|}{0.3} &  \multicolumn{1}{|c|}{0.4} & \multicolumn{1}{|c|}{0.5} \\
			\hline
			$ \delta $ & k=0 & 1.14286 & 1.07874 & 1.01463 & 0.950515 & 0.886401 & 0.822286  \\
			\hline
			$ \delta $ & k=0.2 & 1.15683 & 1.09266 & 1.02849 & 0.964316 & 0.900145 & 0.835973   \\
			\hline
			$ \delta $ & k=0.4 & 1.1708 & 1.10658 & 1.04235 & 0.978118 & 0.913889 & 0.84966  \\
			\hline
			$ \delta $ & k=0.6 &1.18478 & 1.12049 & 1.05621 & 0.991919 & 0.927633 & 0.863347   \\
			\hline
			$ \delta $ & k=0.8 &1.19875 & 1.13441 & 1.07006 & 1.00572 & 0.941377 & 0.877034  \\
			\hline
		\end{tabular}
	\caption{Variation of deflection angle for M = 1, b =3.5 and Q = 0.}\label{table2}
	\end{table}

\begin{figure*}[h]
	\centering
		\subfigure[]{\includegraphics[width=6cm,height=4cm]{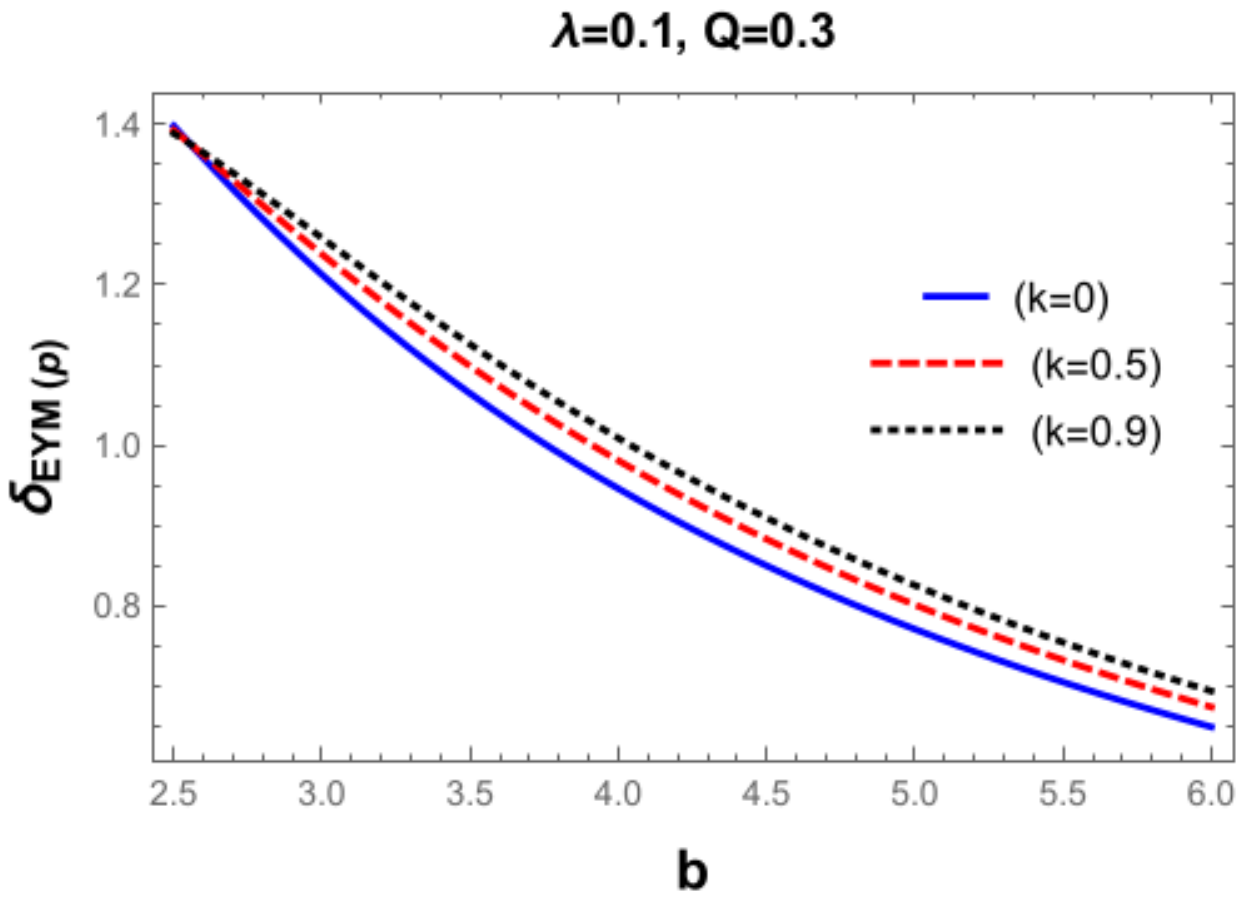}} 
		\subfigure[]{\includegraphics[width=6cm,height=4cm]{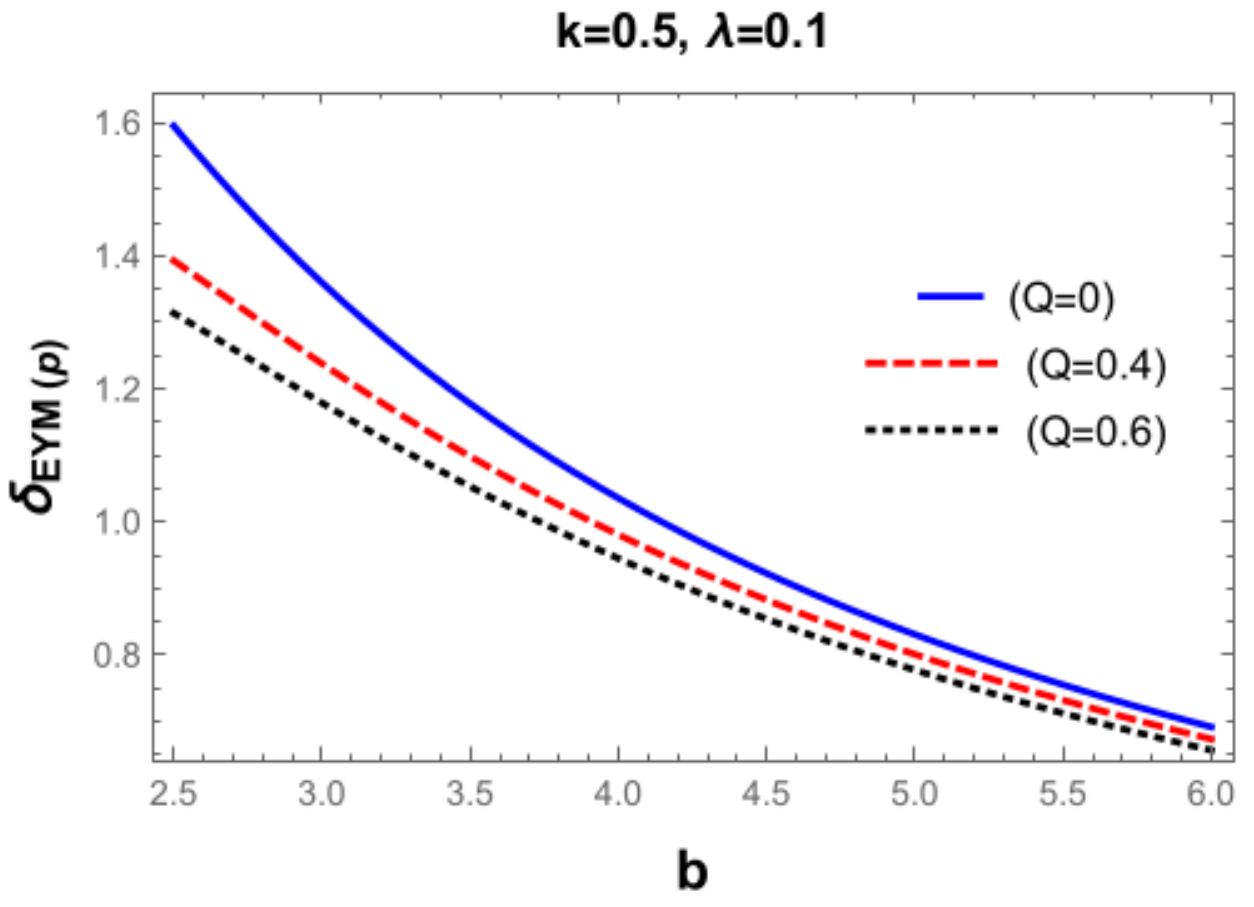}} \\
		\subfigure[]{\includegraphics[width=6cm,height=4cm]{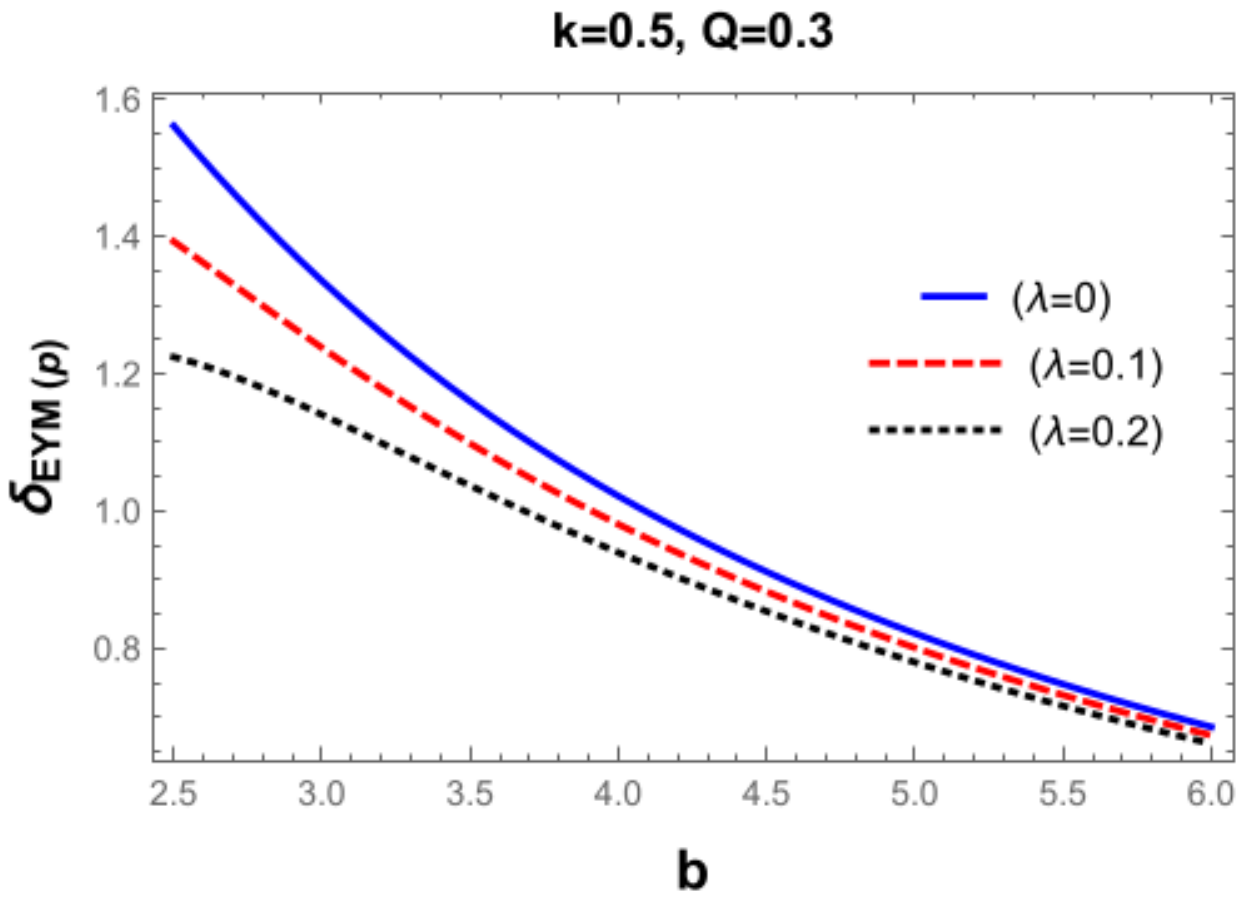}} 
		\subfigure[]{\includegraphics[width=6cm,height=4cm]{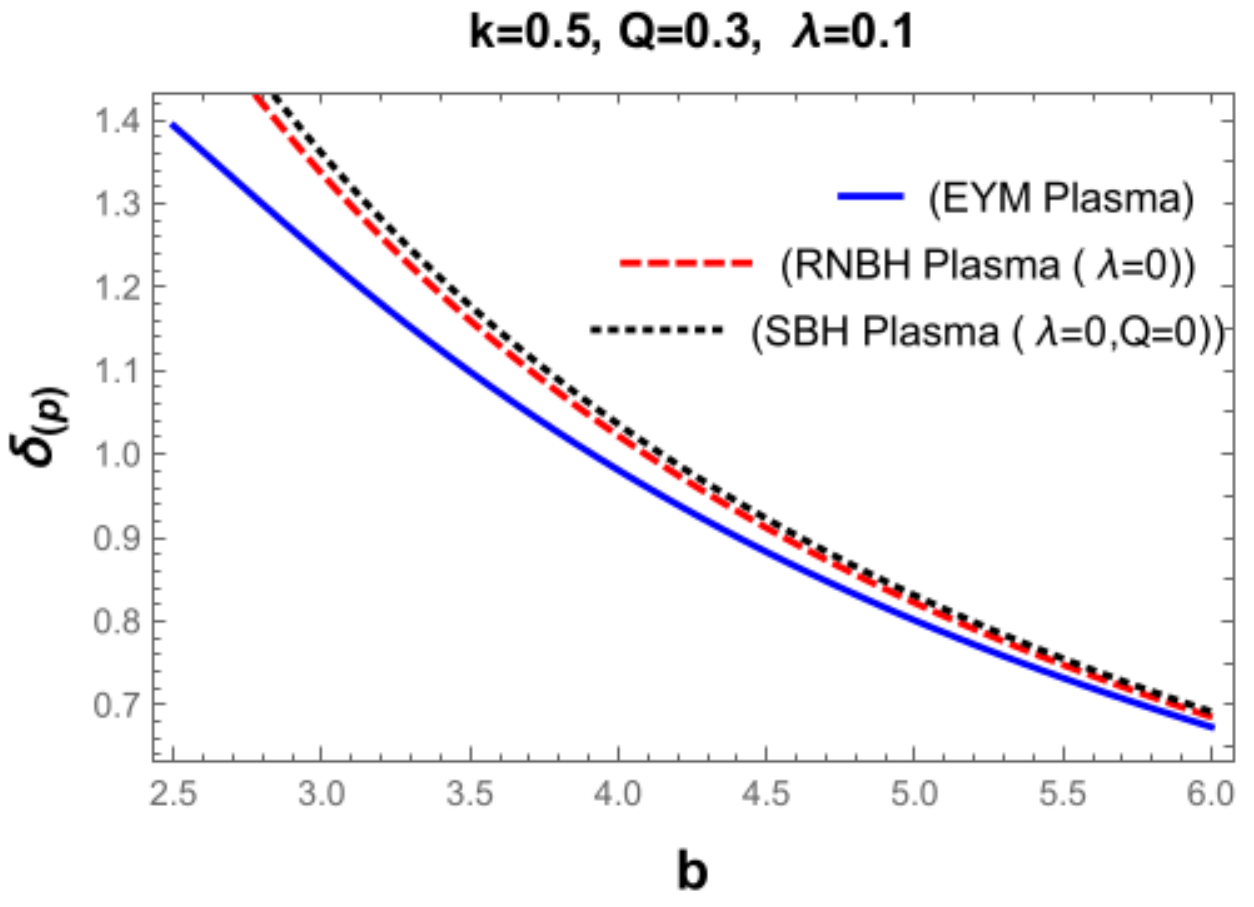}}
	\caption{Variation of deflection angle with the impact parameter.} \label{da}
\end{figure*}
 Eq. \ref{dangle} represents the deflection angle of EYM BH surrounded by a plasma medium. In this equation, the first term is an additive correction to the gravitational deflection due to plasma medium while in the absence of plasma medium, the first term gives vacuum gravitational deflection.  The second and third terms correspond to the charge parameter and YM parameter in presence of plasma parameter respectively. The expression of deflection angle (Eq. \ref{dangle}) reduces to the usual {\bf Reissner–Nordstr{\"o}m} BH case in GR with a magnetic charge in the prescribed limit $\lambda=0$, and it further reduces to Schwarzschild BH if $\lambda=0$ \& $Q=0$. {\bf It can observed from \tablename{ \ref{table1}}, that the value deflection angle decreases with increase in the value of $ \lambda $}. The comparison of these BHs through pictorial representation are depicted in  \figurename{ \ref{da}} (d). It can be easily seen that the presence of additional parameters {\it i.e.} YM and charge parameter, in plasma medium leads to a decrease in the deflection angle when we compared it to the Schwarzschild BH in plasma medium. The dependence of the deflection angle on the impact parameter for different values of charge, YM and plasma parameter is demonstrated in  \figurename{ \ref{da}}. In the upper left panel, the deflection angle increases with an increase in plasma parameter and the deviation of light {\it i.e.} photons is smaller in the vacuum case ($k=0$). In the upper right panel and lower left panel, the deflection angle for different values of charge and YM parameters are depicted receptively. In both cases, we have observed that the deflection angle decreases with an increase in charge as well as the YM parameter.  Further, the small difference of the YM parameter shows the large deviation in deflection angle which is the effect of magnetic charge and indicates the photons are reflected abruptly in the plasma medium.
\section{Conclusions}
\label{sec:7}
\textbf{In this paper, we have studied the effect of homogeneous plasma medium on the shadow and of a rotating BH in a non-minimally coupled EYM theory}. First, we have obtained the null geodesic equations by solving the Hamilton-Jacobi equation. The effective potential and radius of the photon sphere are calculated and their typical variation with different parameters is depicted in \figurename{ \ref{f1}}. Further, we have derived the necessary analytical expression to obtain the shadow of BH in the plasma medium. The location of the observer is an important aspect of the BH shadow and we have examined the images of BH shadow for different inclination angles and it is observed that if the observer locates near the equatorial plane, the size of shadow decreases while the deformation parameter ($\delta_{s}$) increases. Furthermore, it is observed from \figurename{ \ref{f4}} that the presence of plasma medium makes the size of shadow radius enhanced. However, shadow radius decreases with an increase in charge parameter and shows the opposite behavior in the case of a spin parameter (see \figurename{ \ref{f5}}). The effect of the YM parameter on BH shadow has depicted in \figurename{ \ref{f6}} which has a trivial effect and in absence of the YM parameter, our results reduce to Kerr-Newman BH case. We have also obtained from the observables that shadow radius decreases with the increase of the YM parameter and the deformation parameter increases significantly. We have further observed that the energy emission rate increase with a decrease in the value of plasma parameter and the emission rate attains maximum value for the vacuum case {\it i.e.} $k=0$. It is noticeable that the appearance of the YM parameter does not affect significantly the rate of energy emission. Finally, we have computed the deflection angle of massless particles by considering the weak field limit around this BH spacetime in presence of the homogeneous plasma medium and represented them graphically in \figurename{ \ref{da}}. It is also investigated that, the presence of homogeneous plasma parameter increases the deflection angle along with the impact parameter attaining its saturation. While on the other hand, the deflection angle decreases with an increase in the charge and YM parameter and deviation of photons due to both of theses parameters are higher than in the case of plasma medium. The obtained results also reduce to well-known BHs in GR within the prescribed limit as mentioned previously and the difference between them can be seen in \figurename{ \ref{da}} (d). \textbf{In presence of homogeneous plasma, the deflection angle of this particular BH spacetime in the prescribed limits is compared with other well-known BH spacetime in the GR}. The present investigations comprise a more general view of the optical studies of this BH carried out in past in presence of the plasma medium. Since the astronomical BHs in the universe are generally surrounded by the gases accreting around them and therefore, this study will provide a mathematical model to learn about the behavior of photons in these extreme conditions, {\it i.e.} in the background of the plasma. This model can further be improvised by comparing the results obtained with the observed data of supermassive BH in M87 galaxy and it is a matter of separate discussion. We intend to report on this issue in near future. 


\section{acknowledgements}
The authors HN and SK are thankful to the Uttarakhand State Council of Science and Technology (UCOST), Dehradun for financial assistance through R\&D grant number UCS\&T/RD-18/18-19/16038/4. The authors HN and PS acknowledge the financial support provided by Science and Engineering Research Board (SERB), New Delhi through the grant number EMR/2017/000339. All authors also acknowledge the facilities at ICARD, Gurukula Kangri (Deemed to be University), Haridwar, India.

%
%
%

%
%
\bibliographystyle{unsrt} 

\bibliography{eymbib}

\begin{thebibliography}{100}

\bibitem{Hartle:2003yu}
J.~B. Hartle.
\newblock {\em {An introduction to Einstein's general relativity}}.
\newblock 2003.

\bibitem{joshi1993global}
Pankaj~S Joshi.
\newblock Global aspects in gravitation and cosmology.
\newblock {\em Int. Ser. Monogr. Phys}, 87, 1993.

\bibitem{chandrasekhar1998mathematical}
Subrahmanyan Chandrasekhar.
\newblock {\em The mathematical theory of black holes}, volume~69.
\newblock Oxford University Press, 1998.

\bibitem{Akiyama:2019cqa}
Kazunori Akiyama et~al.
\newblock {First M87 Event Horizon Telescope Results. I. The Shadow of the
  Supermassive Black Hole}.
\newblock {\em Astrophys. J.}, 875(1):L1, 2019.

\bibitem{Akiyama:2019brx}
Kazunori Akiyama et~al.
\newblock {First M87 Event Horizon Telescope Results. II. Array and
  Instrumentation}.
\newblock {\em Astrophys. J. Lett.}, 875(1):L2, 2019.

\bibitem{Akiyama:2019sww}
Kazunori Akiyama et~al.
\newblock {First M87 Event Horizon Telescope Results. III. Data Processing and
  Calibration}.
\newblock {\em Astrophys. J. Lett.}, 875(1):L3, 2019.

\bibitem{Akiyama:2019bqs}
Kazunori Akiyama et~al.
\newblock {First M87 Event Horizon Telescope Results. IV. Imaging the Central
  Supermassive Black Hole}.
\newblock {\em Astrophys. J. Lett.}, 875(1):L4, 2019.

\bibitem{Akiyama:2019fyp}
Kazunori Akiyama et~al.
\newblock {First M87 Event Horizon Telescope Results. V. Physical Origin of the
  Asymmetric Ring}.
\newblock {\em Astrophys. J. Lett.}, 875(1):L5, 2019.

\bibitem{Akiyama:2019eap}
Kazunori Akiyama et~al.
\newblock {First M87 Event Horizon Telescope Results. VI. The Shadow and Mass
  of the Central Black Hole}.
\newblock {\em Astrophys. J. Lett.}, 875(1):L6, 2019.

\bibitem{synge1960relativity}
John~Lighton Synge.
\newblock Relativity: the general theory.
\newblock 1960.

\bibitem{synge1966escape}
JL~Synge.
\newblock The escape of photons from gravitationally intense stars.
\newblock {\em Monthly Notices of the Royal Astronomical Society},
  131(3):463--466, 1966.

\bibitem{takahashi2005black}
Rohta Takahashi.
\newblock Black hole shadows of charged spinning black holes.
\newblock {\em Publications of the Astronomical Society of Japan},
  57(2):273--277, 2005.

\bibitem{amir2016shapes}
Muhammed Amir and Sushant~G Ghosh.
\newblock Shapes of rotating nonsingular black hole shadows.
\newblock {\em Physical Review D}, 94(2):024054, 2016.

\bibitem{amir2018shadows}
Muhammed Amir, Balendra~Pratap Singh, and Sushant~G Ghosh.
\newblock Shadows of rotating five-dimensional charged emcs black holes.
\newblock {\em The European Physical Journal C}, 78(5):1--15, 2018.

\bibitem{cunha2017fundamental}
Pedro~VP Cunha, Carlos~AR Herdeiro, and Eugen Radu.
\newblock Fundamental photon orbits: black hole shadows and spacetime
  instabilities.
\newblock {\em Physical Review D}, 96(2):024039, 2017.

\bibitem{gralla2019black}
Samuel~E Gralla, Daniel~E Holz, and Robert~M Wald.
\newblock Black hole shadows, photon rings, and lensing rings.
\newblock {\em Physical Review D}, 100(2):024018, 2019.

\bibitem{stuchlik2018light}
Zden{\v{e}}k Stuchl{\'\i}k, Daniel Charbul{\'a}k, and Jan Schee.
\newblock Light escape cones in local reference frames of kerr--de sitter black
  hole spacetimes and related black hole shadows.
\newblock {\em The European Physical Journal C}, 78(3):1--32, 2018.

\bibitem{moffat2015modified}
JW~Moffat.
\newblock Modified gravity black holes and their observable shadows.
\newblock {\em The European Physical Journal C}, 75(3):1--4, 2015.

\bibitem{guo2020innermost}
Minyong Guo and Peng-Cheng Li.
\newblock Innermost stable circular orbit and shadow of the 4 d
  einstein--gauss--bonnet black hole.
\newblock {\em The European Physical Journal C}, 80(6):1--8, 2020.

\bibitem{sharif2016shadow}
M~Sharif and Sehrish Iftikhar.
\newblock Shadow of a charged rotating non-commutative black hole.
\newblock {\em The European Physical Journal C}, 76(11):1--9, 2016.

\bibitem{zeng2020shadows}
Xiao-Xiong Zeng, Hai-Qing Zhang, and Hongbao Zhang.
\newblock Shadows and photon spheres with spherical accretions in the
  four-dimensional gauss--bonnet black hole.
\newblock {\em The European Physical Journal C}, 80(9):1--11, 2020.

\bibitem{Papnoi:2014aaa}
Uma Papnoi, Farruh Atamurotov, Sushant~G. Ghosh, and Bobomurat Ahmedov.
\newblock {Shadow of five-dimensional rotating Myers-Perry black hole}.
\newblock {\em Phys. Rev. D}, 90(2):024073, 2014.

\bibitem{Sharma:2021von}
Prateek Sharma, Hemwati Nandan, Uma Papnoi, and Arindam~Kumar Chatterjee.
\newblock {Optical and Thermodynamic Properties of a Rotating Dyonic Black Hole
  Spacetime in $\mathcal{N} = 2, U(1)^2$ gauged supergravity}.
\newblock 5 2021.

\bibitem{Kala:2020prt}
Shubham Kala, Saurabh, Hemwati Nandan, and Prateek Sharma.
\newblock {Deflection of light and shadow cast by a dual-charged stringy black
  hole}.
\newblock {\em Int. J. Mod. Phys. A}, 35(28):2050177, 2020.

\bibitem{Peng:2020wun}
Jun Peng, Minyong Guo, and Xing-Hui Feng.
\newblock {Influence of quantum correction on black hole shadows, photon rings,
  and lensing rings}.
\newblock {\em Chin. Phys. C}, 45(8):085103, 2021.

\bibitem{Wang:2021ara}
Mingzhi Wang, Songbai Chen, and Jiliang Jing.
\newblock {Kerr Black hole shadows in Melvin magnetic field with stable photon
  orbits}.
\newblock 4 2021.

\bibitem{Junior:2021dyw}
Haroldo C. D.~Lima Junior, Pedro V.~P. Cunha, Carlos A.~R. Herdeiro, and
  Lu\'\i{}s C.~B. Crispino.
\newblock {Shadows and lensing of black holes immersed in strong magnetic
  fields}.
\newblock {\em Phys. Rev. D}, 104(4):044018, 2021.

\bibitem{Hioki:2009na}
Kenta Hioki and Kei-ichi Maeda.
\newblock {Measurement of the Kerr Spin Parameter by Observation of a Compact
  Object's Shadow}.
\newblock {\em Phys. Rev. D}, 80:024042, 2009.

\bibitem{Amarilla:2010zq}
Leonardo Amarilla, Ernesto~F. Eiroa, and Gaston Giribet.
\newblock {Null geodesics and shadow of a rotating black hole in extended
  Chern-Simons modified gravity}.
\newblock {\em Phys. Rev. D}, 81:124045, 2010.

\bibitem{Kumar:2019pjp}
Rahul Kumar, Sushant~G. Ghosh, and Anzhong Wang.
\newblock {Shadow cast and deflection of light by charged rotating regular
  black holes}.
\newblock {\em Phys. Rev. D}, 100(12):124024, 2019.

\bibitem{khodadi2020black}
Mohsen Khodadi, Alireza Allahyari, Sunny Vagnozzi, and David~F Mota.
\newblock Black holes with scalar hair in light of the event horizon telescope.
\newblock {\em Journal of Cosmology and Astroparticle Physics}, 2020(09):026,
  2020.

\bibitem{contreras2019black}
Ernesto Contreras, JM~Ramirez-Velasquez, {\'A}ngel Rinc{\'o}n, Grigoris
  Panotopoulos, and Pedro Bargue{\~n}o.
\newblock Black hole shadow of a rotating polytropic black hole by the
  newman--janis algorithm without complexification.
\newblock {\em The European Physical Journal C}, 79(9):1--10, 2019.

\bibitem{contreras2021geodesic}
Ernesto Contreras, {\'A}ngel Rinc{\'o}n, Grigoris Panotopoulos, and Pedro
  Bargue{\~n}o.
\newblock Geodesic analysis and black hole shadows on a general non-extremal
  rotating black hole in five-dimensional gauged supergravity.
\newblock {\em Annals of Physics}, 432:168567, 2021.

\bibitem{wang2020shadow}
Mingzhi Wang, Songbai Chen, Jieci Wang, and Jiliang Jing.
\newblock Shadow of a schwarzschild black hole surrounded by a bach--weyl ring.
\newblock {\em The European Physical Journal C}, 80(2):1--11, 2020.

\bibitem{kumar2020black}
Rahul Kumar and Sushant~G Ghosh.
\newblock Black hole parameter estimation from its shadow.
\newblock {\em The Astrophysical Journal}, 892(2):78, 2020.

\bibitem{jusufi2021black}
Kimet Jusufi.
\newblock Black hole shadows in verlinde’s emergent gravity.
\newblock {\em Monthly Notices of the Royal Astronomical Society},
  503(1):1310--1318, 2021.

\bibitem{atamurotov2016horizon}
Farruh Atamurotov, Sushant~G Ghosh, and Bobomurat Ahmedov.
\newblock Horizon structure of rotating einstein--born--infeld black holes and
  shadow.
\newblock {\em The European Physical Journal C}, 76(5):1--16, 2016.

\bibitem{zhang2020optical}
He-Xu Zhang, Cong Li, Peng-Zhang He, Qi-Qi Fan, and Jian-Bo Deng.
\newblock Optical properties of a brane-world black hole as photons couple to
  the weyl tensor.
\newblock {\em The European Physical Journal C}, 80:1--11, 2020.

\bibitem{dastan2016shadow}
Sara Dastan, Reza Saffari, and Saheb Soroushfar.
\newblock Shadow of a kerr-sen dilaton-axion black hole.
\newblock {\em arXiv preprint arXiv:1610.09477}, 2016.

\bibitem{jha2021bumblebee}
Sohan~Kumar Jha and Anisur Rahaman.
\newblock Bumblebee gravity with a kerr--sen like solution and its shadow.
\newblock {\em The European Physical Journal C}, 81(4):1--14, 2021.

\bibitem{atamurotov2021axion}
Farruh Atamurotov, Kimet Jusufi, Mubasher Jamil, Ahmadjon Abdujabbarov, and
  Mustapha Azreg-A{\"\i}nou.
\newblock Axion-plasmon or magnetized plasma effect on an observable shadow and
  gravitational lensing of a schwarzschild black hole.
\newblock {\em Physical Review D}, 104(6):064053, 2021.

\bibitem{perlick2015influence}
Volker Perlick, Oleg~Yu Tsupko, and Gennady~S Bisnovatyi-Kogan.
\newblock Influence of a plasma on the shadow of a spherically symmetric black
  hole.
\newblock {\em Physical Review D}, 92(10):104031, 2015.

\bibitem{atamurotov2015optical}
Farruh Atamurotov, Bobomurat Ahmedov, and Ahmadjon Abdujabbarov.
\newblock Optical properties of black holes in the presence of a plasma: The
  shadow.
\newblock {\em Physical Review D}, 92(8):084005, 2015.

\bibitem{perlick2017light}
Volker Perlick and Oleg~Yu Tsupko.
\newblock Light propagation in a plasma on kerr spacetime: Separation of the
  hamilton-jacobi equation and calculation of the shadow.
\newblock {\em Physical Review D}, 95(10):104003, 2017.

\bibitem{yan2019influence}
Haopeng Yan.
\newblock Influence of a plasma on the observational signature of a high-spin
  kerr black hole.
\newblock {\em Physical Review D}, 99(8):084050, 2019.

\bibitem{ahmedov2019optical}
Bobomurat Ahmedov, Bobur Turimov, Zden{\v{e}}k Stuchl{\'\i}k, and Arman
  Tursunov.
\newblock Optical properties of magnetized black hole in plasma.
\newblock In {\em International Journal of Modern Physics: Conference Series},
  volume~49, page 1960018. World Scientific, 2019.

\bibitem{abdujabbarov2017shadow}
Ahmadjon Abdujabbarov, Bobir Toshmatov, Zden{\v{e}}k Stuchl{\'\i}k, and
  Bobomurat Ahmedov.
\newblock Shadow of the rotating black hole with quintessential energy in the
  presence of plasma.
\newblock {\em International Journal of Modern Physics D}, 26(06):1750051,
  2017.

\bibitem{babar2020optical}
Gulmina~Zaman Babar, Abdullah~Zaman Babar, and Farruh Atamurotov.
\newblock Optical properties of kerr--newman spacetime in the presence of
  plasma.
\newblock {\em The European Physical Journal C}, 80(8):1--10, 2020.

\bibitem{das2020shadow}
Anish Das, Ashis Saha, and Sunandan Gangopadhyay.
\newblock Shadow of charged black holes in gauss--bonnet gravity.
\newblock {\em The European Physical Journal C}, 80(3):1--15, 2020.

\bibitem{zaman2020optical}
Gulmina Zaman~Babar, Abdullah Zaman~Babar, and Farruh Atamurotov.
\newblock Optical properties of kerr-newman spacetime in the presence of
  plasma.
\newblock {\em arXiv e-prints}, pages arXiv--2008, 2020.

\bibitem{huang2018revisiting}
Yang Huang, Yi-Ping Dong, and Dao-Jun Liu.
\newblock Revisiting the shadow of a black hole in the presence of a plasma.
\newblock {\em International Journal of Modern Physics D}, 27(12):1850114,
  2018.

\bibitem{li2021gravitational}
Qiang Li and Towe Wang.
\newblock Gravitational effect of a plasma on the shadow of schwarzschild black
  holes.
\newblock {\em arXiv preprint arXiv:2102.00957}, 2021.

\bibitem{atamurotov2016observing}
Farruh Atamurotov.
\newblock Observing shadow of the schwarzschild black hole in presence of a
  plasma.
\newblock {\em Proceedings of the International Astronomical Union},
  12(S324):351--352, 2016.

\bibitem{liu2019probing}
Feng-Yuan Liu, Yi-Fan Mai, Wen-Yu Wu, and Yi~Xie.
\newblock Probing a regular non-minimal einstein-yang-mills black hole with
  gravitational lensings.
\newblock {\em Physics Letters B}, 795:475--481, 2019.

\bibitem{schneider1992gravitational}
Peter Schneider, J{\"u}rgen Ehlers, and Emilio~E Falco.
\newblock Gravitational lenses as astrophysical tools.
\newblock In {\em Gravitational Lenses}, pages 467--515. Springer, 1992.

\bibitem{schneider2006gravitational}
Peter Schneider, Christopher Kochanek, and Joachim Wambsganss.
\newblock {\em Gravitational lensing: strong, weak and micro: Saas-Fee advanced
  course 33}, volume~33.
\newblock Springer Science \& Business Media, 2006.

\bibitem{petters2012singularity}
Arlie~O Petters, Harold Levine, and Joachim Wambsganss.
\newblock {\em Singularity theory and gravitational lensing}, volume~21.
\newblock Springer Science \& Business Media, 2012.

\bibitem{keeton2005formalism}
Charles~R Keeton and AO~Petters.
\newblock Formalism for testing theories of gravity using lensing by compact
  objects: Static, spherically symmetric case.
\newblock {\em Physical Review D}, 72(10):104006, 2005.

\bibitem{collett2018precise}
Thomas~E Collett, Lindsay~J Oldham, Russell~J Smith, Matthew~W Auger, Kyle~B
  Westfall, David Bacon, Robert~C Nichol, Karen~L Masters, Kazuya Koyama, and
  Remco van~den Bosch.
\newblock A precise extragalactic test of general relativity.
\newblock {\em Science}, 360(6395):1342--1346, 2018.

\bibitem{cao2018weak}
Wei-Guang Cao and Yi~Xie.
\newblock Weak deflection gravitational lensing for photons coupled to weyl
  tensor in a schwarzschild black hole.
\newblock {\em The European Physical Journal C}, 78(3):1--18, 2018.

\bibitem{synge1965relativity}
John~Lighton Synge.
\newblock Relativity: the special theory.
\newblock 1965.

\bibitem{bicak1975general}
J~Bicak and P~Hadrava.
\newblock General-relativistic radiative transfer theory in refractive and
  dispersive media.
\newblock {\em Astronomy and Astrophysics}, 44:389--399, 1975.

\bibitem{bisnovatyi2010gravitational}
GS~Bisnovatyi-Kogan and O~Yu Tsupko.
\newblock Gravitational lensing in a non-uniform plasma.
\newblock {\em Monthly Notices of the Royal Astronomical Society},
  404(4):1790--1800, 2010.

\bibitem{bisnovatyi2017gravitational}
Gennady~S Bisnovatyi-Kogan and Oleg~Yu Tsupko.
\newblock Gravitational lensing in presence of plasma: strong lens systems,
  black hole lensing and shadow.
\newblock {\em Universe}, 3(3):57, 2017.

\bibitem{crisnejo2018weak}
Gabriel Crisnejo and Emanuel Gallo.
\newblock Weak lensing in a plasma medium and gravitational deflection of
  massive particles using the gauss-bonnet theorem. a unified treatment.
\newblock {\em Physical Review D}, 97(12):124016, 2018.

\bibitem{javed2020weak}
Wajiha Javed, Muhammad~Bilal Khadim, and Ali {\"O}vg{\"u}n.
\newblock Weak gravitational lensing by
  bocharova--bronnikov--melnikov--bekenstein black holes using gauss--bonnet
  theorem.
\newblock {\em The European Physical Journal Plus}, 135(7):1--6, 2020.

\bibitem{atamurotov2021weak}
Farruh Atamurotov, Ahmadjon Abdujabbarov, and Javlon Rayimbaev.
\newblock Weak gravitational lensing schwarzschild-mog black hole in plasma.
\newblock {\em The European Physical Journal C}, 81(2):1--10, 2021.

\bibitem{matsuno2021light}
Ken Matsuno.
\newblock Light deflection by squashed kaluza-klein black holes in a plasma
  medium.
\newblock {\em Physical Review D}, 103(4):044008, 2021.

\bibitem{babar2021gravitational}
Gulmina~Zaman Babar, Farruh Atamurotov, and Abdullah~Zaman Babar.
\newblock Gravitational lensing in 4-d einstein--gauss--bonnet gravity in the
  presence of plasma.
\newblock {\em Physics of the Dark Universe}, page 100798, 2021.

\bibitem{babar2021particle}
Gulmina~Zaman Babar, Farruh Atamurotov, Shafqat~Ul Islam, and Sushant~G Ghosh.
\newblock Particle acceleration around rotating einstein-born-infeld black hole
  and plasma effect on gravitational lensing.
\newblock {\em Physical Review D}, 103(8):084057, 2021.

\bibitem{jha2021optical}
Sohan~Kumar Jha, Sahazada Aziz, and Anisur Rahaman.
\newblock Optical properties of lorentz violating kerr-sen-like spacetime in
  the presence of plasma.
\newblock {\em arXiv preprint arXiv:2103.17021}, 2021.

\bibitem{tsupko2021deflection}
Oleg~Yu Tsupko.
\newblock Deflection of light rays by a spherically symmetric black hole in a
  dispersive medium.
\newblock {\em Physical Review D}, 103(10):104019, 2021.

\bibitem{rogers2015frequency}
Adam Rogers.
\newblock Frequency-dependent effects of gravitational lensing within plasma.
\newblock {\em Monthly Notices of the Royal Astronomical Society},
  451(1):17--25, 2015.

\bibitem{jin2020strong}
Xing-Hua Jin, Yuan-Xing Gao, and Dao-Jun Liu.
\newblock Strong gravitational lensing of a 4-dimensional
  einstein--gauss--bonnet black hole in homogeneous plasma.
\newblock {\em International Journal of Modern Physics D}, 29(09):2050065,
  2020.

\bibitem{fathi2020gravitational}
Mohsen Fathi and JR~Villanueva.
\newblock Gravitational lensing of a charged weyl black hole surrounded by
  plasma.
\newblock {\em arXiv e-prints}, pages arXiv--2009, 2020.

\bibitem{hensh2019gravitational}
Sudipta Hensh, Ahmadjon Abdujabbarov, Jan Schee, and Zden{\v{e}}k
  Stuchl{\'\i}k.
\newblock Gravitational lensing around kehagias--sfetsos compact objects
  surrounded by plasma.
\newblock {\em The European Physical Journal C}, 79(6):1--14, 2019.

\bibitem{kimpson2019spatial}
Tom Kimpson, Kinwah Wu, and Silvia Zane.
\newblock Spatial dispersion of light rays propagating through a plasma in kerr
  space--time.
\newblock {\em Monthly Notices of the Royal Astronomical Society},
  484(2):2411--2419, 2019.

\bibitem{rezzolla2014new}
Luciano Rezzolla and Alexander Zhidenko.
\newblock New parametrization for spherically symmetric black holes in metric
  theories of gravity.
\newblock {\em Physical Review D}, 90(8):084009, 2014.

\bibitem{sullivan2020numerical}
Andrew Sullivan, Nicol{\'a}s Yunes, and Thomas~P Sotiriou.
\newblock Numerical black hole solutions in modified gravity theories:
  Spherical symmetry case.
\newblock {\em Physical Review D}, 101(4):044024, 2020.

\bibitem{nashed2019charged}
Gamal~GL Nashed and Salvatore Capozziello.
\newblock Charged spherically symmetric black holes in f (r) gravity and their
  stability analysis.
\newblock {\em Physical Review D}, 99(10):104018, 2019.

\bibitem{aghmohammadi2010spherical}
A~Aghmohammadi, Kh~Saaidi, MR~Abolhassani, and A~Vajdi.
\newblock Spherical symmetric solution in f (r) model around charged black
  hole.
\newblock {\em International Journal of Theoretical Physics}, 49(4):709--716,
  2010.

\bibitem{carames2009spherically}
TRP Caram{\^e}s and ER~Bezerra de~Mello.
\newblock Spherically symmetric vacuum solutions of modified gravity theory in
  higher dimensions.
\newblock {\em The European Physical Journal C}, 64(1):113--121, 2009.

\bibitem{moffat2015black}
JW~Moffat.
\newblock Black holes in modified gravity (mog).
\newblock {\em The European Physical Journal C}, 75(4):1--9, 2015.

\bibitem{newman1965note}
Ezra~T Newman and AI~Janis.
\newblock Note on the kerr spinning-particle metric.
\newblock {\em Journal of Mathematical Physics}, 6(6):915--917, 1965.

\bibitem{azreg2014generating}
Mustapha Azreg-A{\"\i}nou.
\newblock Generating rotating regular black hole solutions without
  complexification.
\newblock {\em Physical Review D}, 90(6):064041, 2014.

\bibitem{azreg2014regular}
Mustapha Azreg-A{\"\i}nou.
\newblock Regular and conformal regular cores for static and rotating
  solutions.
\newblock {\em Physics Letters B}, 730:95--98, 2014.

\bibitem{azreg2014static}
Mustapha Azreg-Ainou.
\newblock From static to rotating to conformal static solutions: rotating
  imperfect fluid wormholes with (out) electric or magnetic field.
\newblock {\em The European Physical Journal C}, 74(5):1--11, 2014.

\bibitem{toshmatov2017rotating}
Bobir Toshmatov, Zden{\v{e}}k Stuchl{\'\i}k, and Bobomurat Ahmedov.
\newblock Rotating black hole solutions with quintessential energy.
\newblock {\em The European Physical Journal Plus}, 132(2):1--21, 2017.

\bibitem{xu2017kerr}
Zhaoyi Xu and Jiancheng Wang.
\newblock Kerr-newman-ads black hole in quintessential dark energy.
\newblock {\em Physical Review D}, 95(6):064015, 2017.

\bibitem{toshmatov2017comments}
Bobir Toshmatov, Zden{\v{e}}k Stuchl{\'\i}k, and Bobomurat Ahmedov.
\newblock Comments on “casimir effect in the kerr spacetime with
  quintessence”.
\newblock {\em Modern Physics Letters A}, 32(21):1775001, 2017.

\bibitem{shaikh2019black}
Rajibul Shaikh.
\newblock Black hole shadow in a general rotating spacetime obtained through
  newman-janis algorithm.
\newblock {\em Physical Review D}, 100(2):024028, 2019.

\bibitem{contreras2020black}
Ernesto Contreras, {\'A}ngel Rinc{\'o}n, Grigoris Panotopoulos, Pedro
  Bargue{\~n}o, and Benjamin Koch.
\newblock Black hole shadow of a rotating scale-dependent black hole.
\newblock {\em Physical Review D}, 101(6):064053, 2020.

\bibitem{kumar2020rotating}
Rahul Kumar and Sushant~G Ghosh.
\newblock Rotating black holes in the novel $4 d $ einstein-gauss-bonnet
  gravity.
\newblock {\em arXiv preprint arXiv:2003.08927}, 2020.

\bibitem{jusufi2020rotating}
Kimet Jusufi, Mubasher Jamil, Hrishikesh Chakrabarty, Qiang Wu, Cosimo Bambi,
  and Anzhong Wang.
\newblock Rotating regular black holes in conformal massive gravity.
\newblock {\em Physical Review D}, 101(4):044035, 2020.

\bibitem{benavides2020rotating}
Carlos~A Benavides-Gallego, Ahmadjon Abdujabbarov, and Cosimo Bambi.
\newblock Rotating and nonlinear magnetic-charged black hole surrounded by
  quintessence.
\newblock {\em Physical Review D}, 101(4):044038, 2020.

\bibitem{yang1954conservation}
Chen-Ning Yang and Robert~L Mills.
\newblock Conservation of isotopic spin and isotopic gauge invariance.
\newblock {\em Physical review}, 96(1):191, 1954.

\bibitem{bartnik1988particlelike}
Robert Bartnik and John McKinnon.
\newblock Particlelike solutions of the einstein-yang-mills equations.
\newblock {\em Physical Review Letters}, 61(2):141, 1988.

\bibitem{bjoraker2000monopoles}
Jeff Bjoraker and Yutaka Hosotani.
\newblock Monopoles, dyons, and black holes in the four-dimensional
  einstein-yang-mills theory.
\newblock {\em Physical Review D}, 62(4):043513, 2000.

\bibitem{jaffe2006quantum}
Arthur Jaffe and Edward Witten.
\newblock Quantum yang-mills theory.
\newblock {\em The millennium prize problems}, 1:129, 2006.

\bibitem{Altas:2021htf}
Emel Altas, Ercan Kilicarslan, and Bayram Tekin.
\newblock {Einstein\textendash{}Yang\textendash{}Mills theory: gauge invariant
  charges and linearization instability}.
\newblock {\em Eur. Phys. J. C}, 81(7):648, 2021.

\bibitem{Kleihaus:1996vk}
Burkhard Kleihaus, Jutta Kunz, and Abha Sood.
\newblock {Sequences of Einstein Yang-Mills dilaton black holes}.
\newblock {\em Phys. Rev. D}, 54:5070--5092, 1996.

\bibitem{Kleihaus:1997rb}
Burkhard Kleihaus, Jutta Kunz, and Abha Sood.
\newblock {Charged SU(N) Einstein Yang-Mills black holes}.
\newblock {\em Phys. Lett. B}, 418:284--293, 1998.

\bibitem{Kleihaus:2002ee}
Burkhard Kleihaus, Jutta Kunz, and Francisco Navarro-Lerida.
\newblock {Rotating Einstein-Yang-Mills black holes}.
\newblock {\em Phys. Rev. D}, 66:104001, 2002.

\bibitem{Nandan:2009kt}
Hemwati Nandan, Nils~M. Bezares-Roder, and Heinz Dehnen.
\newblock {Black Hole Solutions and Pressure Terms in Induced Gravity with
  Higgs Potential}.
\newblock {\em Class. Quant. Grav.}, 27:245003, 2010.

\bibitem{Bezares-Roder:2009lns}
Nils~M. Bezares-Roder, Hemwati Nandan, and Heinz Dehnen.
\newblock {Scalar-field Pressure in Induced Gravity with Higgs Potential and
  Dark Matter}.
\newblock {\em JHEP}, 10:113, 2010.

\bibitem{smoller1993existence}
JA~Smoller, AG~Wasserman, and Shing-Tung Yau.
\newblock Existence of black hole solutions for the einstein-yang/mills
  equations.
\newblock {\em Communications in mathematical physics}, 154(2):377--401, 1993.

\bibitem{volkov1997slowly}
Mikhail~S Volkov and Norbert Straumann.
\newblock Slowly rotating non-abelian black holes.
\newblock {\em Physical Review Letters}, 79(8):1428, 1997.

\bibitem{kleihaus2002rotating}
Burkhard Kleihaus, Jutta Kunz, and Francisco Navarro-L{\'e}rida.
\newblock Rotating einstein-yang-mills black holes.
\newblock {\em Physical Review D}, 66(10):104001, 2002.

\bibitem{kleihaus2011rotating}
Burkhard Kleihaus, Jutta Kunz, and Eugen Radu.
\newblock Rotating black holes in dilatonic einstein-gauss-bonnet theory.
\newblock {\em Physical review letters}, 106(15):151104, 2011.

\bibitem{ghosh2010radiating}
Sushant~G Ghosh and Naresh Dadhich.
\newblock Radiating black holes in einstein-yang-mills theory and cosmic
  censorship.
\newblock {\em Physical Review D}, 82(4):044038, 2010.

\bibitem{jusufi2021quasinormal}
Kimet Jusufi, Mustapha Azreg-A{\"\i}nou, Mubasher Jamil, Shao-Wen Wei, Qiang
  Wu, and Anzhong Wang.
\newblock Quasinormal modes, quasiperiodic oscillations, and the shadow of
  rotating regular black holes in nonminimally coupled einstein-yang-mills
  theory.
\newblock {\em Physical Review D}, 103(2):024013, 2021.

\bibitem{perlick2021calculating}
Volker Perlick and Oleg~Yu Tsupko.
\newblock Calculating black hole shadows: review of analytical studies.
\newblock {\em arXiv preprint arXiv:2105.07101}, 2021.

\end{thebibliography}

\end{document}